\documentclass[journal]{IEEEtran}
\usepackage{amsmath,amssymb,subfigure,epsfig,color,fmtcount,bm,amsthm,multirow,array}
\usepackage[noadjust]{cite}
\newcommand{\p}{\text{P}}
\newcommand{\E}{\text{E}}
\newcommand{\ed}{\stackrel{\text{d}}{=}}
\newcommand{\deq}{\triangleq}

\IEEEoverridecommandlockouts
\newtheorem{definition}{Definition}
\newtheorem{theorem}{\textbf{Theorem}}

\newtheorem{lemma}{Lemma}

\newtheorem{remark}{Remark}
\theoremstyle{plain}

\makeatletter
\def\ps@headings{%
\def\@oddhead{\mbox{}\scriptsize\rightmark \hfil \thepage}%
\def\@evenhead{\scriptsize\thepage \hfil \leftmark\mbox{}}%
\def\@oddfoot{}%
\def\@evenfoot{}}
\makeatother

\begin{document}

\title{On the Generalized Delay-Capacity Tradeoff of Mobile Networks with L\'{e}vy Flight Mobility}
\author{Yoora Kim$^\dag$, Kyunghan Lee$^\ddag$, Ness B. Shroff$^\dag$, Injong Rhee$^\ddag$, and Song Chong$^\sharp$\\
$^\dag$\{kimy, shroff\}@ece.osu.edu, $^\ddag$\{klee8, rhee\}@ncsu.edu, $^\sharp$songchong@kaist.edu
\thanks{This work was supported in part by NSF grants CNS-1065136, CNS-1012700, CNS-0910868 and CNS-1016216, and ARO MURI Award W911NF-08-1-0238.}
}
\maketitle

\begin{abstract}
In the literature, scaling laws for wireless mobile networks have been characterized under various models of node mobility and several assumptions on how communication occurs between nodes. To improve the realism in the analysis of scaling laws, we propose a new analytical framework. The framework is the first to consider a L\'{e}vy flight mobility pattern, which is known to closely mimic human mobility patterns. Also, this is the first work that allows nodes to communicate while being mobile. Under this framework, delays ($\bar{D}$) to obtain various levels of per-node throughput $(\lambda)$ for L\'{e}vy flight are suggested as $\bar{D}(\lambda) = O(\sqrt{\min (n^{1+\alpha} \lambda, n^2 ) })$, where L\'{e}vy flight is a random walk of a power-law flight distribution with an exponent $\alpha \in (0,2]$. The same framework presents a new tighter tradeoff $\bar{D}(\lambda) = O(\sqrt{ \max ( 1,n\lambda^3 ) })$ for \textit{i.i.d.} mobility, whose delays are lower than existing results for the same levels of per-node throughput.
\end{abstract}

\section{Introduction}
Since the work in~\cite{grossglauser:mobility} that showed that mobility can be exploited to improve network throughput, there has been a plethora of work on this subject. A major effort in this direction has been in the design of delay tolerant networks (DTNs). However, this benefit in throughput comes at a significant delay cost. The amount of delays required to achieve a level of throughput for various mobility models such as \textit{i.i.d.} mobility, random waypoint (RWP), random direction (RD), and Brownian motion (BM) have been extensively studied in~\cite{SL14Neely,SL16Neely,SL18Sharma,SL20Lin,SL22Gamal}. Specifically, the delay required for constant per-node throughput has been shown to grow as $\Theta(n)$, which scales as fast as the network size $n$, for most mobility models including \textit{i.i.d.} mobility, RWP, RD, and BM~\cite{SL17Toumpis, SL14Neely, SL20Lin, SL21Lin}. Despite significant advances in the development of delay-capacity scaling laws, there has been considerable skepticism regarding the applicability of the results to real mobile networks because of various simplifying assumptions used in the analysis.
%the significant benefit in throughput, the huge amount of delay has resulted a high barrier in realizing networking applications that exploit mobility.

\begin{figure}[t!]
\centering
\subfigure[Contacts while being mobile] {\epsfig{figure=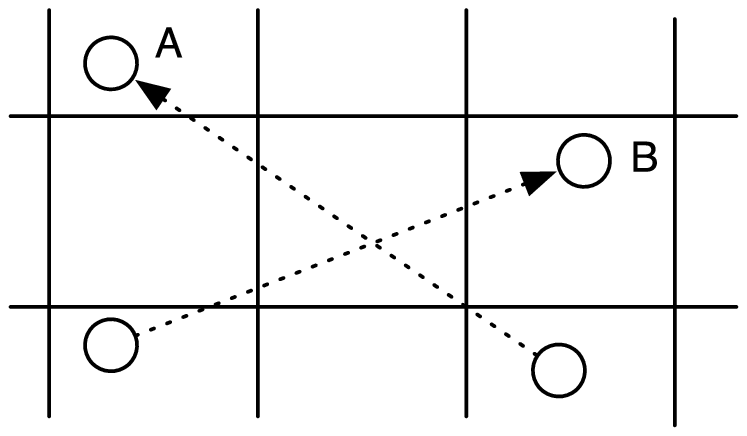,width=0.27\textwidth}}
\subfigure[A trajectory of L\'{e}vy flight] {\epsfig{figure=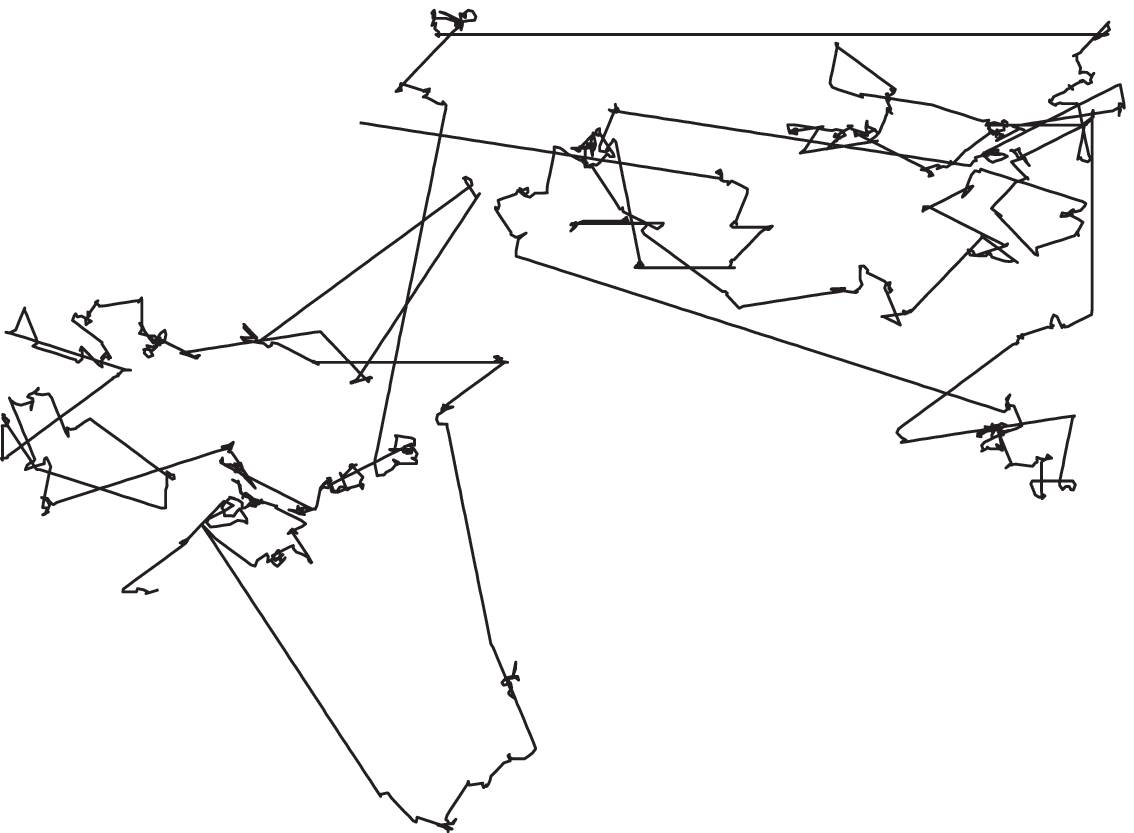, width=0.21\textwidth}}
\caption{Our considerations for improving the realism of delay-capacity tradeoff: (a) Conventionally, nodes $A$ and $B$ are assumed to have no contact opportunity, but we consider they can meet each other even during their movements. (b) We also consider that nodes follow L\'{e}vy flight mobility.}
\label{fig:assumptions}
\end{figure}

In this paper, we address two issues towards making the delay-capacity tradeoff analysis more realistic: 1) contacts among nodes in the middle of their movements and 2) L\'{e}vy mobility patterns of nodes in the network. In the literature, for mathematical simplicity, existing results have assumed that nodes show slotted movements, and they do not communicate with each other while being mobile. Thus, they make contacts with other nodes and transfer data only at the edge of time slots. In other words, as shown in Fig.~\ref{fig:assumptions} (a), the opportunity for meeting other nodes during mobility has been ignored, although such opportunities can substantially reduce packet delivery delays. Also, in this work we focus on the L\'{e}vy flight model, which is widely accepted to closely mimic the actual movement of humans~\cite{rhee:levymobility, lee:slaw09}. The trajectory of this model is illustrated in Fig.~\ref{fig:assumptions} (b). To enhance the realism in the analysis of delay-capacity tradeoff, we develop a new analytical framework which takes both of these factors into account by developing a technique that characterizes the distribution of ``first meeting time'' among nodes conforming to L\`{e}vy flight mobility in a two-dimensional space. It is important to note that the exact distribution of the first meeting time of L\'{e}vy flight even in a one-dimensional space has been an open problem even though it has applicability in a diverse set of research problems (e.g., characterization of particle movements and animal movements) in physics and mathematics. It is also informative to note that the distribution of the first meeting time of BM, which can be considered as an extreme case of L\'{e}vy flight, is also an open problem as noted in \cite{Sharma04scaling,Sharma04TR}.

 %In our framework, we handle a number of mathematical challenges using theories of stochastic geometry and probability.

%find that there is an assumption widely used for mathematical simplicity in the literature which leads to an inflated estimate of the delay, even in an order sense. The assumption is that the nodes do not communicate with each other while being mobile. In other words, contact opportunities in the middle of movements are not considered as shown in Fig.~\ref{fig:assumptions}. To capture such intermediate contacts, we develop a new framework that allows nodes to freely move around an area without discretized locations and meet other nodes even while being mobile.
%Under this framework, we study an upper bound of the delay-capacity tradeoff for L\'{e}vy flight mobility model and further identify the exact delay-capacity tradeoff for i.i.d. mobility model.
%Under this framework, we explicitly show that delays to obtain various levels of throughput in mobile networks are overestimated for the i.i.d. mobility model, and provide a new tighter delay-capacity tradeoff. Then, we further extend the framework to study a delay-capacity tradeoff for a more realistic mobility model.

%L\'{e}vy flight is a mobility model known to closely resemble the movement patterns of humans~\cite{rhee:levymobility, lee:slaw09} and %animals~\cite{viswanathan:levy, ramos-fernandez:levy} in reality.
%We adopt L\'{e}vy flight as a realistic mobility model, as it is known to be statistically similar to human movement patterns~\cite{rhee:levymobility, lee:slaw09}.

L\'{e}vy flight, the mobility model we focus on in this paper, is a subset of L\'{e}vy mobility in which a node moves from position to position in a constant time. Another special case, L\'{e}vy walk, in which a node moves from one position to another in time proportional to the distance between the positions.
Except for the notion of the time required for each movement, L\'{e}vy flight and L\'{e}vy walk are fundamentally the same random walk whose flight length distribution asymptotically follows a power-law $f_\alpha (z) \approx 1/z^{\alpha+1}$, where $z$ and $\alpha$ denote the flight length (i.e., moving distance of each slotted movement) and the power-law slope ranging $0 < \alpha < 2$, respectively. The heavy-tailed movements of L\'{e}vy mobility render the delay characterization extremely challenging. Our framework addresses these challenges using theories from stochastic geometry and probability, and provides a delay-capacity tradeoff for L\'{e}vy flight. Also, for a simpler \textit{i.i.d.} mobility model, we provide a tighter delay-capacity tradeoff compared to existing studies using the same framework.

\begin{figure}[t!]
\centering
\subfigure[L\'{e}vy flight] {\epsfig{figure=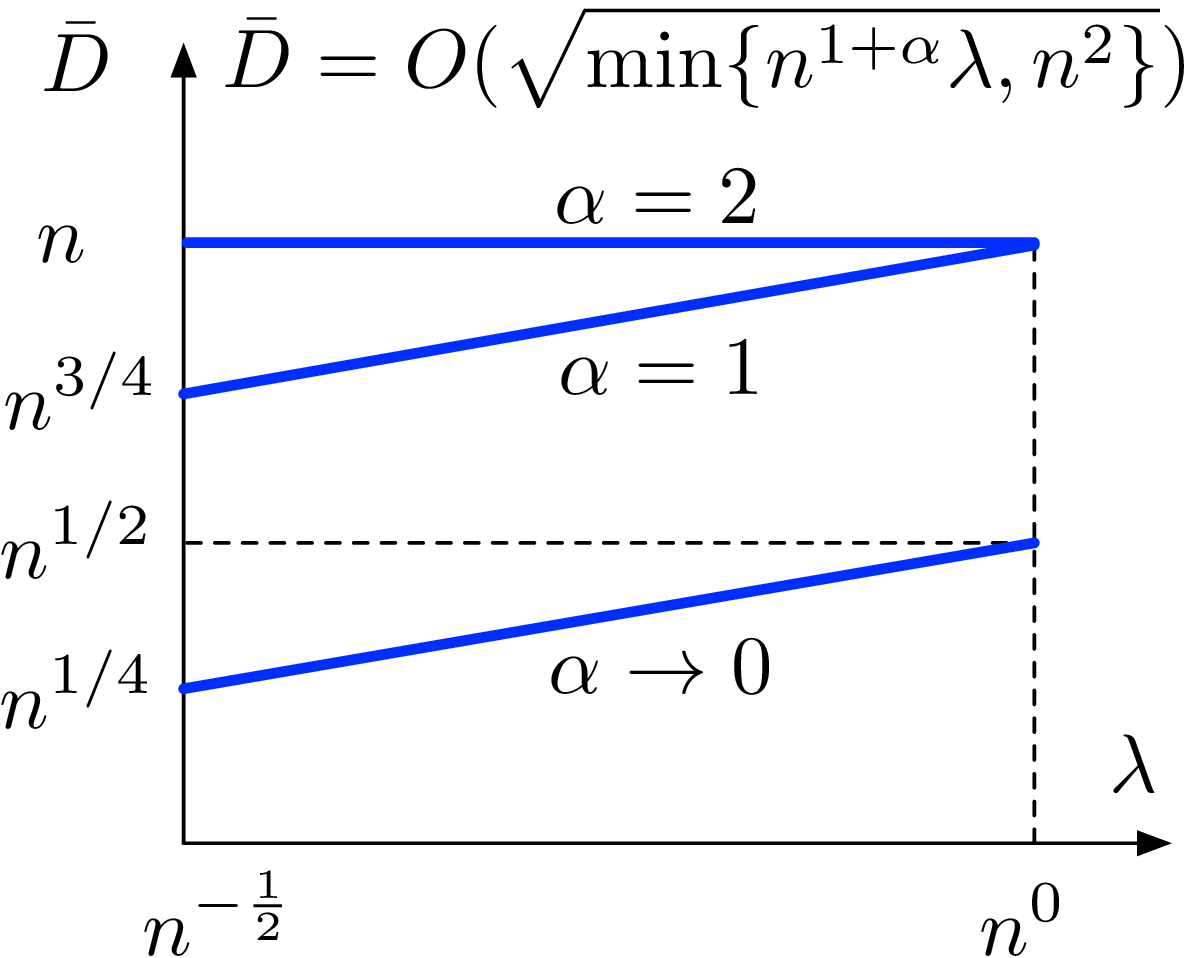,width=0.4\columnwidth}}
\subfigure[\textit{i.i.d.} mobility] {\epsfig{figure=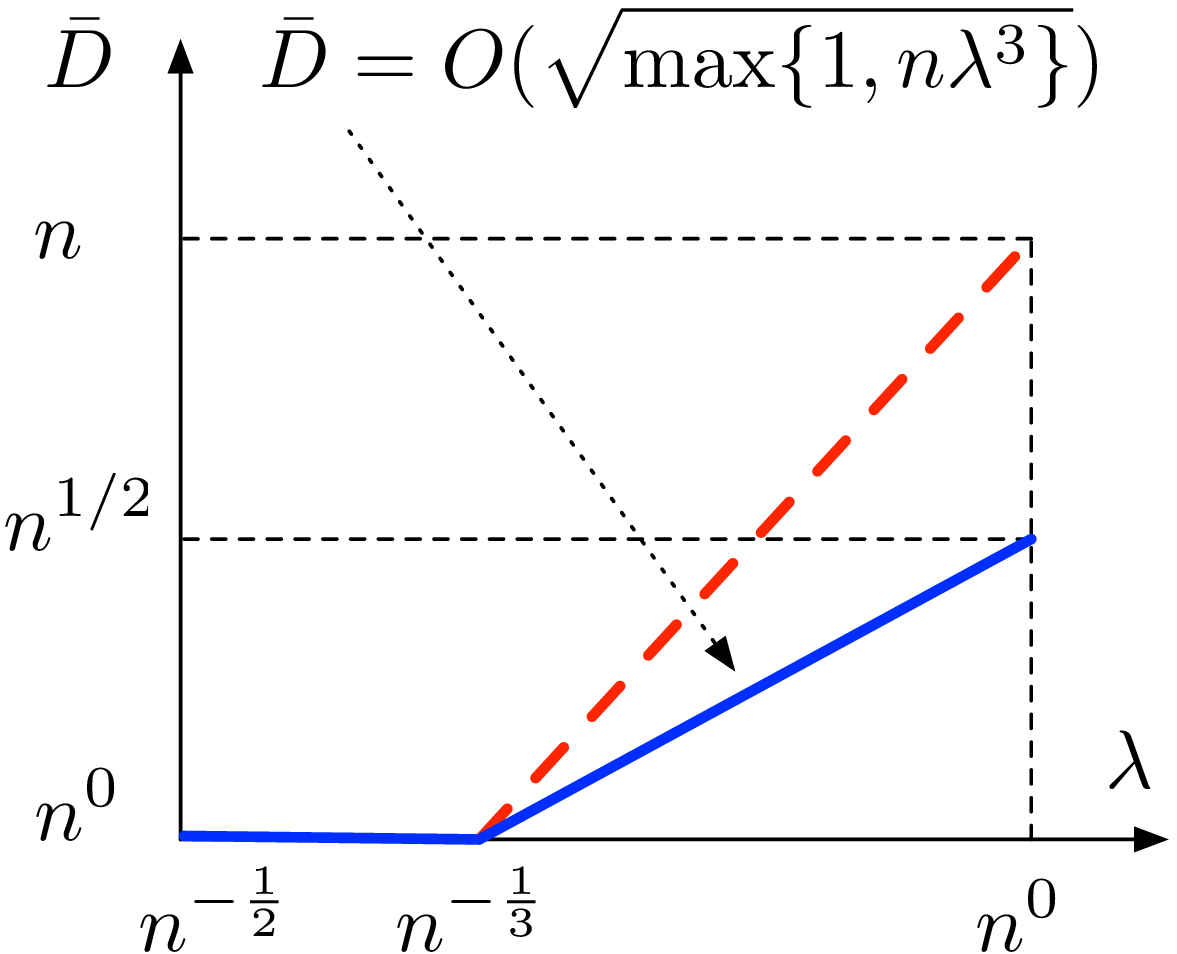,width=0.4\columnwidth}}
\caption{(a) The delay-capacity tradeoffs for L\'{e}vy flight with various $\alpha$ obtained from our analytical framework and (b) the new tradeoff
for \textit{i.i.d.} mobility (solid line) shown together with the previously known tradeoff (dotted line).}
\label{fig:tradeoffs}
\end{figure}

\begin{figure}[t!]
\centering
\subfigure[$\lambda = \Theta(1/\sqrt{n})$]{\epsfig{figure=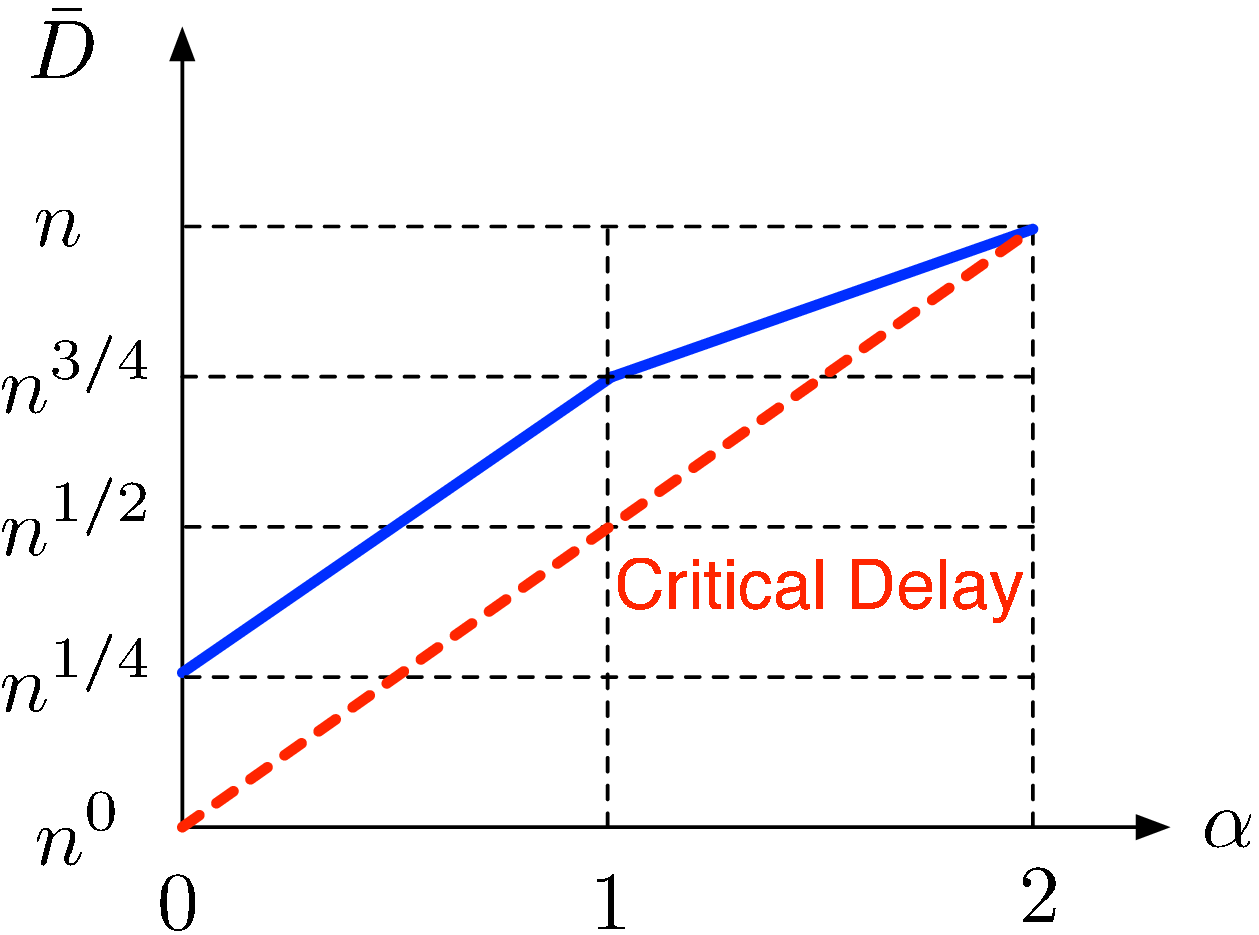,width=0.48\columnwidth}}
\subfigure[$\lambda = \Theta(1)$]{\epsfig{figure=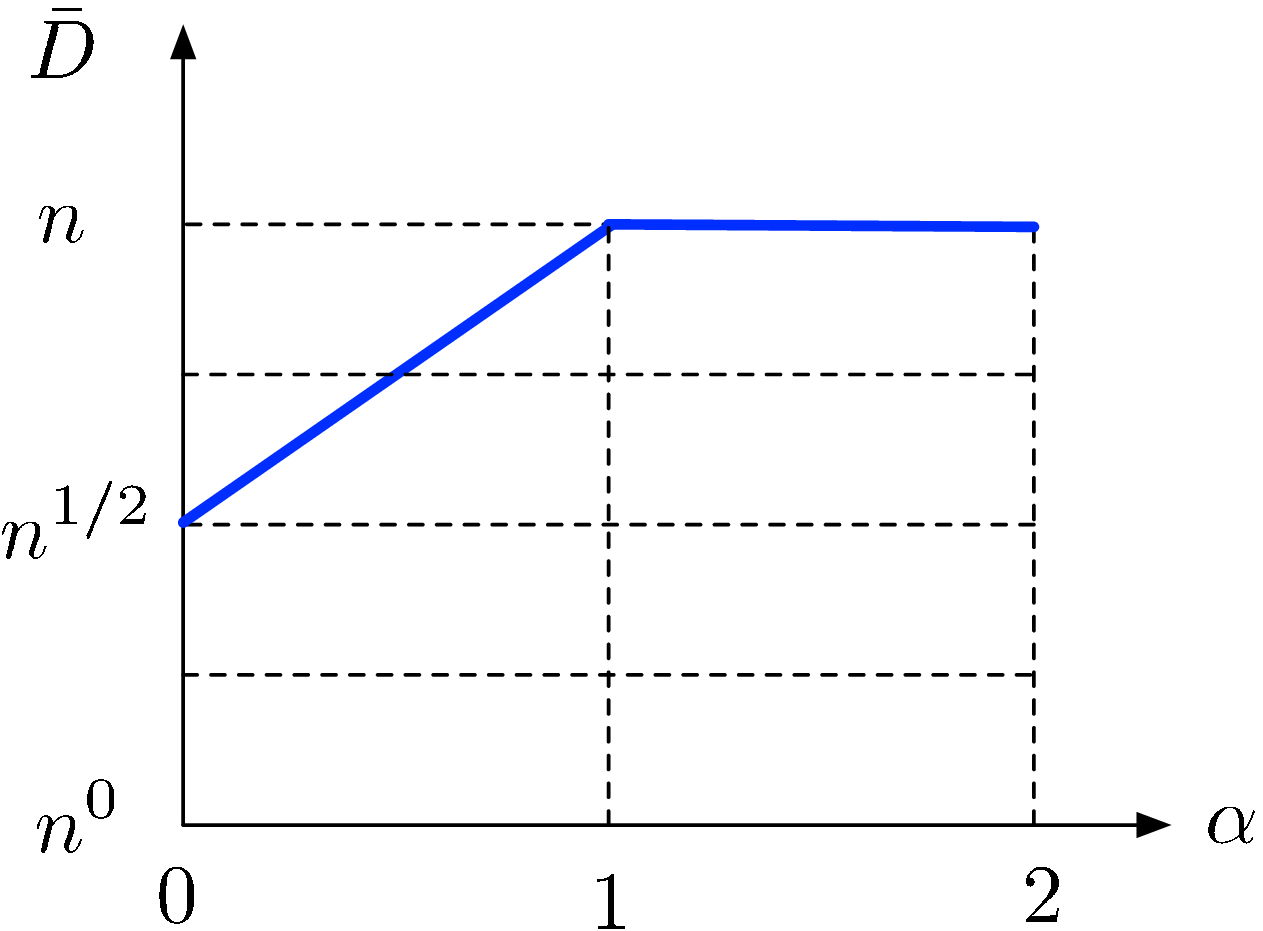,width=0.48\columnwidth}}
\caption{The upper bounds on $\bar{D}(\lambda)$ for L\'{e}vy flight with $\alpha \in (0,2]$ to obtain (a) $\lambda = \Theta(1/\sqrt{n})$ and (b) $\lambda = \Theta(1)$. The critical delay for L\'{e}vy flight derived in \cite{lee11scaling} is depicted in (a).}
\label{fig:levyalpha}
\end{figure}

Fig.~\ref{fig:tradeoffs} and Table~\ref{tab:main} summarize the new tradeoffs identified using our analytical framework. The results show that the tradeoff for L\'{e}vy flight follows $\bar{D}(\lambda) = O(\sqrt{\min\{n^{1+\alpha} \lambda, n^2\}})$ to obtain a per-node throughput of  $\lambda = \Theta(n^{-\eta}) \, (0\leq \eta \leq 1/2)$ as shown in Figs.~\ref{fig:levyalpha} (a) and (b). These results are well aligned with the critical delay (i.e., minimum delay required to achieve $\lambda = \omega (1/\sqrt{n})$) suggested in~\cite{lee11scaling}. Our tradeoffs show an important finding that the delay required to obtain constant per-node throughput (i.e., $\lambda = \Theta(1)$) can be smaller than $\Theta(n)$ in mobile networks with mobility models such as L\'{e}vy flight with $\alpha < 1$ and \textit{i.i.d.} mobility. This is an important observation given that most of the existing studies present the delay required to obtain constant per-node throughput to be $\Theta(n)$ for almost all mobility models including the \textit{i.i.d.} mobility.

Our tradeoff for L\'{e}vy flight becomes especially more interesting when we input $\alpha$ values from measurements presented in Table~\ref{tbl:realalpha} into the tradeoff. This gives us a hint on how the performance of the network will scale in reality when the network consists of devices mainly carried (or driven) by humans.  For $\alpha$ values between 0.53 and 1.81, the delays to obtain $\lambda = \Theta(1)$ are expected to lie between $O(n^{0.77})$ and $O(n)$. This implies that in reality, a DTN mainly operated by human mobility may indeed experience less than $\Theta(n)$ delay in some areas.  This observation of smaller delay suggests that mobile networks relying on opportunistic transmissions may have higher practical values in reality given that the delays have been overestimated by mobility and contact models with less realism.

%For i.i.d. mobility, the blue solid line showing the tighter delay-capacity tradeoff is significantly lower than that of the previously known model~\cite{SL20Lin} in the red dotted line. For L\'{e}vy flight, the tradeoffs when $\alpha \rightarrow 0$ and $\alpha = 1,2$ are exemplified. Note that when $\alpha = 2$, L\'{e}vy flight boils down to BM~\cite{lee11scaling} and our tradeoff becomes identical to that for BM identified in~\cite{SL21Lin}. It is important to note that the delay identified in our framework for constant per-node throughput is less than or equal to $\Theta(n)$ which has been reported in many literatures previously as the delay required to obtain constant per-node throughput for most of mobility models.

%\addtocounter{footnote}{1}
%\footnotetext[\value{footnote}]{$\lambda$ and $\bar{D}$ denote, respectively, per-node throughput and the optimal delay, whose definitions are given in Section~\ref{sec:model}.}

\begin{table}[h]
\caption{The new delay-capacity tradeoffs identified using our analytical framework for L\'{e}vy flight and \textit{i.i.d.} mobility}
  \begin{center}
    {%\scriptsize
  \begin{tabular}{|c|c|c|} \hline
    Mobility &  Tradeoff $\bar{D}(\lambda)$ & $\bar{D}(\lambda)$ for $\lambda= \Theta(1)$  \\ \hline \hline
    L\'{e}vy flight & $O(\sqrt{\min(n^{1+\alpha} \lambda, n^2)})$ & $ O(\sqrt{\min(n^{1+\alpha}, n^2)})$ \\\hline \hline
    \textit{i.i.d.}   &		$O(\sqrt{ \max (1,n\lambda^3) })$ & $O(n^{1/2})$ \\ \hline
  \end{tabular}
}
\end{center}
\label{tab:main}
\end{table}

\begin{table}[t!]

	\caption{Experimental $\alpha$ values from five sites presented in~\cite{rhee:levymobility}. These values are from experimental GPS traces from five sites with more than 100 participants in total.}
	\centering
	\begin{tabular}{|c|c|c|c|c|c|}
	\hline
	Site 		& $\alpha$ 	& Site 		& $\alpha$ & Site 		& $\alpha$ \\
	\hline\hline
	KAIST 		& 0.53  	& New York 	& 1.62 & State fair 	& 1.81 \\
	NCSU 		& 1.27  	& Disney World 		& 1.20  &  & \\
	\hline
	\end{tabular}
	\label{tbl:realalpha}
	\vspace{-0.2cm}
\end{table}

The rest of the paper is organized as follows. We overview a list of related work in Section~\ref{sec:related} and introduce our system models and definitions of performance metrics in Section~\ref{sec:model}. We then provide the intuition on how our analytical framework evaluates delay-capacity tradeoff using the properties of first meeting time of random walks in Section~\ref{sec:analysis}. Based on the understanding in Section~\ref{sec:analysis}, we analyze the tradeoffs of L\'{e}vy flight and \textit{i.i.d.} mobility models in Sections~\ref{sec:analysis:LF} and~\ref{sec:analysis:iid}, respectively. After briefly concluding our work in Section~\ref{sec:conclusion}, we provide full details of proofs used in Sections~\ref{sec:analysis:LF} and~\ref{sec:analysis:iid} through Appendices. 
\section{Related Work}\label{sec:related}
In \cite{SL1Gupta}, it was shown that the per-node throughput of random wireless networks with $n$ static nodes scales as $O(1/\sqrt{n})$. The result was later enhanced to $\Theta (1/\sqrt{n})$ by exercising individual power control \cite{SL5Ashish}. Grossglauser and Tse~\cite{grossglauser:mobility} proved that constant per-node throughput is achievable by using mobility when nodes follow ergodic and stationary mobility models. This contradicted the conventional belief that node mobility negatively impacts network capacity due to interruptions in connectivity. %It is later shown that the gain comes at the cost of larger delay~\cite{SL8Bansal}.

Many follow-up studies \cite{SL8Bansal, SL12Perevalov,SL13Tsingos,SL14Neely,SL17Toumpis,SL18Sharma,SL20Lin}
have been devoted to characterize and exploit the delay-capacity tradeoffs. In particular, the delay required to obtain constant per-node throughput has been studied under various mobility models \cite{SL14Neely,SL16Neely,SL18Sharma,SL20Lin,SL22Gamal}. The key message is that the delay of 2-hop relaying proposed in \cite{grossglauser:mobility} is $\Theta(n)$ for most  mobility models such as \textit{i.i.d.} mobility, RD, RWP, and BM.

The delay-capacity tradeoff for per-node throughput $\lambda = \Theta(n^{-\eta})\,(0\le\eta\le1/2)$ is first presented in \cite{SL19Sharma} as $\bar{D}(\lambda) = O(n \lambda)$ for RWP model. In~\cite{SL20Lin}, the authors identified that $\bar{D}(\lambda) = \Theta(  \max ( 1,n\lambda^3 ) )$ holds for \textit{i.i.d.} mobility. Later, \cite{SL21Lin} showed that $\bar{D}(\lambda) = \Theta(n)$ holds for BM irrespective of $\lambda$.

More realistic mobility models, L\'{e}vy mobility models known to closely capture human movement patterns, were first analyzed in~\cite{lee11scaling} for a special case of the tradeoff. Using spatio-temporal probability density functions, the critical delay defined by the minimum delay required to achieve larger throughput than $\Theta(1/\sqrt{n})$ is identified for L\'{e}vy flight as well as L\'{e}vy walk.

Existing results on delay-capacity tradeoffs for mobile networks have been built under the assumption that nodes are able to communicate with each other only at the edge of time slots for slotted movements. Also, there has been no framework which can fully understand the delay-capacity tradeoff for L\'{e}vy mobility. In this paper, we develop an analytical framework which handles both of these issues and are able to use this framework to characterize the delay-capacity tradeoff.

%Unfortunately, existing results on delay-capacity tradeoffs of mobile networks have been built on the assumption that nodes are able to communicate with each other only at the edge of time slots for slotted movements. Also, there has been no framework that can fully understand the delay-capacity tradeoff under L\'{e}vy mobility. In this paper, we develop an analytical framework which handles both of these issues and are able to use this framework to characterize the delay capacity tradeoff.

%To obtain more realistic delay-capacity tradeoff of a network in which nodes are driven by humans (e.g., mobile social networks, pocket-switched networks), we design a framework handling the \textit{first meeting time} of L\'{e}vy mobility with contacts during movements and first present the delay-capacity tradeoff under L\'{e}vy flight.

%Unfortunately the framework proposed in \cite{lee11scaling} is designed to handle such a special case and not able to answer the full scale of delay-capacity tradeoff of a mobile network with L\'{e}vy mobility. Our paper is the first of its kind addressing the delay-capacity tradeoff under L\'{e}vy flight mobility. 
\section{System Model}\label{sec:model}

\subsection{Network Model}\label{sec:model:network model}
We consider a wireless mobile network indexed by $n\in\mathbb{N}\deq\{1,2,\ldots\}$, where, in the $n$th network, $n$~nodes move on a completely wrapped-around disc $\bm{\mathcal{D}}\,(\subset \mathbb{R}^2)$ whose radius scales as $\Theta(\sqrt{n})$.\footnote{In all notations, a bold font symbol is used to denote a value or a set of values in $\mathbb{R}^2$.} Without loss of generality, we set the radius and the center of the disc~$\bm{\mathcal{D}}$ as $\sqrt{n}$ and $\bm{0}\deq(0,0)$, respectively, i.e., $\bm{\mathcal{D}} = \{\bm{x}\in\mathbb{R}^2\,|\, |\bm{x}| \le \sqrt{n}\}$. We assume that the density of the network is fixed to~1 as $n$ increases.\footnote{This model is often referred to as an extended network model. In another model, called a unit network model, the network area is fixed to 1 and the density increases as $n$ while the spacing and velocity of nodes scale as $1/\sqrt{n}$.} We also assume that all nodes are homogeneous in that each node generates data with the same intensity to its own destination. The packet generation process at each node is independent of node mobility. The generated packets are assumed to have no expiration until their delivery and the size of each node's buffer is assumed to be unlimited. Each packet can be delivered by either direct one-hop transmission or over multiple hops using relay nodes.

To model interference in wireless networks, we adopt the following protocol model as in \cite{SL22Gamal,SL1Gupta}. Let $\bm{X}_i(t)$ denote the location of node $i\,(=1,\ldots,n)$ at time $t\,(\ge 0)$. Let $L_{(i,j)}(t) \deq |\bm{X}_{i}(t)-\bm{X}_j(t)|$ denote the Euclidean distance between nodes~$i$ and $j$ at time $t$. Under the protocol model, nodes transmit packets successfully at a constant rate $W$ bits/sec, if and only if the following is satisfied: for a transmitter $i$, a receiver $j$ and every other node $u \neq i,j$ transmitting simultaneously,
\begin{align*}
L_{(u,j)}(t) &\geq (1+\Delta) \, L_{(i,j)}(t),
\end{align*}
where $\Delta$ is some positive constant. In addition, the distance between nodes~$i$ and $j$ at time $t$ should satisfy $L_{(i,j)}(t) \le r$, where $r\,(>0)$ denotes the maximum communication range. We assume the fluid packet model~\cite{SL22Gamal}, which allows concurrent transmissions of node pairs (with the rate divided by the number of pairs) interfering each other. We denote by~$\Pi$ the class of all feasible scheduling schemes conforming our descriptions.

\subsection{Mobility Model}\label{sec:model:mobility model}
In this subsection, we mathematically describe the L\'{e}vy flight model and the \textit{i.i.d.} mobility model. At time $t=0$, node~$i$ chooses its location uniformly on the disc~$\bm{\mathcal{D}}$ (i.e., $\bm{X}_i(0)\sim\text{Uniform}(\bm{\mathcal{D}})$), which is independent of the others $\bm{X}_j(0)$ for $j \neq i$. We assume that time is divided into slots of unit length and is indexed by $k\in\mathbb{N}$. At the beginning of the $k$th slot (i.e., at time $t = k-1$), node $i$ chooses its next slotted location $\bm{X}_i(k)$ according to the associated mobility model. During the $k$th slot (i.e., during time $t \in (k-1, k]$), node~$i$ moves from $\bm{X}_i(k-1)$ to $\bm{X}_i(k)$ with a constant velocity. Thus, $\bm{X}_i(k-1+\delta)$ for $\delta \in (0,1)$ is determined by $\bm{X}_i(k-1)$ and $\bm{X}_i(k)$ as follows:
\begin{align}\label{eqn:model:LF:location any time}
\bm{X}_i(k-1+\delta) &= (1-\delta)\bm{X}_i(k-1) + \delta\bm{X}_i(k).
\end{align}

\smallskip
\noindent{\bf \em L\'{e}vy Flight Model.} At the beginning of the $k$th slot (i.e., at time $t = k-1$), node $i$ chooses flight angle and flight length, denoted by $\theta_i(k)\,(\in(0,2\pi])$ and $Z_i(k)\,(>0)$, respectively. During the $k$th slot (i.e., during time $t\in(k-1,k]$), node~$i$ moves from $\bm{X}_i(k-1)$ to the selected direction $\theta_i(k)$ of the distance $Z_i(k)$. Thus, the location $\bm{X}_i(k)$ is determined as
\begin{align}\label{eqn:model:mobility model:LF}
\bm{X}_i(k) &= \bm{X}_i(k-1) + \bm{V}_i(k),
\end{align}
where
\begin{align}\label{eqn:model:mobility model:LF:flight vector}
\bm{V}_i(k) \deq \big(Z_i(k)\cos\theta_i(k), Z_i(k)\sin\theta_i(k)\big).
\end{align}
The flight angle $\theta_i(k)$ and the flight length $Z_i(k)$ are independent of each other and also independent of the previous locations $\bm{X}_i(t)$ for the times $t \in [0, k-1]$ before they are generated. Hence, $\bm{V}_i(k)$ is independent of $\bm{X}_{i}(t)$ for all $t \in [0, k-1]$.

Each flight angle $\theta_i(k)$ and flight length $Z_i(k)$ are independent and identically distributed across node index~$i$ and slot index~$k$. Let $\theta$ and $Z$ be a generic random variable for $\theta_i(k)$ and $Z_i(k)$, respectively. Then, the flight angle $\theta$ is uniformly distributed over $(0,2\pi]$, and the flight length $Z$ is generated from a random variable $Z^\star$ having the L\'{e}vy $\alpha$-stable distribution~\cite{nolan:2012} by the relation $Z = |Z^\star|$. The probability density function of $Z^\star$ is given by
\begin{align}\label{eqn:model:mobility model:LF:exact_pdf}
f_{Z^\star}(z) = \frac{1}{2\pi} \int_{-\infty}^{\infty} e^{-izt}\varphi_{Z^\star}(t) \, \text{d}t,
\end{align}
where $\varphi_{Z^\star}(t) \deq \E[e^{itZ^\star}]$ is the characteristic function of~$Z^\star$ and is given by $\varphi_{Z^\star}(t)= e^{-|st|^\alpha}$. Here, $|s|> 0$ is a scale factor determining the width of the distribution, and $\alpha \in (0,2]$ is a distribution parameter that specifies the shape (i.e., heavytail-ness) of the distribution. The flight length~$Z$ for $\alpha \in (0,1)$ has infinite mean and variance, while $Z$ for $\alpha \in [1,2)$ has finite mean but infinite variance. For $\alpha= 2$, the L\'{e}vy $\alpha$-stable distribution reduces to a Gaussian distribution with a mean of zero and variance of $2s^2$, for which the flight length $Z$ has finite mean and variance.

Due to the complex form of the distribution, the L\'{e}vy $\alpha$-stable distribution for $\alpha \in (0,2)$ is often treated as a power-law type of asymptotic form:
\begin{align}\label{eqn:model:mobility model:LF:app_pdf}
f_{Z^\star}(z) \sim \frac{1}{|z|^{1+\alpha}},
\end{align}
where we use the notation $a(z)\sim b(z)$ for any two real functions $a(z)$ and $b(z)$ to denote $\lim_{z\to\infty}[a(z)/b(z)]=1$~\cite{nolan:2012}. The form (\ref{eqn:model:mobility model:LF:app_pdf}) is known to closely approximate the tail part of the distribution in~(\ref{eqn:model:mobility model:LF:exact_pdf}), and a number of papers in mathematics and physics, e.g.,~\cite{RW01Drysdale,PhysRevE.73.057102}, analyze L\'{e}vy mobility using the form~(\ref{eqn:model:mobility model:LF:app_pdf}). For mathematical tractability, in our analysis we will also use the asymptotic form~(\ref{eqn:model:mobility model:LF:app_pdf}). Specifically, we assume that there exist constants $c\,(>0)$ and $z_{\text{th}}\,(>0)$ such that
\begin{align}\label{eqn:model:power law ccdf}
\p\{Z>z\}
&= \frac{c}{z^\alpha}, \quad \text{ for all } z \ge z_{\text{th}}.
\end{align}

\smallskip
\noindent{\bf \em \textit{i.i.d.} Mobility Model.} At the beginning of the $k$th slot (i.e., at time $t = k-1$), node~$i$ chooses $\bm{X}_i(k)$ uniformly on the disc~$\bm{\mathcal{D}}$, which is independent of its previous locations $\bm{X}_i(t)$ for the times $t \in [0,k-1]$ as well as the others $\bm{X}_j(k)$ for $j \neq i$. Thus, $\bm{X}_i(k)$ is independent and identically distributed across node index~$i$ and slot index~$k$.

\subsection{Contact Model}\label{sec:model:contact model}
In our contact model, nodes are allowed to meet while being mobile. Hence, for a time $t^\star$ in a domain $\{t\,|\,t \ge 0\}$, we say that \emph{nodes $i$ and $j$ meet at time $t^\star$} (or are \emph{in contact at time~$t^\star$}) if they satisfy
\begin{align*}
L_{(i,j)}(t^\star)\le r.
\end{align*}

In the widely adopted contact model where nodes are allowed to meet only at the end of their movements (i.e., at slot boundaries), a meeting event can occur for a time $k^\star$ in a domain $\{k\,|\,k\in\{0\} \cup \mathbb{N}\}$ satisfying $L_{(i,j)}(k^\star) \le r$. We call this class of contact model \emph{slotted contact model} throughout the paper.

Mobile nodes are exposed to more contact opportunities in our contact model compared to the slotted contact model.

\subsection{Performance Metrics}\label{sec:model:performance metrics}
The key performance metrics of our interest are per-node throughput and average delay as defined next:

\begin{definition}[Per-node throughput] Let $\Lambda_{\pi:i}(t)$ be the total number of bits received at the destination node $i$ up to time~$t$ under a scheduling scheme $\pi\in\Pi$. Let $\lambda_\pi$ be the per-node throughput under $\pi$. Then,
\begin{align*}
\lambda_{\pi} &\deq \liminf_{t \rightarrow \infty} \frac{1}{n} \sum_{i=1}^{n} \frac{\Lambda_{\pi:i} (t)}{t}.
\end{align*}
\end{definition}

\begin{definition}[Average delay] Let $D_{\pi:i,v}$ be the time taken for the $v$th packet generated from the source node $i$ to arrive at its destination node under a scheduling scheme $\pi\in\Pi$. Let $\bar{D}_\pi$ be the average delay under $\pi$. Then,
\begin{align*}
\bar{D}_{\pi} &\deq \lim_{w \rightarrow \infty} \frac{1}{n} \sum_{i=1}^{n} \frac{1}{w} \sum_{v=1}^{w} D_{\pi:i,v}.
\end{align*}
\end{definition}

 %In this paper, we are interested in analyzing the scaling property of the minimum average delay achieving per-node throughput $\lambda$ ranging from $\Theta(1/\sqrt{n})$ to $\Theta(1)$.
In this paper, we focus on analyzing the scaling property of the smallest average delay achieving per-node throughput $\lambda$.
We call this minimum average delay \emph{optimal delay} throughout this paper.
We focus on the throughput only in the range from $\Theta(1/\sqrt{n})$ to $\Theta(1)$, since this range corresponds to the case where mobility can be used to improve the per-node throughput.

\begin{definition}[Optimal delay] Let $\bar{D}(\lambda)$ be the optimal delay to achieve per-node throughput $\lambda$. It is then given by
\begin{align*}
\bar{D}(\lambda) &\deq \inf_{\{\pi\in\Pi \,|\, \lambda_\pi = \lambda\}} \bar{D}_{\pi}.
\end{align*}
\end{definition}

\subsection{Throughput Achieving Scheme}\label{sec:model:scheme}
We now consider a scheme~$\hat{\pi}$ that can realize per-node throughput~$\lambda_{\hat{\pi}}$ scaling from $\Theta(1/\sqrt{n})$ to $\Theta(1)$. The scheme~$\hat{\pi}$ operates as follows:
\begin{itemize}
\item When a packet is generated from a source node and the destination of the packet is within the communication range of the source node, the packet is transmitted to the destination node immediately.
\item Otherwise, the source node broadcasts the packet to all neighboring nodes within its communication range. Note that this broadcast is only performed by the source node when the packet is generated.
\item Any nodes carrying the packet can deliver the packet to the destination node when they are within the communication range of the destination node.
\item When one of the packets (including the original packet and the duplicated ones) reaches the destination node, all others are not considered for delivery.
\end{itemize}
By appropriately scaling $r$ as a function of $n$, the scheme $\hat{\pi}$ can achieve per-node throughput $\lambda_{\hat{\pi}}$ ranging from $\Theta(1/\sqrt{n})$ to $\Theta(1)$, as shown in the following lemma.

\begin{lemma}\label{lemma:capacity scaling} Let the communication range $r$ scale as $\Theta(n^{\beta})$ $(0 \le \beta \le 1/4)$. Then, the per-node throughput $\lambda_{\hat{\pi}}$ under the scheme $\hat{\pi}$ scales as $\Theta(n^{-2\beta})$.
\end{lemma}

\noindent\textit{Proof:}
If the network has been running for a long enough time, all nodes become to work as relay nodes and begin to have packets for all other nodes. Therefore, for a network with $n/a_n$ disjoint area where $a_n = \Theta(r^2)$, all areas with more than two nodes can always be activated. Let $b_n$ denote the probability of having more than two nodes in an area. We then have
\begin{align*}
b_n &= 1-\left(1-\frac{a_n}{n}\right)^n - n\frac{a_n}{n}\left(1-\frac{a_n}{n}\right)^{n-1}.
\end{align*}
In addition, the total network throughput becomes $b_n n / a_n$ and accordingly the per-node throughput is $\lambda_{\hat{\pi}} = b_n/a_n$. Without loss of generality, we assume $r = n^{\beta}\,(0 \le \beta \le 1/4)$. Then, the per-node throughput $\lambda_{\hat{\pi}}$ is given by
\begin{align*}
\lambda_{\hat{\pi}} &= n^{-2\beta}- n^{-2\beta} \left(1-n^{2\beta-1}\right)^n - \left(1-n^{2\beta-1}\right)^{n-1}.
\end{align*}
In the following, we will show that
\begin{subequations}
\begin{align}
& \lim_{n\to\infty}\frac{\lambda_{\hat{\pi}}}{n^{-2\beta}} \nonumber \\
& \quad = \lim_{n\to\infty} \big\{1 \!-\! \left(1\!-\!n^{2\beta\!-\!1}\right)^n \!-\! n^{2\beta} \left(1\!-\!n^{2\beta\!-\!1}\right)^{n\!-\!1} \big\} \label{eqn:model:scaling(3)} \\
& \quad = \begin{cases}
1 - 2\exp(-1), & \text{ if } \beta = 0,\label{eqn:model:scaling(4)} \\
1, & \text{ if } \beta \in (0, 1/4].
\end{cases}
\end{align}
\end{subequations}
Hence, the per-node throughput $\lambda_{\hat{\pi}}$ under the scheme $\hat{\pi}$ scales as $\Theta(n^{-2\beta})$.
Note that
\begin{align*}
\lim_{n\to\infty} \log (1-n^{2\beta-1})^n
&= \lim_{n\to\infty} n^{2\beta} \log (1-n^{2\beta-1})^{n^{1-2\beta}} \\
&= \lim_{n\to\infty} n^{2\beta} \log (\exp(-1)) \\
&= \begin{cases}
-1, & \text{ if } \beta = 0, \\
-\infty, & \text{ if } \beta \in (0, 1/4],
\end{cases}
\end{align*}
which gives
\begin{align}\label{eqn:model:scaling(1)}
\lim_{n\to\infty} (1-n^{2\beta-1})^n
&= \begin{cases}
\exp(-1), & \text{ if } \beta = 0, \\
0, & \text{ if } \beta \in (0, 1/4].
\end{cases}
\end{align}
Similarly,
\begin{align*}
&\lim_{n\to\infty} \log \big(n^{2\beta}(1-n^{2\beta-1})^{n-1}\big) \\
&= \lim_{n\to\infty} \big\{2\beta\log n  + (n-1) n^{2\beta-1} \log (1-n^{2\beta-1})^{n^{1-2\beta}}\big\} \\
&= \lim_{n\to\infty} \big\{2\beta\log n  + n^{2\beta} \log (\exp(-1)) \big\}\\
&= \begin{cases}
-1, & \text{ if } \beta = 0, \\
-\infty, & \text{ if } \beta \in (0, 1/4],
\end{cases}
\end{align*}
which gives
\begin{align}\label{eqn:model:scaling(2)}
\lim_{n\to\infty} \!\! n^{2\beta}(1\!-\!n^{2\beta\!-\!1})^{n\!-\!1}
&= \begin{cases}
\exp(-1), & \text{ if } \beta = 0, \\
0, & \text{ if } \beta \in (0, 1/4].
\end{cases}
\end{align}
By applying (\ref{eqn:model:scaling(1)}) and (\ref{eqn:model:scaling(2)}) to (\ref{eqn:model:scaling(3)}), we have (\ref{eqn:model:scaling(4)}). This complete the proof.
 \hfill $\blacksquare$

\smallskip
Let $\bar{D}_{\hat{\pi}}(\beta)$ be the average delay under $\hat{\pi}$ when $r = \Theta(n^{\beta})$. Lemma~\ref{lemma:capacity scaling} implies that by setting $\beta = -\log_n(\sqrt{\lambda})$, the scheme~$\hat{\pi}$ achieves the per-node throughput $\lambda_{\hat{\pi}} = \lambda$. Since the scheme $\hat{\pi}$ is of the class $\Pi$, the order of $\bar{D}(\lambda)$ can be obtained from $\bar{D}_{\hat{\pi}}(\beta)$ with the use of $\beta = -\log_n(\sqrt{\lambda})$, i.e.,
\begin{align}\label{eqn:model:logic}
\bar{D}(\lambda) &=\inf_{\{\pi\in\Pi \,|\, \lambda_\pi = \lambda\}} \bar{D}_{\pi}\le \bar{D}_{\hat{\pi}}(-\log_n(\sqrt{\lambda})).
\end{align}

\section{Preliminaries}\label{sec:analysis}
In this section, we provide the key intuition to understand how our analytical framework utilizes the properties of \textit{first meeting time} in the derivation of delay-capacity tradeoffs under the L\'{e}vy flight and the \emph{i.i.d.} mobility models. We then sketch the challenges residing in our framework and briefly describe our approach to address these challenges.

\subsection{Delay Analysis with First Meeting Time}
The first meeting time of two nodes moving in a two-dimensional space, which is directly connected to $\bar{D}_{\hat{\pi}}$, is defined below:
\begin{definition}[First meeting time]\label{def:first meeting time} For $i \neq j$, the first meeting time of nodes~$i$ and~$j$, denoted by $T_{(i,j)}$, is defined as
\begin{align*}
T_{(i,j)} &\deq \inf \big\{t \ge 0 \,\big|\, L_{(i,j)}(t) \le r \big\}.
\end{align*}
Since $T_{(i,j)}$ is independent and identically distributed across pair index $(i,j)$, we use~$T$ to denote a generic random variable for $T_{(i,j)}$.
\end{definition}

Let $D_{(s,d)}$ be a random variable representing the time taken by a packet generated from a source node~$s$ to arrive at a destination node~$d$. Since the packet generation process is independent of node mobility, we consider that each packet is generated at time $t=0$ without loss of generality. Then, the packet delay under the scheme~$\hat{\pi}$, denoted by $D_{\hat{\pi}:(s,d)}$, can be expressed in terms of the first meeting time as
\begin{align*}
\begin{split}
D_{\hat{\pi}:(s,d)} &=
\begin{cases}
0, & \text{if } d \in \mathcal{I}(s), \\
\min \big(T_{(i,d)}; i \in \mathcal{I}(s)\big), & \text{if } d \notin \mathcal{I}(s),
\end{cases}
\end{split}
\end{align*}
where $\mathcal{I}(s) \deq \{i \,|\, L_{(s,i)}(0) \le r\}$ denotes a set of node indices that are within the communication range of the node~$s$ at time $t=0$. Note that $s\in\mathcal{I}(s)$ by definition. Hence, the following equation represents the scheme $\hat{\pi}$ described in Section~\ref{sec:model}:
\begin{align*}
D_{\hat{\pi}:(s,d)} &=
\min \big(T_{(i,d)}; i \in \mathcal{I}(s)\setminus\{d\}\big).
\end{align*}
From Definition~2, the average delay $\bar{D}_{\hat{\pi}}$ can be obtained by
\begin{align}\label{eqn:analysis:LF:avg delay under hat pi}
\bar{D}_{\hat{\pi}}
&= \E\big[D_{\hat{\pi}:(s,d)}\big] \nonumber \\
&= \E \big[\min \big(T_{(i,d)}; i \in \mathcal{I}(s)\setminus\{d\} \big) \big].
\end{align}

\subsection{Distribution of the First Meeting Time}
In order to evaluate (\ref{eqn:analysis:LF:avg delay under hat pi}), the distribution of the first meeting time $\p\{T_{(i,d)} < \tau\}$ is essential. To obtain the distribution, we start from defining the following: let $I_{(i,j)}(k)\,(i \neq j, k\in\mathbb{N})$ be a random variable indicating the occurrence of a meeting event between nodes~$i$ and~$j$ during the $k$th slot (i.e., time $t\in(k-1,k]$), i.e.,
\begin{align*}
I_{(i,j)}(k) &\deq
\begin{cases}
0, & \text{if } L_{(i,j)}(t) > r~\text{for all } t \in (k-1,k], \\
1, & \text{if } L_{(i,j)}(t) \le r~\text{for some } t \in (k-1,k].
\end{cases}
\end{align*}
For notational simplicity, throughout this paper, we omit $(i,j)$ in $I_{(i,j)}(\cdot)$ and $L_{(i,j)}(\cdot)$, unless there is confusion. We then define a function $H(k, l_0)$ for $k\in\mathbb{N}$ and $l_0\in(r,2\sqrt{n}]$, which denotes the probability that nodes~$i$ and~$j$ are not in contact during the $k$th slot, conditioned on the fact that the initial distance between the nodes was $l_0$ and after that the nodes have not been in contact by time $t = k-1$, i.e.,
\begin{align}\label{eqn:definition of H}
\begin{split}
H(k,l_0)
&\deq
\begin{cases}
\p\{I(1) = 0 \,|\, L(0) = l_0\}, &\text{if } k = 1, \\
\p\{I(k) = 0 \,|\, I(k-1) =  \ldots \\
\hspace{4 mm}  = I(1) = 0, L(0) = l_0\}, &\text{if } k = 2,3,\cdots.
\end{cases}
\end{split}
\end{align}
Note that $l_0$ is upper bounded by $2\sqrt{n}$ since the radius of the disc $\bm{\mathcal{D}}$ is set to $\sqrt{n}$.

We find that the distribution of the first meeting time $T$ can be obtained from the function $H(k,l_0)$ as shown in the following lemma.

\begin{lemma}\label{lemma:analysis:CCDF of T}
For $\tau = 0, 1,2,\ldots$, the complementary cumulative distribution function (CCDF) of the first meeting time $T$ can be obtained by
\begin{align}\label{eqn:ccdf in lemma}
\p\{T > \tau \}
= \int_{r^+}^{2\sqrt{n}} \Big(\prod_{k=1}^{\tau} H(k,l_0)\Big)\, \text{d}F_{L(0)} (l_0),
\end{align}
where $F_{L(0)}(\cdot)$ denotes the cumulative distribution function (CDF) of $L(0)$, and we use the convention $\prod_{k=1}^{0}(\cdot) \deq 1$.
\end{lemma}

\noindent\textit{Proof:} For $\tau=0$, the event $\{T > 0\}$ implies the event $\{L(0) >r\}$, and vice versa. Hence, we have
\begin{align*}
\p\{T > 0 \} = \p\{L(0) >r\} = \int_{r^+}^{2\sqrt{n}} 1\, \text{d}F_{L(0)} (l_0).
\end{align*}
For $\tau=1,2,\ldots$, the CCDF $\p \{T > \tau \}$ can be obtained by
\begin{align*}
&\p\{T  > \tau\} \nonumber \\
&= \p\{L(0) > r, I(1) = \ldots = I(\tau) = 0\} \nonumber \\
&= \int_{r^+}^{2\!\sqrt{n}} \!\!\p\{I(1) \!=\! \ldots = I(\tau) \,=\, 0| L(0) \!=\! l_0\}\, \text{d}F_{L(0)} (l_0) \nonumber \\
&= \int_{r^+}^{2\sqrt{n}} \Big(\prod_{k=1}^{\tau} H(k,l_0)\Big)\, \text{d}F_{L(0)} (l_0),
\end{align*}
which completes the proof. \hfill $\blacksquare$

\smallskip
The identity $\p\{T > 0\} = \p\{L(0) >r\}$ shown in Lemma~\ref{lemma:analysis:CCDF of T} has the following implications for $\p\{T > 0\}$: (i) It is determined by the spatial distribution of nodes at time $t=0$ (which is assumed to be uniform on the disc~$\bm{\mathcal{D}}$). Hence, it is invariant for both the L\'{e}vy flight and the \emph{i.i.d.} mobility models. (ii) It represents the probability that two arbitrary nodes are out of the communication range at time $t=0$. Since $\p\{T>0\}$ is frequently used throughput this paper, we define it as $P_{o}$ and summarize its implications using the following lemma.

\begin{lemma}\label{lemma:analysis:CCDF of T(0)} Suppose that at time $t=0$, nodes are distributed uniformly on a disc of radius $\sqrt{n}$. Define
\begin{align}\label{eqn:analysis:def:P_out}
P_{o} \deq \p\{T > 0\} \, (= \p\{L(0) >r\}).
\end{align}
Then, $P_{o}$ is bounded by
\begin{align*}
1-\frac{r^2}{n} \le P_{o} \le 1-\frac{r^2}{3n}.
\end{align*}
\end{lemma}

\noindent\textit{Proof:} See Appendix~A. \hfill $\blacksquare$

\subsection{Technical Challenge and Approach}
In our framework, characterizing the function $H(k,l_0)$ in~(\ref{eqn:definition of H}) which appears in the expression for $\p\{T>\tau\}$ in~(\ref{eqn:ccdf in lemma}) is the key to analyze the optimal delay. The major technical challenge arises from tracking meeting events in the middle of a time slot. The meetings over time are heavily correlated irrespective of mobility models. The correlation can be understood as follows: let us consider two consecutive slots, say the $k$th and the $(k+1)$th slots, for ease of explanation. The occurrence of a meeting event during the $k$th slot (resp. the $(k+1)$th slot) is determined by the locations of nodes $i$ and $j$ at the slot boundaries, i.e., at times $t=k-1,k$ (resp. at times $t=k,k+1$). Hence, both $I(k)$ and $I(k+1)$ depend on the values of $\bm{X}_i(k)$ and $\bm{X}_j(k)$, and accordingly the sequence $\{I(k)\}_{k\in\mathbb{N}}$ is \emph{correlated} in our contact model. Due to the complexity involved in the correlation, deriving the exact form of $H(k,l_0)$ appears to be mathematically intractable. To address this challenge, we take a detour to derive a bound on $H(k,l_0)$ using theories from stochastic geometry and probability. The detailed analysis of $H(k,l_0)$ for the L\'{e}vy flight model and the \textit{i.i.d.} mobility model is presented in Lemmas~\ref{lemma:analysis:LF:property H(k,l)} and \ref{lemma:analysis:iid:property H(k,l)}, respectively, which allow us to reach the conclusions of this paper.

\section{Delay Analysis for the L\'{e}vy Flight Model}\label{sec:analysis:LF}
In this section, we analyze the optimal delay under the L\'{e}vy flight model. We use the following four steps in our analysis:
\begin{itemize}
\item In Step~1, the average delay under our L\'{e}vy flight model is formulated explicitly using the distribution of the first meeting time~$T$.
\item In Step~2, we derive a bound on the distribution of~$T$ by characterizing the function $H(k,l_0)$ under the L\'{e}vy flight model. The difficulty of handling contacts while being mobile is addressed in this step.
\item In Step~3, we connect the result of Step 2 to the delay scaling under the L\'{e}vy flight model.
\item In Step~4, we study the delay-capacity tradeoff by combining the capacity scaling in Lemma~\ref{lemma:capacity scaling} and the delay scaling obtained in Step~3.
\end{itemize}

%==========================================
%
%  Step 1
%
%==========================================
\smallskip\smallskip
\noindent{\bf \em Step 1 (Formulation of the average delay using the first meeting time distribution):} As shown in~(\ref{eqn:analysis:LF:avg delay under hat pi}), $\bar{D}_{\hat{\pi}} = \E [\min (T_{(i,d)}; i \in \mathcal{I}(s)\setminus\{d\})]$. Under the L\'{e}vy flight model, $T_{(i,d)}$ for $i \in \mathcal{I}(s)$ are heavily correlated since the next slotted location $\bm{X}_i(k+1)$ depends on the current location $\bm{X}_i(k)$. Note that all the nodes $i\in I(s)$ are in proximity of the node~$s$, and thus $\min (T_{(i,d)}; i \in \mathcal{I}(s)\setminus\{d\})$ is not easily tractable. Therefore, we use the following bound to describe $\bar{D}_{\hat{\pi}}$ using~$T$.
\begin{align}\label{eqn:analysis:LF:avg delay under hat pi bound}
\bar{D}_{\hat{\pi}} \le \E \big[T_{(s,d)}\big] = \E\big[T\big].
\end{align}
For the simpler \textit{i.i.d.} mobility model, we are able to derive a tighter bound on $\min (T_{(i,d)}; i \in \mathcal{I}(s)\setminus\{d\})$. We present the result in Step 1 of Section~\ref{sec:analysis:iid}.

Let $\lceil T \rceil$ denote the smallest integer greater than or equal to~$T$. Then, since $\E[T] \le \E[\lceil T \rceil] \le \E[T]+1$, the order of $\E[\lceil T \rceil]$ is the same as that of $\E[T]$, and $\E[\lceil T \rceil]$ is an upper bound on $\bar{D}_{\hat{\pi}}$, as shown in the following lemma.

\begin{lemma}\label{lemma:analysis:LF:opt delay bound} The average delay $\bar{D}_{\hat{\pi}}$ of the scheme $\hat{\pi}$ under the L\'{e}vy flight model is bounded by
\begin{align}\label{eqn:analysis:LF:avg delay bound}
\bar{D}_{\hat{\pi}} \le \E\big[\lceil T \rceil\big],
\end{align}
where $T$ is the generic random variable for the first meeting time $T_{(i,j)}$ defined in~Definition~\ref{def:first meeting time}. The expectation $\E[\lceil T \rceil]$ can be obtained from the distribution of~$T$ by
\begin{align*}
\E\big[\lceil T \rceil\big] = \sum_{\tau=0}^{\infty} \p\{T > \tau \}.
\end{align*}
\end{lemma}

\noindent\textit{Proof:} From~(\ref{eqn:analysis:LF:avg delay under hat pi bound}), we have $\bar{D}_{\hat{\pi}} \le \E[T].$ Since $T \le \lceil T \rceil$, we have $\E[T] \le \E[\lceil T \rceil]$, which gives~(\ref{eqn:analysis:LF:avg delay bound}). Since the random variable $\lceil T \rceil$ takes on only nonnegative integer values, the expectation $\E[\lceil T \rceil]$ can be obtained by
\begin{align*}
\E\big[\lceil T \rceil\big] = \sum_{\hat\tau=1}^{\infty} \p \big\{\lceil T \rceil \ge \hat\tau \big\} = \sum_{\hat\tau=1}^{\infty} \p \big\{T> \hat\tau-1 \big\},
\end{align*}
where the second equality comes from the property that $\p \{\lceil T \rceil \ge \hat\tau \} = \p\{T > \hat\tau-1\}$ for all $\hat{\tau} = 1,2,\ldots$. Replacing $\hat\tau-1$ with~$\tau$ gives the lemma. \hfill $\blacksquare$

%==========================================
%
%  Step 2
%
%==========================================
\smallskip\smallskip
\noindent{\bf \em Step 2 (Characterization of the first meeting time distribution):} In this step, we first analyze the characteristics of the function $H(k,l_0)$ under the L\'{e}vy flight model (See Lemma~\ref{lemma:analysis:LF:property H(k,l)}). By exploiting the characteristics, we then derive a bound on the first meeting time distribution (See Lemma~\ref{lemma:analysis:LF:ccdf}). This bound enables us to derive a formula for the expectation $\E[\lceil T \rceil]$ used in Lemma~\ref{lemma:analysis:LF:opt delay bound} (See Lemma~\ref{lemma:analysis:LF:bound for U}).

\begin{figure}[t!]
\centering
{\epsfig{figure=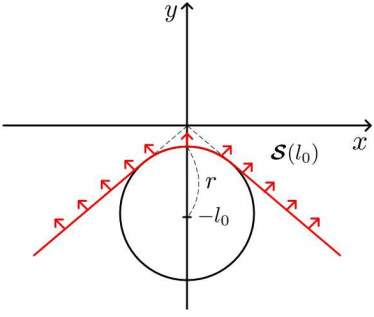,width=0.27\textwidth}}
\caption{The illustration of $\bm{\mathcal{S}}(l_0)$.}
\label{fig:example_S}
\end{figure}

\begin{lemma}\label{lemma:analysis:LF:property H(k,l)} Under the L\'{e}vy flight model, the function $H(k,l_0)$ in~(\ref{eqn:definition of H}) has the following characteristics: \\
(i) Let $\Delta\bm{V}$ be a generic random variable for $\bm{V}_i(k)-\bm{V}_j(k)$ representing a flight differential between nodes~$i$ and $j$.\footnote{The existence of the generic random variable for $\bm{V}_i(k)-\bm{V}_j(k)$ is proven in Lemma~\ref{lemma:appendix:LF:property of delta V} in Appendix~B.} Then, geometrically the function $H(1,l_0)$ can be viewed as the probability of the flight differential falling into a set $\bm{\mathcal{S}}(l_0)\,(\subset\mathbb{R}^2)$ defined as follows. Let $\bm{\mathcal{D}}_{r}(\bm{u})\,(\subset \mathbb{R}^2)$ denote a disc of radius~$r$ centered at $\bm{u} \in\mathbb{R}^2$, i.e., $\bm{\mathcal{D}}_{r}(\bm{u}) \deq \{\bm{x}\in\mathbb{R}^2\,|\, |\bm{x}-\bm{u}| \le r\}$. Let $\overline{(\bm{v},\bm{w})}$ denote a line connecting two points $\bm{v},\bm{w}\in\mathbb{R}^2$. For a fixed $l_0\in(r,2\sqrt{n}]$, define a set $\bm{\mathcal{S}}(l_0)$ as
\begin{align}\label{eqn:analysis:LF:definition S(l)}
\bm{\mathcal{S}}(l_0) &\deq \big\{\bm{x}\in\mathbb{R}^2 \,\big|\, \overline{(\bm{0}, \bm{x})} \cap \bm{\mathcal{D}}_{r}((0,-l_0)) = \varnothing\big\}.
\end{align}
An example of $\bm{\mathcal{S}}(l_0)$ is shown in Fig.~\ref{fig:example_S}. The set $\bm{\mathcal{S}}(l_0)$ has a connection with the function $H(1,l_0)$ as follows:
\begin{align*}
H(1,l_0) &= \p\big\{\Delta\bm{V} \in \bm{\mathcal{S}}(l_0)\big\}.
\end{align*}
(ii) The function $H(1,\cdot)$ is nondecreasing. \\
(iii) From (ii), we have $H(1,l_0) \le \hat{P}$ for all $l_0 \in (r, 2\sqrt{n}]$,
where
\begin{align}\label{eqn:analysis:LF:definition of P^o}
\hat{P} \deq H(1,2\sqrt{n}).
\end{align}
(iv) For $k=2,3,\ldots$, each function $H(k, l_0)$ is also bounded above by $\hat{P}$. Thus, for all $k\in\mathbb{N}$ and $l_0 \in (r, 2\sqrt{n}]$, we have
\begin{align*}
H(k,l_0) &\le \hat{P}.
\end{align*}
(v) There exist constants $c_l, c_u\,(>0)$, and $n_{\text{th}}\in\mathbb{N}$ such that for all $n \ge n_{\text{th}}$, $\hat{P}$ is bounded above and below by
\begin{subequations}
\begin{align}
\hat{P} &\le 1- \frac{2c_l}{\pi}  \left(\frac{1}{2\sqrt{n} + r}\right)^\alpha \sin^{-1}\left(\frac{r}{2\sqrt{n}}\right), \label{eqn:analysis:LF:property P^o-2} \\
\hat{P} &\ge 1- \frac{2^{\alpha/2+2}c_u}{\pi}  \left(\frac{1}{2\sqrt{n} - r}\right)^\alpha \sin^{-1}\left(\frac{r}{2\sqrt{n}}\right). \label{eqn:analysis:LF:property P^o-1}
\end{align}
\end{subequations}
\end{lemma}

\noindent\textit{Proof:} See Appendix~B. \hfill $\blacksquare$

\smallskip
Based on the formula for $\p\{T > \tau \}$ in Lemma~\ref{lemma:analysis:CCDF of T} and (iv) in Lemma~\ref{lemma:analysis:LF:property H(k,l)}, we derive a bound on $\p\{T > \tau \}$ in terms of $\hat{P}$ as follows: for $\tau=1,2,\ldots$, we have from (iv) in Lemma~\ref{lemma:analysis:LF:property H(k,l)} that
\begin{align}\label{eqn:analysis:LF:bound on hat H c}
0 \le \prod_{k=1}^{\tau}H(k,l_0) \le (\hat{P})^{\tau}.
\end{align}
Since~(\ref{eqn:analysis:LF:bound on hat H c}) holds for all $l_0 \in (r, 2\sqrt{n}]$, by integrating~(\ref{eqn:analysis:LF:bound on hat H c}) over $l_0 \in (r, 2\sqrt{n}]$, we have
\begin{align}\label{eqn:analysis:LF:geo}
&\int_{r^+}^{2\sqrt{n}} \Big(\prod_{k=1}^{\tau}H(k,l_0)\Big)\, \text{d}F_{L(0)} (l_0) \nonumber \\
& \quad \le \int_{r^+}^{2\sqrt{n}} (\hat{P})^{\tau} \, \text{d}F_{L(0)} (l_0) = (\hat{P})^{\tau}P_{o}.
\end{align}
By combining (\ref{eqn:analysis:LF:geo}) and Lemma~\ref{lemma:analysis:CCDF of T}, we have
\begin{align}\label{eqn:analysis:LF:ccdf}
\p\{T > \tau \} &\le (\hat{P})^{\tau}P_{o}, \quad \text{ for } \tau = 1,2,\ldots.
\end{align}
Since $\p\{T > \tau \} = P_{o}$ for $\tau = 0$, the bound in (\ref{eqn:analysis:LF:ccdf}) also holds for $\tau = 0$. The above result is summarized in Lemma~\ref{lemma:analysis:LF:ccdf}.

\begin{lemma}\label{lemma:analysis:LF:ccdf} Under the L\'{e}vy flight model, the CCDF of the first meeting time~$T$ is bounded by
\begin{align*}
\p\{T > \tau \} &\le (\hat{P})^{\tau}P_{o}, \quad \text{ for } \tau = 0,1,\ldots,
\end{align*}
where $\hat{P}$ and $P_{o}$ are defined in (\ref{eqn:analysis:LF:definition of P^o}) and (\ref{eqn:analysis:def:P_out}), respectively.
\end{lemma}

\noindent\textit{Proof:} Combining Lemma~\ref{lemma:analysis:CCDF of T} and (iv) in Lemma~\ref{lemma:analysis:LF:property H(k,l)} gives Lemma~\ref{lemma:analysis:LF:ccdf}. The detailed derivation was described earlier in~(\ref{eqn:analysis:LF:bound on hat H c})-(\ref{eqn:analysis:LF:ccdf}). \hfill $\blacksquare$

\begin{lemma}\label{lemma:analysis:LF:bound for U} The expectation $\E[\lceil T \rceil]$ under the L\'{e}vy flight model is bounded by
\begin{align*}
\E\big[\lceil T \rceil\big] \le \frac{P_{o}}{1-\hat{P}}.
\end{align*}
\end{lemma}

\noindent\textit{Proof:} Using Lemma~\ref{lemma:analysis:LF:ccdf}, we can give a bound on $\E[\lceil T \rceil]$ in Lemma~\ref{lemma:analysis:LF:opt delay bound} as
\begin{align*}
\E\big[\lceil T \rceil\big] = \sum_{\tau=0}^{\infty}\p\{T > \tau \} \le P_{o}\sum_{\tau=0}^{\infty} (\hat{P})^{\tau}.
\end{align*}
By (v) in Lemma~\ref{lemma:analysis:LF:property H(k,l)}, we have $\hat{P}<1$ for any $r>0$. Thus, the expectation $\E[\lceil T \rceil]$ is bounded by the geometric series which converges to $\frac{P_{o}}{1-\hat{P}}$. \hfill $\blacksquare$

%==========================================
%
%  Step 3
%
%==========================================
\smallskip\smallskip
\noindent{\bf \em Step 3 (Analysis of the delay scaling):} In Lemma~\ref{lemma:capacity scaling}, we have analyzed the order of the per-node throughput $\lambda_{\hat{\pi}}$ of the scheme $\hat{\pi}$. The results in Lemma~\ref{lemma:analysis:LF:opt delay bound}, (v) in Lemma~\ref{lemma:analysis:LF:property H(k,l)}, and Lemma~\ref{lemma:analysis:LF:bound for U} allow us to analyze the order of the average delay~$\bar{D}_{\hat{\pi}}$, which is shown in Lemma~\ref{lemma:analysis:LF:delay scaling}.

\begin{lemma}\label{lemma:analysis:LF:delay scaling}
Let the communication range $r$ scale as $\Theta(n^\beta)$ $(0 \le \beta \le 1/4)$. Then, the average delay $\bar{D}_{\hat{\pi}}$ of the scheme~$\hat{\pi}$ under the L\'{e}vy flight model with parameter $\alpha\in(0,2]$ scales as follows:
\begin{align*}
\bar{D}_{\hat{\pi}} &= O(\min(n^{(1+\alpha)/2-\beta}, n)).
\end{align*}
\end{lemma}

\noindent\textit{Proof:} Here, we provide a sketch of the proof with details given in Appendix~B. Under the L\'{e}vy flight model with parameter $\alpha\in(0,2]$, we have $(1-\hat{P})^{-1}=\Theta(n^{(1+\alpha)/2 - \beta})$ by (v) in Lemma~\ref{lemma:analysis:LF:property H(k,l)}. In addition, $P_{o} = \Theta(1)$ by Lemma~\ref{lemma:analysis:CCDF of T(0)}. Hence, from Lemma~\ref{lemma:analysis:LF:opt delay bound} and~Lemma~\ref{lemma:analysis:LF:bound for U}, we have
\begin{align}\label{eqn:analysis:LF:delay order under hat pi}
\bar{D}_{\hat{\pi}} \le \E\big[\lceil T \rceil\big] \le \frac{P_{o}}{1-\hat{P}} = \Theta(n^{(1+\alpha)/2-\beta}).
\end{align}
In addition, under the L\'{e}vy flight model, we have a trivial upper bound for all $\alpha\in(0,2]$ as
\begin{align}\label{eqn:analysis:LF:delay order under hat pi-1}
\bar{D}_{\hat{\pi}} = O(n).
\end{align}
Combining (\ref{eqn:analysis:LF:delay order under hat pi}) and (\ref{eqn:analysis:LF:delay order under hat pi-1}) yields our lemma.
\hfill $\blacksquare$

%==========================================
%
%  Step 4
%
%==========================================
\smallskip\smallskip
\noindent{\bf \em Step 4 (Analysis of the delay-capacity tradeoff):} In the last step, we derive the delay-capacity tradeoff under the L\'{e}vy flight model. By combining the capacity scaling in Lemma~\ref{lemma:capacity scaling} and the delay scaling in Lemma~\ref{lemma:analysis:LF:delay scaling}, we get the following theorem.

\begin{theorem}\label{thm:analysis:LF:tradeoff} Under the L\'{e}vy flight model with parameter $\alpha\in(0,2]$, the delay-capacity tradeoff $\bar{D}(\lambda)$ for per-node throughput $\lambda = \Theta(n^{-\eta})~(0 \le \eta \le 1/2)$ is given by
\begin{align*}
\bar{D}(\lambda) = O(\sqrt{\min(n^{1+\alpha}\lambda, n^2)}).
\end{align*}
\end{theorem}

\noindent\textit{Proof:} With the use of $\beta = -\log_n \sqrt{\lambda}$, the scheme $\hat{\pi}$ can achieve the per-node throughput $\lambda_{\hat{\pi}} = \lambda$ and the average delay $\bar{D}_{\hat{\pi}} = O(\sqrt{\min(n^{1+\alpha}\lambda, n^2)})$ by Lemma~\ref{lemma:capacity scaling} and Lemma~\ref{lemma:analysis:LF:delay scaling}, respectively. Therefore, from (\ref{eqn:model:logic}), we have our theorem. \hfill $\blacksquare$

\section{Delay Analysis for the \emph{i.i.d.} Mobility Model}\label{sec:analysis:iid}
In this section, we provide detailed analytical steps for obtaining the optimal delay under the \textit{i.i.d.} mobility model. We again follow the four steps analogous to those used for the L\'{e}vy flight model.

%==========================================
%
%  Step 1
%
%==========================================
\smallskip\smallskip
\noindent{\bf \em Step 1 (Formulation of the average delay using the first meeting time distribution):} From~(\ref{eqn:analysis:LF:avg delay under hat pi}), the average delay $\bar{D}_{\hat{\pi}}$ under the scheme $\hat{\pi}$ is obtained by
\begin{align}\label{eqn:analysis:iid:avg delay under hat pi}
\bar{D}_{\hat{\pi}}
&= \p\{d \!\notin\! \mathcal{I}(s)\} \!\cdot\! \E \big[\!\min \!\big(T_{(i,d)}; i \in \mathcal{I}(s)\big)\big|\,d \notin \mathcal{I}(s)\big].
\end{align}
As pointed out in Step~1 of Section~\ref{sec:analysis:LF}, the random variables $T_{(i,d)}$ for $i \in \mathcal{I}(s)$ are \emph{dependent}. However, the dependency disappears when the nodes move to the next locations after a single time slot under the \textit{i.i.d.} mobility model. The property of choosing a completely independent location at every time slot in the \textit{i.i.d.} mobility enables this independence to occur. By applying this observation, we derive a bound on $T_{(i,d)}$ for $i \in \mathcal{I}(s)$ as follows: let $|\mathcal{I}(s)|$ denote the cardinality of the set $\mathcal{I}(s)$. We condition on the values of $|\mathcal{I}(s)|$ and rewrite the expectation on the right-hand side of~(\ref{eqn:analysis:iid:avg delay under hat pi}) as
\begin{align}\label{eqn:analysis:iid:rhs of avg delay}
\begin{split}
&\E\big[\min \big(T_{(i,d)}; i \in \mathcal{I}(s)\big)\,\big|\,d \notin \mathcal{I}(s)\big] \\
&= \sum_{m=1}^{n-1} \p\{|\mathcal{I}(s)| = m \,|\, d \notin \mathcal{I}(s)\} \\
& \qquad \quad  \cdot \E \big[\min \big(T_{(i,d)}; i \in \mathcal{I}(s)\big)\,\big|\,|\mathcal{I}(s)| = m, d \notin \mathcal{I}(s)\big]\\
&= \sum_{m=1}^{n-1} \p\{|\mathcal{I}(s)| = m \,|\, d \notin \mathcal{I}(s)\} \cdot \E\big[\min \big(T^\star_{1}, T^\star_{2}, \ldots, T^\star_{m}\big)\big],
\end{split}
\end{align}
where $T^\star_{v}\,(v=1,\ldots,m)$ denotes the first meeting time of the node $d$ and the $v$th node in the set $\mathcal{I}(s)$, provided that $|\mathcal{I}(s)|=m$ and $d \notin \mathcal{I}(s)$. Let $T_1, \ldots, T_m$ be $m$ independent copies of the generic random variable $T$. Then, we can derive a bound on $T^\star_{v}$ in terms of $T_v$ as follows:
\begin{align}\label{eqn:analysis:iid:iid bound}
T_{v}^\star
&\deq \inf\{t \ge 0 \,|\, L_{(i_v,d)}(t)\le r, i_v \in \mathcal{I}(s), d \notin \mathcal{I}(s)\} \nonumber\\
&\le \inf\{t \ge 1 \,|\, L_{(i_v,d)}(t)\le r, i_v \in \mathcal{I}(s), d \notin \mathcal{I}(s)\} \nonumber\\
&\ed 1 + T_v.
\end{align}
Here, $i_v$ denotes the $v$th index in the set $\mathcal{I}(s)$ and $\ed$ denotes ``equal in distribution". The last equation comes from the aforementioned nature of the \emph{i.i.d.} mobility model in which the locations of nodes are reshuffled at every time slot.

We define a function $\bar{U}:\{1,2,\ldots,n-1\} \to \mathbb{R}$ by
\begin{align}\label{eqn:analysis:iid:def:U}
\bar{U}(m) \deq
\E\big[\min\big(\lceil T_{v} \rceil; v=1,\ldots,m\big)\big].
\end{align}
Note that discretization of a random variable $T_{v}$ to $\lceil T_{v} \rceil$ is for mathematical simplicity and it does not affect the result (i.e., order of the optimal delay) of this paper. The function $\bar{U}(m)$ works as a tight upper bound on $\bar{D}_{\hat{\pi}}$ as shown in the following lemma.

\begin{lemma}\label{lemma:analysis:iid:opt delay bound} The average delay $\bar{D}_{\hat{\pi}}$ of the scheme $\hat{\pi}$ under the \emph{i.i.d.} mobility model is bounded by
\begin{align}\label{eqn:analysis:iid:avg delay bound}
\bar{D}_{\hat{\pi}} & \le P_{o}  + P_{o} \cdot \E\big[\bar{U}(B_{(n-2,P_{o}^c)}+1)\big],
\end{align}
where $P_{o}$ is defined in~(\ref{eqn:analysis:def:P_out}), $P_{o}^c \deq 1 - P_{o}$, and $B_{(n-2,P_{o}^c)}$ denotes a binomial random variable with parameters $n-2$ (trial) and $P_{o}^c$ (probability). The function $\bar{U}(m)\,(m=1,\ldots,n-1)$ can be obtained from the distribution of~$T$ by
\begin{align*}
\bar{U}(m) =  \sum_{\tau=0}^{\infty} \big(\p\{T > \tau \}\big)^m.
\end{align*}
\end{lemma}

\noindent\textit{Proof:} Since $T_{v}^\star \le 1 + T_v \le 1+ \lceil T_v \rceil$ for $v=1,\ldots,m$ by~(\ref{eqn:analysis:iid:iid bound}), we have
\begin{align*}
\min\big(T_1^\star, \ldots, T_m^\star\big)
&\le 1 + \min\big(\lceil T_1 \rceil, \ldots, \lceil T_m \rceil\big).
\end{align*}
By taking expectations, we have
\begin{align}\label{eqn:analysis:iid:bound U}
\E\big[\min\big(T_{1}^\star, \ldots, T_m^\star \big)\big]
&\le  1 + \bar{U}(m).
\end{align}
Since $\bm{X}_i(0)$ is independent and identically distributed across node index~$i$, each node $i\,(\neq s)$ belongs to the set $\mathcal{I}(s)$ independently of each other with probability $P_{o}^c$. Thus, the random variable $|\mathcal{I}(s)|-1$ (here, 1 is subtracted to exclude the case $s\in\mathcal{I}(s)$) subjected to the condition $d\notin\mathcal{I}(s)$ follows a binomial distribution with parameters $n-2$ and $P_{o}^c$, i.e.,
\begin{align}\label{eqn:analysis:iid:binomial}
\p\{|\mathcal{I}(s)| = m \,|\, d \notin \mathcal{I}(s)\} =  \p\{B_{(n-2, P_{o}^c)} = m-1\}.
\end{align}
By applying~(\ref{eqn:analysis:iid:bound U}) and~(\ref{eqn:analysis:iid:binomial}) to~(\ref{eqn:analysis:iid:rhs of avg delay}), we have
\begin{align}\label{eqn:analysis:iid:bound U2}
&\E\big[\min \big(T_{(i,d)}; i \in \mathcal{I}(s)\big)\,\big|\,d \notin \mathcal{I}(s)\big] \nonumber\\
&\quad \le 1 +  \E \big[\bar{U}(B_{(n-2,P_{o}^c)}+1) \big].
\end{align}
Combining (\ref{eqn:analysis:iid:avg delay under hat pi}) and (\ref{eqn:analysis:iid:bound U2}) yields~(\ref{eqn:analysis:iid:avg delay bound}).

Since the random variable $\min(\lceil T_v\rceil; v=1,\ldots,m)$ takes on only nonnegative integer values, $\bar{U}(m)$ can be obtained by
\begin{align}\label{eqn:analysis:iid:exact formula for U 1}
\bar{U}(m)
= \sum_{\hat{\tau}=1}^{\infty} \p\big\{\min\big(\lceil T_v\rceil; v=1,\ldots,m\big) \ge \hat{\tau} \big\}.
\end{align}
By noting that $\lceil T_v\rceil$ is independent and identically distributed across $v=1,\ldots,m$, we have
\begin{align}\label{eqn:analysis:iid:exact formula for U 2}
&\p\big\{\min\big(\lceil T_v \rceil; v=1,\ldots,m \big) \ge \hat{\tau} \big\} \nonumber \\
&\quad = \big(\p \big\{\lceil T\rceil \ge \hat{\tau} \big\}\big)^m = \big(\p \big\{T > \hat{\tau}-1 \big\}\big)^m,
\end{align}
where the second equality comes from the property that $\p \{\lceil T \rceil \ge \hat\tau \} = \p\{T > \hat\tau-1\}$ for all $\hat{\tau}=1,2,\ldots$. Hence, applying~(\ref{eqn:analysis:iid:exact formula for U 2}) to~(\ref{eqn:analysis:iid:exact formula for U 1}) and replacing $\hat\tau-1$ with~$\tau$ give the lemma. \hfill $\blacksquare$

%==========================================
%
%  Step 2
%
%==========================================
\smallskip\smallskip
\noindent{\bf \em Step 2 (Characterization of the first meeting time distribution):} In this step, similarly to the approach for the L\'{e}vy flight model, we first analyze the characteristics of the function $H(k,l_0)$ in (\ref{eqn:definition of H}) under the \emph{i.i.d.} mobility model (See Lemma~\ref{lemma:analysis:iid:property H(k,l)}). By exploiting the characteristics, we then derive a bound on the first meeting time distribution (See Lemma~\ref{lemma:analysis:iid:ccdf}). This bound enables us to derive a formula for the function $\bar{U}(\cdot)$ used in Lemma~\ref{lemma:analysis:iid:opt delay bound} (See Lemma~\ref{lemma:analysis:iid:bound for U(m)}).

As will be shown below, the characteristics of $H(k,l_0)$ under the \emph{i.i.d.} mobility model are similar to those under the L\'{e}vy flight model. Hence, an upper bound on $\p\{T >\tau\}$ can be derived using the probabilities $\hat{P}\,(\deq H(1,2\sqrt{n}))$ and $P_{o}$ also for the \emph{i.i.d.} mobility model. The main difference is that the formula for $H(1,l_0)$ is of different form and has a different scaling property when $l_0 = 2\sqrt{n}$.

\begin{figure}[t!]
\centering
{\epsfig{figure=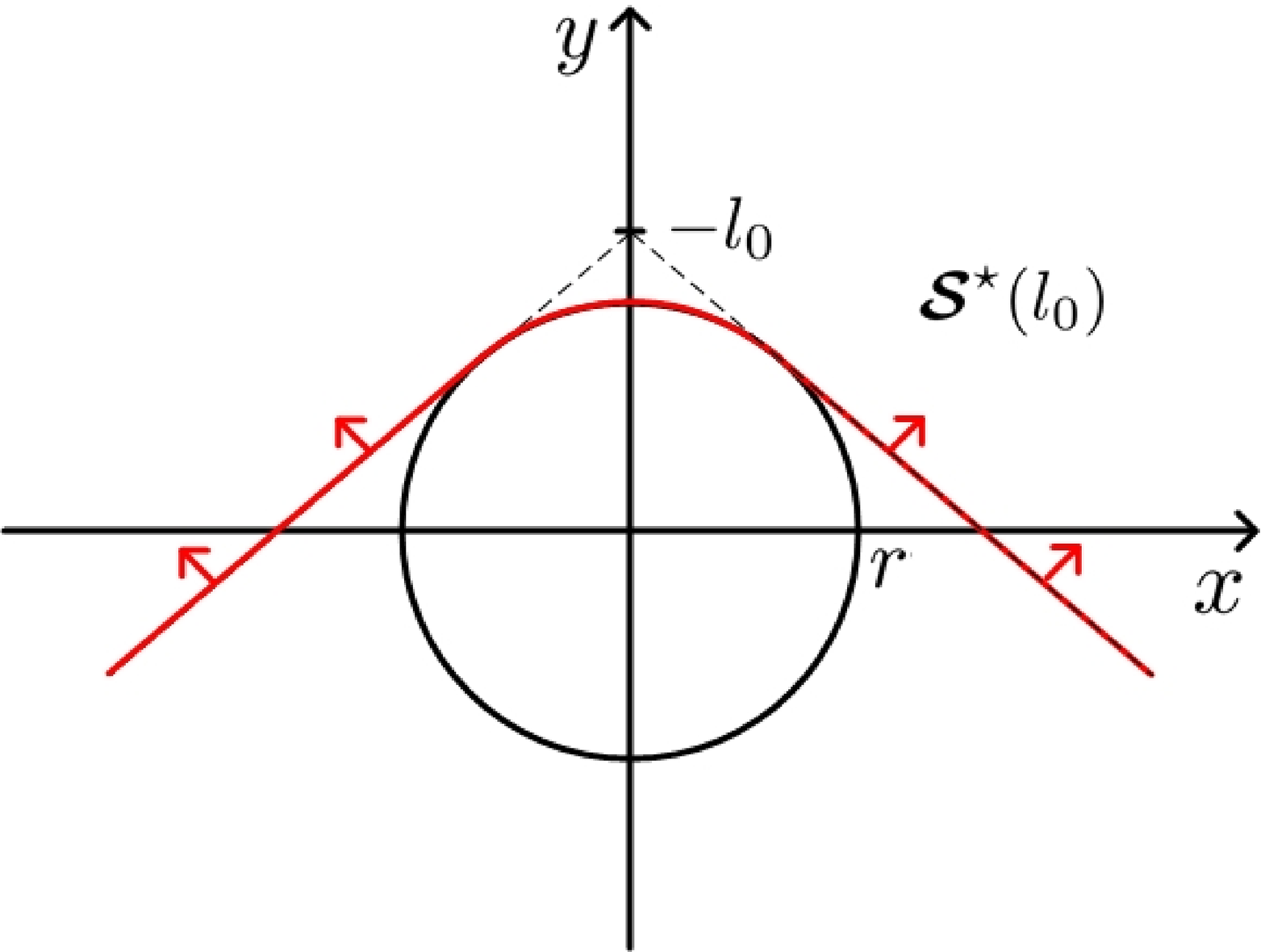,width=0.27\textwidth}}
\caption{The illustration of $\bm{\mathcal{S}}^\star(l_0)$.}
\label{fig:example_S_star}
\end{figure}

\begin{lemma}\label{lemma:analysis:iid:property H(k,l)} Under the \emph{i.i.d.} mobility model, the function $H(k,l_0)$ in (\ref{eqn:definition of H}) has the following characteristics: \\
(i) Let $\Delta\bm{X}$ be a generic random variable for $\bm{X}_i(t)-\bm{X}_j(t)$ representing a location differential between nodes~$i$ and $j$.\footnote{The existence of the generic random variable for $\bm{X}_i(t)-\bm{X}_j(t)$ is proven in Lemma~\ref{lemma:appendix:iid:property of delta X} in Appendix~C.} Then, geometrically the function $H(1,l_0)$ can be viewed as the probability of the location differential falling into a set $\bm{\mathcal{S}}^\star(l_0)\,(\subset\mathbb{R}^2)$ defined for $l_0\in(r,2\sqrt{n}]$ as
\begin{align}\label{eqn:analysis:iid:definition S(l)}
\bm{\mathcal{S}}^\star(l_0) &\deq \big\{\bm{x}\in\mathbb{R}^2  \,\big|\, \overline{((0,l_0), \bm{x})} \cap \bm{\mathcal{D}}_{r}(\bm{0}) = \varnothing\big\},
\end{align}
where the definitions of $\overline{(\cdot,\cdot)}$ and $\bm{\mathcal{D}}_\cdot(\cdot)$ can be found in Lemma~\ref{lemma:analysis:LF:property H(k,l)}. An example of $\bm{\mathcal{S}}^\star(l_0)$ is shown in Fig.~\ref{fig:example_S_star}. The set $\bm{\mathcal{S}}^\star(l_0)$ has a connection with the function $H(1,l_0)$ as follows:
\begin{align*}
H(1,l_0) &= \p\big\{\Delta\bm{X} \in \bm{\mathcal{S}}^\star(l_0)\big\}.
\end{align*}
(ii) The function $H(1,\cdot)$ is nondecreasing. \\
(iii) From (ii), we have $H(1,l_0) \le \hat{P}$ for all $l_0 \in (r, 2\sqrt{n}]$. \\
(iv) For $k=2,3,\ldots$, each function $H(k, l_0)$ is also bounded above by $\hat{P}$. Thus, for all $k\in\mathbb{N}$ and $l_0 \in (r, 2\sqrt{n}]$, we have
\begin{align*}
H(k, l_0) \le \hat{P}.
\end{align*}
(v) $\hat{P}$ is bounded above and below for all $n\in\mathbb{N}$ by
\begin{subequations}
\begin{align}
\hat{P} &\le 1- \frac{1}{\pi}\sin^{-1}\Big(\frac{r}{2\sqrt{n}}\Big), \label{eqn:analysis:iid:property P^o-1} \\
\hat{P} &\ge 1- \frac{r^2}{2n}- \frac{2r}{\pi\sqrt{n}} - \frac{5}{\pi}\sin^{-1}\Big(\frac{r}{2\sqrt{n}}\Big). \label{eqn:analysis:iid:property P^o-2}
\end{align}
\end{subequations}
\end{lemma}

\noindent\textit{Proof:} See Appendix~C. \hfill $\blacksquare$

\smallskip
Similarly to Step 2 in Section~\ref{sec:analysis:LF}, we derive a bound on $\p\{T > \tau \}$ in terms of $\hat{P}$ as follows: from (iv) in Lemma~\ref{lemma:analysis:iid:property H(k,l)} and Lemma~\ref{lemma:analysis:CCDF of T}, we have
\begin{align}\label{eqn:analysis:iid:ccdf}
\p\{T > \tau \} &\le (\hat{P})^{\tau}P_{o}, \quad \text{ for } \tau = 1,2,\ldots.
\end{align}
Since $\p\{T > \tau \} = P_{o}$ for $\tau = 0$, the bound in (\ref{eqn:analysis:iid:ccdf}) also holds for $\tau = 0$. The above result is summarized in Lemma~\ref{lemma:analysis:iid:ccdf}.

\begin{lemma}\label{lemma:analysis:iid:ccdf} Under the \emph{i.i.d.} mobility model, the CCDF of the first meeting time~$T$ is bounded by
\begin{align*}
\p\{T > \tau \} \le (\hat{P})^{\tau}P_{o}, \quad \text{ for } \tau = 0,1,\ldots,
\end{align*}
where $\hat{P}$ and $P_{o}$ are defined in~(\ref{eqn:analysis:LF:definition of P^o}) and (\ref{eqn:analysis:def:P_out}), respectively.
\end{lemma}

\noindent\textit{Proof:} Combining Lemma~\ref{lemma:analysis:CCDF of T} and (iv) in Lemma~\ref{lemma:analysis:iid:property H(k,l)} gives Lemma~\ref{lemma:analysis:iid:ccdf}. The detailed derivation was described earlier in~(\ref{eqn:analysis:iid:ccdf}). \hfill $\blacksquare$

\smallskip
Using Lemma~\ref{lemma:analysis:iid:ccdf}, we can give a bound on the function $\bar{U}(m)$ in~Lemma~\ref{lemma:analysis:iid:opt delay bound} as
\begin{align}\label{eqn:analysis:iid:bound on U}
\bar{U}(m) = \sum_{\tau=0}^{\infty} (\p\{T > \tau \})^m \le (P_{o})^m \sum_{\tau=0}^{\infty} ((\hat{P})^m)^{\tau}.
\end{align}
By (v) in Lemma~\ref{lemma:analysis:iid:property H(k,l)}, we have $(\hat{P})^m<1$ for any $r>0$. Thus, $\bar{U}(m)$ is bounded by a convergent geometric series and we summarize the result in Lemma~\ref{lemma:analysis:iid:bound for U(m)}.

\begin{lemma}\label{lemma:analysis:iid:bound for U(m)} The function $\bar{U}(m)\,(m=1,\ldots,n-1)$ defined in~(\ref{eqn:analysis:iid:def:U}) is bounded under the \emph{i.i.d.} mobility model by
\begin{align*}
\bar{U}(m) \le \frac{(P_{o})^m}{1-(\hat{P})^m}.
\end{align*}
\end{lemma}

\noindent\textit{Proof:} Combining Lemma~\ref{lemma:analysis:iid:opt delay bound}, (v) in Lemma~\ref{lemma:analysis:iid:property H(k,l)}, and Lemma~\ref{lemma:analysis:iid:ccdf} gives Lemma~\ref{lemma:analysis:iid:bound for U(m)}. The detailed derivation was described earlier in~(\ref{eqn:analysis:iid:bound on U}). \hfill $\blacksquare$

\smallskip
The bound in Lemma~\ref{lemma:analysis:iid:bound for U(m)} is essentially the same format with that of the slotted contact model under the \emph{i.i.d.} mobility model. The only difference is that $\hat{P}$ additionally considers intermediate meetings.

%==========================================
%
%  Step 3
%
%==========================================
\smallskip\smallskip
\noindent{\bf \em Step 3 (Analysis of the delay scaling):} In this step, we analyze the order of the average delay $\bar{D}_{\hat{\pi}}$ under the \emph{i.i.d.} mobility model. To efficiently handle the expectation $\E[\bar{U}(B_{(n-2,P_{o}^c)}+1)]$ in Lemma~\ref{lemma:analysis:iid:opt delay bound}, we derive a bound on the expectation as follows: first, we rewrite $\E[\bar{U}(B_{(n-2,P_{o}^c)}+1)]$ by conditioning on $B_{(n-2,P_{o}^c)}$ as
\begin{align}\label{eqn:analysis:iid:before split}
\E[\bar{U}(B_{(\!n\!-\!2,P_{o}^c)}\!+\!1)] \!=\! \sum_{m=1}^{n-1} \bar{U}(m) \!\cdot\! \p\{B_{(\!n\!-\!2,P_{o}^c)} \!=\! m\!-\!1\}.
\end{align}
We then decompose~(\ref{eqn:analysis:iid:before split}) into two terms as
\begin{align}\label{eqn:analysis:iid:split into two terms}
&\E\big[\bar{U}(B_{(n-2,P_{o}^c)}+1)\big] \nonumber\\
&= \sum_{m=1}^{\lceil\gamma r^2\rceil-1} \bar{U}(m) \cdot \p\{B_{(n-2,P_{o}^c)} = m-1\} \nonumber\\
& \qquad + \sum_{m=\lceil\gamma r^2\rceil}^{n-1} \bar{U}(m) \cdot \p\{B_{(n-2,P_{o}^c)} = m-1\}\nonumber\\
&\le \bar{U}(1) \sum_{m=1}^{\lceil\gamma r^2\rceil-1} \p\{B_{(n-2,P_{o}^c)} = m-1\} + \bar{U}(\lceil\gamma r^2\rceil),
\end{align}
where $\gamma$ is a constant in $(0,1)$ and $\gamma r^2$ implies the $\gamma$ fraction of the average number of nodes within the communication range of a source node. In (\ref{eqn:analysis:iid:split into two terms}), we used the property that $\bar{U}(m)$ is a nonincreasing function of $m$. Hence, by Lemma~\ref{lemma:analysis:iid:opt delay bound} and (\ref{eqn:analysis:iid:split into two terms}), the average delay $\bar{D}_{\hat{\pi}}$ of the scheme $\hat{\pi}$ under the \emph{i.i.d.} mobility model is bounded by:
\begin{align}\label{eqn:analysis:iid:order}
\bar{D}_{\hat{\pi}} &\le P_{o} + P_{o} \cdot \bar{U}(1) \cdot\p\{B_{(n-2,P_{o}^c)} \le \lceil\gamma r^2\rceil-2\} \nonumber \\
& \qquad + P_{o} \cdot \bar{U}(\lceil\gamma r^2\rceil).
\end{align}
The results in (\ref{eqn:analysis:iid:order}), Lemmas~\ref{lemma:analysis:CCDF of T(0)} and~\ref{lemma:analysis:iid:bound for U(m)}, and (v) in Lemma~\ref{lemma:analysis:iid:property H(k,l)} allow us to analyze the order of the average delay $\bar{D}_{\hat{\pi}}$, which is shown in Lemma~\ref{lemma:analysis:iid:delay scaling}.

\begin{lemma}\label{lemma:analysis:iid:delay scaling} Let the communication range $r$ scale as $\Theta(n^\beta)$ $(0 \le \beta \le 1/4)$. Then, the average delay $\bar{D}_{\hat{\pi}}$ of the scheme $\hat{\pi}$ under the \emph{i.i.d.} mobility model scales as follows:
\begin{align*}
\bar{D}_{\hat{\pi}} &= O(n^{\max(0,1/2-3\beta)}).
\end{align*}
\end{lemma}

\noindent\textit{Proof:} Here, we provide a sketch of the proof with details given in Appendix~C. \\
\emph{\underline{Order of $P_{o}$:}} By Lemma~\ref{lemma:analysis:CCDF of T(0)},
\begin{align}\label{eqn:analysis:iid:order1}
P_{o} = \Theta(1).
\end{align}
\emph{\underline{Order of $\bar{U}(1)$:}} By (v) in Lemma~\ref{lemma:analysis:iid:property H(k,l)}, we have $(1-\hat{P})^{-1} = \Theta(n^{1/2-\beta})$. Hence, combining (\ref{eqn:analysis:iid:order1}) and Lemma~\ref{lemma:analysis:iid:bound for U(m)} yields
\begin{align}\label{eqn:analysis:iid:order2}
\bar{U}(1) \le \frac{P_{o}}{1-\hat{P}} = \Theta(n^{1/2-\beta}).
\end{align}
\emph{\underline{Order of $\bar{U}(\lceil\gamma r^2\rceil)$:}} By (\ref{eqn:analysis:iid:order1}), we have $(P_{o})^{\lceil\gamma r^2\rceil} = \Theta(1)$. In addition, by (v) in Lemma~\ref{lemma:analysis:iid:property H(k,l)}, we have $(1-(\hat{P})^{\lceil\gamma r^2\rceil})^{-1} = \Theta(n^{\max(0,1/2-3\beta)})$. Hence, Lemma~\ref{lemma:analysis:iid:bound for U(m)} gives
\begin{align}\label{eqn:analysis:iid:order3}
\bar{U}(\lceil\gamma r^2\rceil) \le \frac{(P_{o})^{\lceil\gamma r^2\rceil}}{1-(\hat{P})^{\lceil\gamma r^2\rceil}} = \Theta(n^{\max(0,1/2-3\beta)}).
\end{align}
\emph{\underline{Order of $\p\{B_{(n-2,P_{o}^c)} \le \lceil\gamma r^2\rceil-2\}$:}}
By using Chernoff's inequality, for any fixed $\gamma\in(0,1/3)$ and $n \ge \frac{2}{1-3\gamma}$, we have
\begin{align*}
\p\{B_{(n-2,P_{o}^c)} \le \gamma r^2\}
&\le \exp\left(-\frac{1}{2}\Big(\frac{n-2-3\gamma n}{3n}\Big)^2 r^2\right)\\
&\le 2\Big(\frac{3n}{n-2-3\gamma n}\Big)^2 r^{-2} = O(n^{-2\beta}),
\end{align*}
which results in
\begin{align}\label{eqn:analysis:iid:order4}
\p\{B_{(n-2,P_{o}^c)} \le \lceil\gamma r^2\rceil-2\} &= O(n^{-2\beta}).
\end{align}
Combining (\ref{eqn:analysis:iid:order})-(\ref{eqn:analysis:iid:order4}) gives the lemma. \hfill $\blacksquare$

%==========================================
%
%  Step 4
%
%==========================================
\smallskip\smallskip
\noindent{\bf \em Step 4 (Analysis of the delay-capacity tradeoff):} In the last step, we derive the delay-capacity tradeoff under the \emph{i.i.d.} mobility model. By combining the capacity scaling in Lemma~\ref{lemma:capacity scaling} and the delay scaling in Lemma~\ref{lemma:analysis:iid:delay scaling}, we get the following theorem.

\begin{theorem}\label{thm:analysis:iid:tradeoff} Under the \emph{i.i.d.} mobility model, the delay-capacity tradeoff $\bar{D}(\lambda)$ for per-node throughput $\lambda = \Theta(n^{-\eta})~(0 \le \eta \le 1/2)$ is given by
\begin{align*}
\bar{D}(\lambda) &= O(\sqrt{\max(1,n\lambda^3)}).
\end{align*}
\end{theorem}

\noindent\textit{Proof:} With the use of $\beta = -\log_n \sqrt{\lambda}$, the scheme $\hat{\pi}$ can achieve the per-node throughput $\lambda_{\hat{\pi}} = \lambda$ and the average delay $\bar{D}_{\hat{\pi}} = O(\sqrt{\max(1,n\lambda^3)})$ by Lemma~\ref{lemma:capacity scaling} and Lemma~\ref{lemma:analysis:iid:delay scaling}, respectively. Therefore, from (\ref{eqn:model:logic}), we have our theorem. \hfill $\blacksquare$

\section{Concluding Remarks}\label{sec:conclusion}
In this paper, we developed a new analytical framework that substantially improves the realism in delay-capacity analysis by considering (i) L\'{e}vy flight mobility, which is known to closely resemble human mobility patterns and (ii) contact opportunities in the middle of movements of nodes. Using our framework, we obtained the first delay-capacity tradeoff for L\'{e}vy flight and derived a new tighter tradeoff for \emph{i.i.d.} mobility. For L\'{e}vy flight, our analysis shows that the tradeoff holds $\bar{D}(\lambda) = O(\sqrt{\min(n^{1+\alpha} \lambda, n^2)})$ for $\lambda = \Theta(n^{-\eta})\,(0\le\eta\le1/2)$ as shown in Figs.~\ref{fig:tradeoffs}~(a),~\ref{fig:levyalpha}~(a), and ~\ref{fig:levyalpha}~(b). Our result is well aligned with the critical delay suggested in~\cite{lee11scaling}. For \emph{i.i.d.} mobility, our analysis provides $\bar{D}(\lambda) = O(\sqrt{\max(1,n\lambda^3) })$ as shown in Fig.~\ref{fig:tradeoffs}~(b). These tradeoffs are especially remarkable in both L\'{e}vy flight and \emph{i.i.d.} mobility for the constant per-node throughput (i.e., $\lambda = \Theta(1)$) as they demonstrate that the delay can be less than $\Theta(n)$, which has been widely accepted for most mobility models. Our future work includes (i) an extension of our framework to analyze the delay-capacity tradeoff under L\'{e}vy walk and (ii) another extension to capture correlated movement patterns among nodes.

\appendices
\section{Proof of Lemma~\ref{lemma:analysis:CCDF of T(0)}}
\begin{figure}[t!]
\centering
{\epsfig{figure=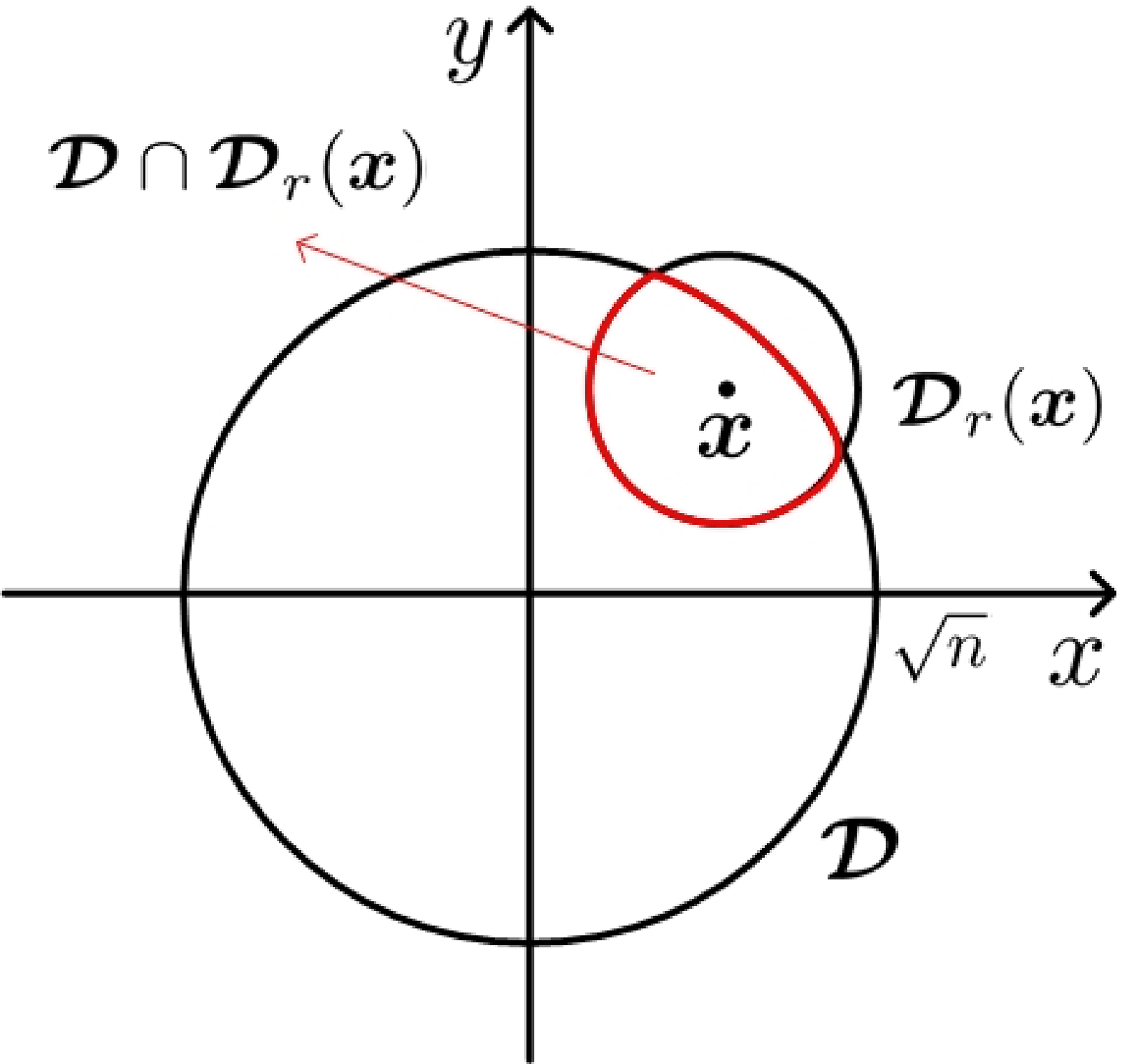,width=0.22\textwidth}}
{\epsfig{figure=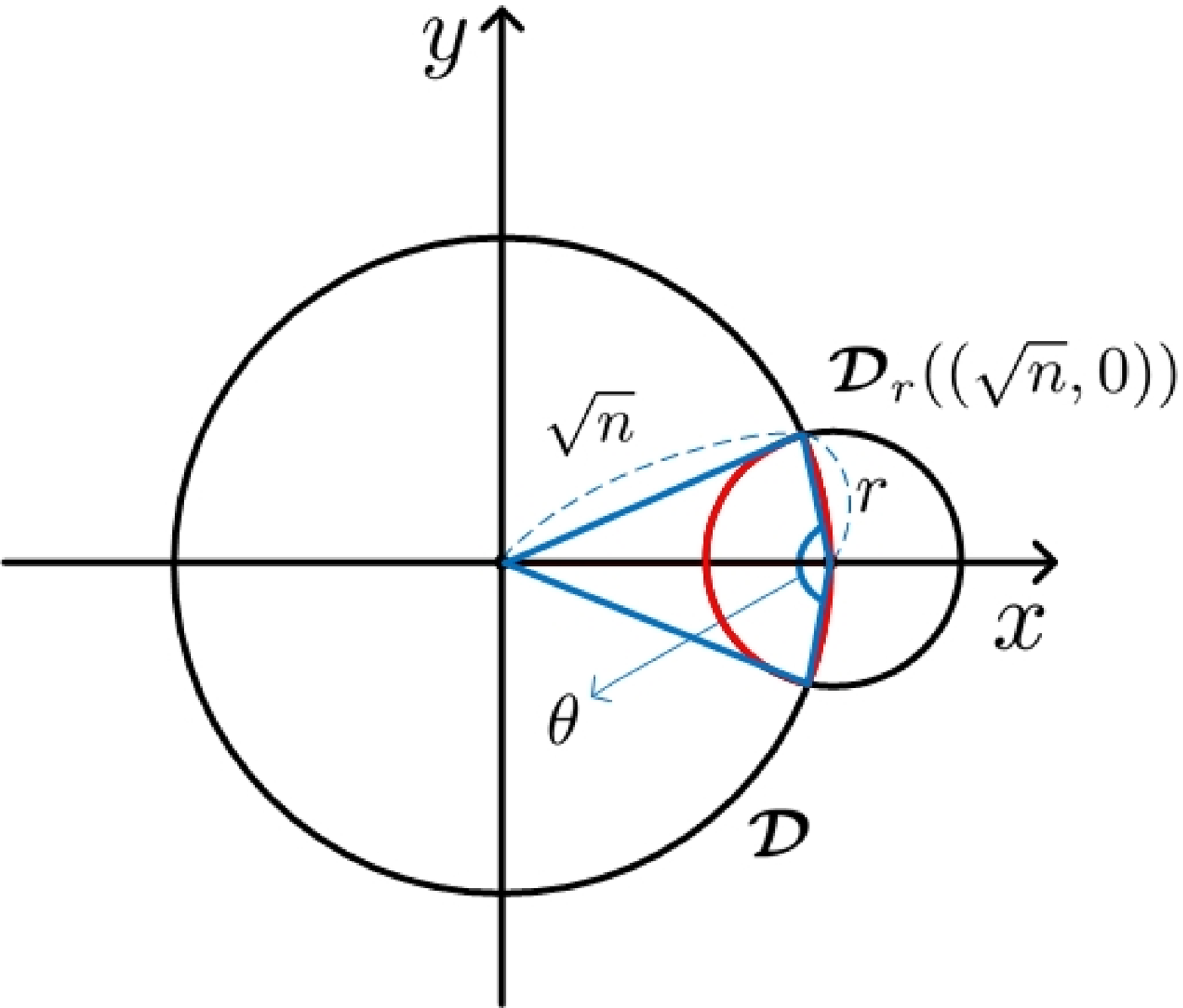,width=0.24\textwidth}}
\caption{Examples of $\bm{\mathcal{D}}\cap\bm{\mathcal{D}}_r(\bm{x})$: $\text{Area}(\bm{\mathcal{D}}\cap\bm{\mathcal{D}}_r(\bm{x}))$ is nonincreasing as $|\bm{x}|$ increases (Left).
When $|\bm{x}| = n$ (e.g., $\bm{x}= (\sqrt{n}, 0)$), $\text{Area}(\bm{\mathcal{D}}\cap\bm{\mathcal{D}}_r(\bm{x}))$ is minimized and a lower bound on $\text{Area}(\bm{\mathcal{D}}\cap\bm{\mathcal{D}}_r((\sqrt{n},0))$ can be obtained in terms of $\theta=2\cos^{-1}(\frac{r}{2\sqrt{n}})$, as given in~(\ref{eqn:appendix:LF:scaling P_out:lower}) (Right).}
\label{fig:probability_out}
\end{figure}

By the definition of $P_{o}$ in (\ref{eqn:analysis:def:P_out}), we have
\begin{align}\label{eqn:appendix:LF:scaling P_out:main}
P_{o} &= \p\{L_{(i,j)}(0) > r\} = 1-\p\{L_{(i,j)}(0) \le r\}.
\end{align}
Let $F_{\bm{X}_i(0)}(\cdot)$ denote the CDF of $\bm{X}_i(0)$. Then, by conditioning on the values of $\bm{X}_i(0)$, the probability $\p\{L_{(i,j)}(0) \le r\}$ in (\ref{eqn:appendix:LF:scaling P_out:main}) can be rewritten as
\begin{align}\label{eqn:appendix:LF:scaling P_out:1-P_out}
&\p\{L_{(i,j)}(0) \le r\} \nonumber \\
&\quad = \int_{\bm{\mathcal{D}}}\p\{L_{(i,j)}(0)\le r| \bm{X}_i(0) = \bm{x}\} \,\text{d}F_{\bm{X}_i(0)}(\bm{x}) \nonumber \\
&\quad = \int_{\bm{\mathcal{D}}}\p\{\bm{X}_j(0) \in \bm{\mathcal{D}}_r(\bm{x}) | \bm{X}_i(0) = \bm{x}\} \, \text{d}F_{\bm{X}_i(0)}(\bm{x}) \nonumber \\
&\quad = \int_{\bm{\mathcal{D}}}\p\{\bm{X}_j(0) \in \bm{\mathcal{D}}_r(\bm{x}) \} \,\text{d}F_{\bm{X}_i(0)}(\bm{x}),
\end{align}
where the last equality comes from the independence between $\bm{X}_i(0)$ and $\bm{X}_j(0)$. Note that, since $\bm{X}_j(0)\in\bm{\mathcal{D}}$ with probability~1 and $\bm{X}_j(0)\sim\text{Uniform}(\bm{\mathcal{D}})$, the probability $\p\{\bm{X}_j(0) \in \bm{\mathcal{D}}_r(\bm{x})\}$ in the integral in (\ref{eqn:appendix:LF:scaling P_out:1-P_out}) is given by
\begin{align}\label{eqn:appendix:analysis:area}
\p\{\bm{X}_j(0) \in \bm{\mathcal{D}}_r(\bm{x})\}
= \frac{\text{Area}(\bm{\mathcal{D}}\cap\bm{\mathcal{D}}_r(\bm{x}))} {\text{Area}(\bm{\mathcal{D}})},
\end{align}
where $\text{Area}(\bm{S})$ denotes the area of a set $\bm{S}\subset\mathbb{R}^2$.
An example of $\bm{\mathcal{D}}\cap\bm{\mathcal{D}}_r(\bm{x})$ is shown in Fig.~\ref{fig:probability_out}. From the figure, it is obvious that $\text{Area}(\bm{\mathcal{D}}\cap\bm{\mathcal{D}}_r(\bm{x}))$ is nonincreasing as~$\bm{x}$ approaches to the boundary of the disc $\bm{\mathcal{D}}$. Hence, (\ref{eqn:appendix:analysis:area}) is bounded above by
\begin{align}\label{eqn:appendix:LF:scaling P_out:upper}
\p\{\bm{X}_j(0) \in \bm{\mathcal{D}}_r(\bm{x})\}
&\le \frac{\text{Area}(\bm{\mathcal{D}}\cap\bm{\mathcal{D}}_r(\bm{0}))}{\pi (\sqrt{n})^2} \nonumber \\
&=\frac{\text{Area}(\bm{\mathcal{D}}_r(\bm{0}))}{\pi n} = \frac{r^2}{n}.
\end{align}
In addition, it is bounded below by
\begin{align}\label{eqn:appendix:LF:scaling P_out:lower}
\p\{\bm{X}_j(0) \in \bm{\mathcal{D}}_r(\bm{x})\} &\ge \frac{ \text{Area}(\bm{\mathcal{D}}\cap\bm{\mathcal{D}}_r((\sqrt{n},0)))}{\pi (\sqrt{n})^2} \nonumber \\
&\ge \frac{\pi r^2}{\pi n} \frac{\theta}{2\pi} = \frac{r^2}{n} \frac{\theta}{2\pi},
\end{align}
where $\theta \deq 2\cos^{-1}(\frac{r}{2\sqrt{n}})$ (See Fig.~\ref{fig:probability_out}). Since $\sqrt{n} \ge r$, we have $\theta \ge 2\cos^{-1}(\frac{r}{2r}) = \frac{2\pi}{3}$. Hence, the inequality in (\ref{eqn:appendix:LF:scaling P_out:lower}) is further bounded by
\begin{align}\label{eqn:appendix:LF:scaling P_out:lower2}
\p\{\bm{X}_j(0) \in \bm{\mathcal{D}}_r(\bm{x})\} &\ge \frac{r^2}{3n}.
\end{align}
By substituting (\ref{eqn:appendix:LF:scaling P_out:upper}) and (\ref{eqn:appendix:LF:scaling P_out:lower2}) into (\ref{eqn:appendix:LF:scaling P_out:1-P_out}), we have
\begin{align*}
\frac{r^2}{3n} \le \p\{L_{(i,j)}(0) \le r\} \le \frac{r^2}{n},
\end{align*}
which, combined with (\ref{eqn:appendix:LF:scaling P_out:main}), gives
\begin{align*}
1-\frac{r^2}{n} \le P_{o} \le 1-\frac{r^2}{3n}.
\end{align*}

\section{Proofs of Lemmas for the L\'{e}vy Flight Model}
Here, we give detailed proofs of Lemmas~\ref{lemma:analysis:LF:property H(k,l)} and~\ref{lemma:analysis:LF:delay scaling}, which are used for analyzing the optimal delay under the L\'{e}vy flight model. To prove Lemma~\ref{lemma:analysis:LF:property H(k,l)}, we need the following Lemmas~\ref{lemma:appendix:LF:property of delta V},~\ref{lemma:appendix:LF:markov}, and~\ref{lemma:appendix:LF:app power law}.

%------------------------------------------------
%
% Lemma
%
%------------------------------------------------
\begin{lemma}\label{lemma:appendix:LF:property of delta V}
For $i \neq j$ and $k\in\mathbb{N}$, let
\begin{align*}
\Delta\bm{V}_{(i,j)}(k) \deq \bm{V}_i(k) - \bm{V}_j(k),
\end{align*}
where $\bm{V}_{\cdot}(k)$ (representing the $k$th flight of a node $\cdot$) is defined in (\ref{eqn:model:mobility model:LF:flight vector}). Then, under the L\'{e}vy flight model, $\Delta\bm{V}_{(i,j)}(k)$ has the following properties: \\
(i) $\Delta\bm{V}_{(i,j)}(k)$ is independent of $\bm{X}_u(t)$ for all $u=1,\ldots,n$ and $t \in [0, k-1]$. \\
(ii) $\Delta\bm{V}_{(i,j)}(k)$ is identically distributed across pair index~$(i,j)$ and slot index~$k$. Hence, we use $\Delta \bm{V}$ to denote a generic random variable for $\Delta\bm{V}_{(i,j)}(k)$. \\
(iii) For $\bm{v}\in\mathbb{R}^2$, let $\angle\bm{v}$ denote the angle at vertex $\bm{0}$ enclosed by the line $\overline{(\bm{0},\bm{v})}$ and the positive $x$-axis. Then, the angle $\angle\Delta\bm{V}$ is a uniform random variable on the interval $(0,2\pi]$ and is independent of the length $|\Delta\bm{V}|$.
\end{lemma}

\noindent\textit{Proof:} (i) For any $u=1,\ldots,n$, $\bm{X}_u(t)\,(0 \le t \le k-1)$ under the L\'{e}vy flight model is completely determined by $\bm{F}_u(k-1) \deq \{\bm{X}_u(0),\bm{V}_u(1),\ldots,\bm{V}_u(k-1)\}$ (by the relations~(\ref{eqn:model:LF:location any time}) and~(\ref{eqn:model:mobility model:LF})). Since $\bm{V}_i(k)$ is independent of $\bm{F}_u(k-1)$, it is independent of $\bm{X}_u(t)$. By the same reason, $\bm{V}_j(k)$ is independent of $\bm{X}_u(t)$. Therefore, the difference $\bm{V}_i(k) - \bm{V}_j(k)$ is independent of $\bm{X}_u(t)$. \\
(ii) Since each of the flight angle $\theta_u(k)$ and the flight length $Z_u(k)$ is independent and identically distributed across node index~$u$ and slot index~$k$, the random variable $\bm{V}_u(k)\,(\deq (Z_u(k)\cos\theta_u(k),Z_u(k)\sin\theta_u(k)))$ is also independent and identically distributed across~$u$ and~$k$. Therefore, the difference $\bm{V}_i(k) - \bm{V}_j(k)$ is identically distributed across pair index~$(i,j)$ and slot index~$k$. However, it is not necessarily independent across~$(i,j)$ while it is independent across $k$ for a fixed $(i,j)$.\\
(iii) To prove (iii), it suffices to show that for any $v\ge 0$,
\begin{align}\label{eqn:appendix:LF:circular sym}
\p\big\{\angle\Delta\bm{V}_{(i,j)}(k) \le \theta \,\big|\, |\Delta\bm{V}_{(i,j)}(k)| = v\big\} = \frac{\theta}{2\pi},
\end{align}
where $0 < \theta \le 2\pi$. In the following, we will prove (\ref{eqn:appendix:LF:circular sym}).

\begin{figure}[t!]
\centering
{\epsfig{figure=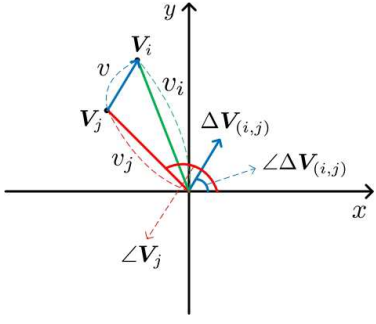,width=0.27\textwidth}}
\caption{An example of vectors satisfying the condition in (\ref{eqn:appendix:LF:circular sym(1)}): for a given $(v_i,v_j,v)$, the angle $\angle\Delta\bm{V}_{(i,j)}$ is determined by the angle $\angle\bm{V}_j$. Since $\angle\bm{V}_j\sim\text{Uniform}[0,2\pi]$, we have  $\angle\Delta\bm{V}_{(i,j)}\sim\text{Uniform}[0,2\pi]$. }
\label{fig:example_circular_sym}
\end{figure}

For simplicity, we omit the slot index~$k$ in $\bm{V}_\cdot(k)$ and $\Delta\bm{V}_{(i,j)}(k)$ in the rest of this proof. By conditioning on the values of $(|\bm{V}_i|,|\bm{V}_j|)$, we can rewrite the probability on the left-hand side of (\ref{eqn:appendix:LF:circular sym}) as follows:
\begin{align}\label{eqn:appendix:LF:circular sym(0)}
&\p\big\{\angle \Delta\bm{V}_{(i,j)} \le \theta \,\big|\, |\Delta\bm{V}_{(i,j)}| = v \big\} \nonumber \\
&= \int_{(v_i,v_j)} \!\!\!\p\big\{\angle \Delta\bm{V}_{(i,j)} \le \theta \big| (|\bm{V}_i|,|\bm{V}_j|,|\Delta\bm{V}_{(i,j)}|) \!=\! (v_i, v_j, v) \big\} \nonumber \\
&\hspace{12 mm}\cdot\p\big\{(|\bm{V}_i|,|\bm{V}_j|) \!=\! (v_i, v_j) \big| |\Delta\bm{V}_{(i,j)}| \!=\! v\big\} \, \text{d}(v_i,v_j).
\end{align}
For a fixed $v \ge 0$, consider an event $\{(|\bm{V}_i|,|\bm{V}_j|) = (v_i, v_j)\}$ such that
\begin{align}\label{eqn:appendix:LF:circular sym(1)}
\p\big\{(|\bm{V}_i|,|\bm{V}_j|) = (v_i, v_j) \,\big|\, |\Delta\bm{V}_{(i,j)}| = v\big\} >0.
\end{align}
An example satisfying (\ref{eqn:appendix:LF:circular sym(1)}) is shown in Fig.~\ref{fig:example_circular_sym}. Under the condition $(|\bm{V}_i|,|\bm{V}_j|,|\Delta\bm{V}_{(i,j)}|) = (v_i,v_j,v)$, the angle $\angle\Delta\bm{V}_{(i,j)}$ is determined by the angle $\angle\bm{V}_j$ as the figure shows. Since $\angle\bm{V}_j\sim\text{Uniform}[0,2\pi]$, we have  $\angle\Delta\bm{V}_{(i,j)}\sim\text{Uniform}[0,2\pi]$. That is, for $0 < \theta \le 2\pi$ we have
\begin{align*}
\p\big\{\angle \Delta\bm{V}_{(i,j)} \le \theta \,\big|\, (|\bm{V}_i|,|\bm{V}_j|,|\Delta\bm{V}_{(i,j)}|) = (v_i, v_j, v) \big\} = \frac{\theta}{2\pi}.
\end{align*}
Since the above equality holds \emph{for any} $(v_i, v_j)$ satisfying (\ref{eqn:appendix:LF:circular sym(1)}) for a given $v$, the probability in (\ref{eqn:appendix:LF:circular sym(0)}) boils down to the following:
\begin{align*}
&\p\big\{\angle \Delta\bm{V}_{(i,j)} \le \theta \,\big|\, |\Delta\bm{V}_{(i,j)}| = v \big\} \\
& = \frac{\theta}{2\pi} \int_{(v_i,v_j)} \!\!\!\p\big\{(|\bm{V}_i|,|\bm{V}_j|) \!=\! (v_i, v_j) \,\big|\, |\Delta\bm{V}_{(i,j)}| \!=\! v\big\} \, \text{d}(v_i,v_j) \\
& = \frac{\theta}{2\pi}.
\end{align*}
This completes the proof. \hfill $\blacksquare$

%------------------------------------------------
%
% Lemma
%
%------------------------------------------------
\begin{lemma}\label{lemma:appendix:LF:markov} Suppose $k\in\mathbb{N}$ and $l\in(r,2\sqrt{n}]$. Then, for any sets $\mathcal{L}(\cdot) \subset [0, 2\sqrt{n}]$ satisfying
\begin{align}\label{eqn:appendix:LF:markov condition}
\p\{L(k-1) = l, L(t) \in \mathcal{L}(t), 0 \le t \le k-1\} > 0,
\end{align}
we have under the L\'{e}vy flight model the following:
\begin{align}\label{eqn:appendix:LF:markov}
& \p\{I(k) = 0 \,|\, L(k-1) = l, L(t) \in \mathcal{L}(t), 0 \le t \le k-1\} \nonumber \\
& \quad = \p\{\Delta\bm{V} \in \bm{\mathcal{S}}(l)\}.
\end{align}
The definitions of $\Delta\bm{V}$ and $\bm{\mathcal{S}}(l)$ can be found in Lemma~\ref{lemma:analysis:LF:property H(k,l)}.
\end{lemma}

\begin{remark}
Before proving the lemma, we give a remark. Lemma~\ref{lemma:appendix:LF:markov} implies that the future states of a meeting process under the L\'{e}vy flight model depend only on the state at the beginning of the current slot, not on the sequence of events that preceded it. In addition, the conditional probability distribution of the future state described above is time homogeneous (i.e., the probability in~(\ref{eqn:appendix:LF:markov}) does not depend on the slot index $k$). This restricted time homogeneous memoryless property enables us to derive a bound on the first meeting time distribution as a \emph{geometric form} (See Lemma~\ref{lemma:analysis:LF:ccdf}).
\end{remark}

\noindent\textit{Proof:} For notational simplicity, we let
\begin{align}\label{eqn:appendix:LF:def_history}
\mathcal{F}(k-1) \deq \{L(t) \in \mathcal{L}(t), 0 \le t \le k-1\}
\end{align}
satisfying~(\ref{eqn:appendix:LF:markov condition}). For $i \neq j$ and $t\ge 0$, let
\begin{align*}
\Delta\bm{X}_{(i,j)}(t) \deq \bm{X}_i(t) - \bm{X}_j(t).
\end{align*}
For simplicity, we omit $(i,j)$ in $\Delta\bm{X}_{(i,j)}(t)$. Then, by conditioning on the values of $\angle \Delta\bm{X}(k-1)$, the left-hand side of (\ref{eqn:appendix:LF:markov}) can be rewritten as
\begin{align}\label{eqn:appendix:LF:markov lhs}
& \p\{I(k) = 0 | L(k-1) = l, \mathcal{F}(k-1)\} \nonumber \\
& = \!\int_{0}^{2\pi}\!\! \p\{I(k) = 0 | \angle \Delta\bm{X}(k\!-\!1\!) = \theta, L(k\!-\!1\!) = l, \!\mathcal{F}(k\!-\!1\!)\} \nonumber \\
& \hspace{11 mm} \text{d} F_{\angle\Delta\bm{X}(k-1)| L(k-1) = l, \mathcal{F}(k-1)}(\theta),
\end{align}
where $F_{\angle\Delta\bm{X}(k-1)| L(k-1) = l, \mathcal{F}(k-1)}(\cdot)$ denotes the CDF of the random variable $\angle\Delta\bm{X}(k-1)$ conditioned that $L(k-1) = l$ and $\mathcal{F}(k-1)$.
Since $L(k-1) = |\Delta\bm{X}(k-1)|$, the joint condition $ \angle \Delta\bm{X}(k-1) = \theta$ and $L(k-1) = l$ is equivalent to $\Delta\bm{X}(k-1)=l\bm{e}^{j\theta}$, where $\bm{e}^{j\theta} \deq (\cos\theta, \sin\theta)$. Hence, the probability in (\ref{eqn:appendix:LF:markov lhs}) can be expressed as
\begin{align}\label{eqn:appendix:LF:markov lhs-(1)}
& \p\{I(k) = 0 \,|\, \angle \Delta\bm{X}(k-1) = \theta, L(k-1) = l, \mathcal{F}(k-1)\} \nonumber \\
&\quad = \p\{I(k) = 0 \,|\, \Delta\bm{X}(k-1)=l\bm{e}^{j\theta}, \mathcal{F}(k-1)\}.
\end{align}
The key idea of the proof is to use the following equality: \emph{for any} $k\in\mathbb{N}$, $l\in(r,2\sqrt{n}]$, $\theta\in(0,2\pi]$, and $\mathcal{F}(k-1)$, we have
\begin{align}\label{eqn:appendix:LF:markov key}
\begin{split}
&\p\{I(k) = 0 \,|\, \Delta\bm{X}(k-1)=l\bm{e}^{j\theta}, \mathcal{F}(k-1)\} \\
& \quad = \p\{\Delta\bm{V} \in \bm{\mathcal{S}}(l)\}.
\end{split}
\end{align}
By substituting the combined result of (\ref{eqn:appendix:LF:markov lhs-(1)}) and (\ref{eqn:appendix:LF:markov key}) into (\ref{eqn:appendix:LF:markov lhs}), we have the lemma.

\begin{figure}[t!]
\centering
{\epsfig{figure=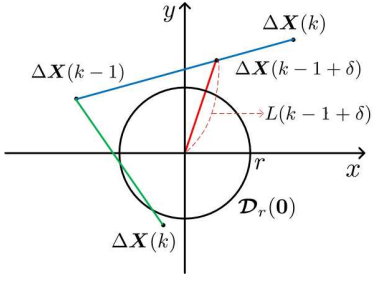,width=0.27\textwidth}}
\caption{An illustration of a meeting event during the $k$th slot: in case of the blue line, $L(k-1+\delta)>r$ for all $\delta\in(0,1]$, i.e., $I(k) = 0$. However, in case of the green line, there exist multiple $\delta\in(0,1]$ such that $L(k-1+\delta)\le r$, i.e., $I(k)=1$.}
\label{fig:meeting_event}
\end{figure}

\begin{figure}[t!]
\centering
{\epsfig{figure=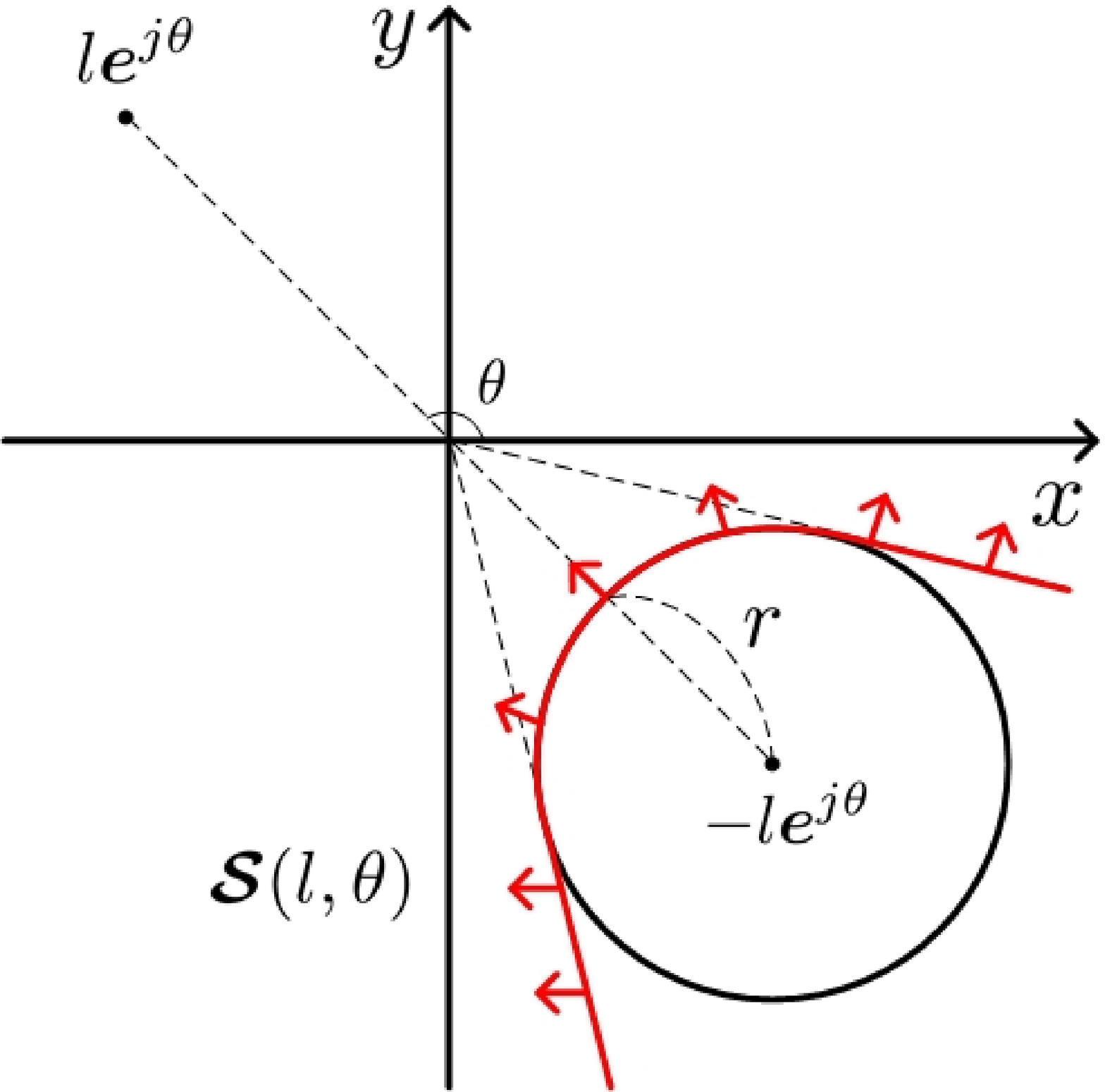,width=0.24\textwidth}}
\caption{An example of $\bm{\mathcal{S}}(l,\theta)$: when $\theta = \frac{\pi}{2}$, $\bm{\mathcal{S}}(l,\theta)$ is identical to $\bm{\mathcal{S}}(l)$.}
\label{fig:example_S_theta}
\end{figure}

In the following, we show (\ref{eqn:appendix:LF:markov key}). We first consider the event $\{I(k) = 0\}$. By definition, the event $\{I(k) = 0\}$ occurs if and only if $L(t)  > r $ for all $t\in(k-1,k]$, equivalently, $L(k-1+\delta)  > r $ for all $\delta\in(0,1]$. Since $\bm{X}_{i}(k-1+\delta) = (1-\delta)\bm{X}_{i}(k-1) + \delta\bm{X}_{i}(k)$ by (\ref{eqn:model:LF:location any time}), we have
\begin{align}\label{eqn:appendix:LF:diff location}
\Delta\bm{X}(k-1+\delta) = (1-\delta)\Delta\bm{X}(k-1) + \delta\Delta\bm{X}(k).
\end{align}
This implies that the event $\{L(k-1+\delta)  > r~\text{for all } \delta\in(0,1]\}$ occurs if and only if the following event occurs (See Fig.~\ref{fig:meeting_event}):
\begin{align}\label{eqn:appendix:LF:markov event}
\Big\{\overline{\big(\Delta\bm{X}(k-1), \Delta\bm{X}(k)\big)} \cap \bm{\mathcal{D}}_{r}(\bm{0}) = \varnothing\Big\}.
\end{align}
We next consider the event $\{I(k) = 0\}$ conditioned by $\Delta\bm{X}(k-1) = l \bm{e}^{j\theta}$ and $\mathcal{F}(k-1)$. Then, since $\bm{X}_i(k)= \bm{X}_i(k-1)+\bm{V}_i(k)$ by (\ref{eqn:model:mobility model:LF}), we have
\begin{align*}
\Delta\bm{X}(k)
&= \Delta\bm{X}(k-1)+ \Delta\bm{V}(k) = l \bm{e}^{j\theta} + \Delta\bm{V}(k).
\end{align*}
Thus, given the conditions $\Delta\bm{X}(k-1) = l\bm{e}^{j\theta}$ and $\mathcal{F}(k-1)$, (\ref{eqn:appendix:LF:markov event}) is reduced to the following:
\begin{align*}
& \Big\{\overline{\big(\Delta\bm{X}(k-1), \Delta\bm{X}(k)\big)} \cap \bm{\mathcal{D}}_{r}(\bm{0}) = \varnothing\Big\}\\
&\quad = \Big\{\overline{\big(l \bm{e}^{j\theta}, l \bm{e}^{j\theta} + \Delta\bm{V}(k)\big)} \cap \bm{\mathcal{D}}_{r}(\bm{0}) = \varnothing\Big\} \\
&\quad =\Big\{\overline{\big(\bm{0}, \Delta\bm{V}(k)\big)} \cap \bm{\mathcal{D}}_{r}(-l\bm{e}^{j\theta}) = \varnothing\Big\} \\
&\quad =\Big\{\Delta\bm{V}(k) \in \bm{\mathcal{S}}(l,\theta)\Big\},
\end{align*}
where
\begin{align*}
\bm{\mathcal{S}}(l,\theta) \deq \big\{\bm{x}\in\mathbb{R}^2  \,\big|\, \overline{(\bm{0}, \bm{x})} \cap \bm{\mathcal{D}}_{r}(-l\bm{e}^{j\theta}) = \varnothing\big\}.
\end{align*}
An example of $\bm{\mathcal{S}}(l,\theta)$ is shown in Fig.~\ref{fig:example_S_theta}. Hence, the probability on the left-hand side of (\ref{eqn:appendix:LF:markov key}) becomes
\begin{align}\label{eqn:appendix:LF:markov key -1}
&\p\{I(k) = 0 \,|\, \Delta\bm{X}(k\!-\!1) = l \bm{e}^{j\theta}, \mathcal{F}(k\!-\!1) \} \nonumber \\
&= \p\{\Delta\bm{V}(k) \in  \bm{\mathcal{S}}(l,\theta)\,|\, \Delta\bm{X}(k\!-\!1) = l \bm{e}^{j\theta}, \mathcal{F}(k\!-\!1)\}.
\end{align}
By (i) in Lemma~\ref{lemma:appendix:LF:property of delta V}, $\Delta\bm{V}(k)$ is independent of $\Delta\bm{X}(k-1)$ and $\mathcal{F}(k-1)$, and thus we have
\begin{align}\label{eqn:appendix:LF:markov (i)}
&\p\{\Delta\bm{V}(k) \in  \bm{\mathcal{S}}(l,\theta)\,|\, \Delta\bm{X}(k-1) = l \bm{e}^{j\theta}, \mathcal{F}(k-1)\} \nonumber \\
&\quad = \p\{\Delta\bm{V}(k) \in  \bm{\mathcal{S}}(l,\theta)\}.
\end{align}
In addition, by (ii) in Lemma~\ref{lemma:appendix:LF:property of delta V},
\begin{align}\label{eqn:appendix:LF:markov (ii)}
\p\{\Delta\bm{V}(k) \in  \bm{\mathcal{S}}(l,\theta)\} = \p\{\Delta\bm{V} \in  \bm{\mathcal{S}}(l,\theta)\}.
\end{align}
Finally, by (iii) in Lemma~\ref{lemma:appendix:LF:property of delta V}, the probability in (\ref{eqn:appendix:LF:markov (ii)}) is invariant for any $\theta\in(0,2\pi]$. When $\theta = \frac{\pi}{2}$, we have $\bm{\mathcal{S}}(l,\frac{\pi}{2}) = \bm{\mathcal{S}}(l)$. Hence, the following holds for any $\theta\in(0,2\pi]$:
\begin{align}\label{eqn:appendix:LF:markov (iii)}
\p\{\Delta\bm{V} \in  \bm{\mathcal{S}}(l,\theta)\} = \p\{\Delta\bm{V} \in  \bm{\mathcal{S}}(l)\}.
\end{align}
Combining (\ref{eqn:appendix:LF:markov key -1}), (\ref{eqn:appendix:LF:markov (i)}), (\ref{eqn:appendix:LF:markov (ii)}), and (\ref{eqn:appendix:LF:markov (iii)}) gives (\ref{eqn:appendix:LF:markov key}). This completes the proof. \hfill $\blacksquare$

%------------------------------------------------
%
% Lemma
%
%------------------------------------------------
\begin{lemma}\label{lemma:appendix:LF:app power law} Let $Z_1$, $Z_2$ and $\theta_1, \theta_2$ be independent copies of the generic random variables $Z$ (flight length) and $\theta$ (flight angle), respectively. Suppose that there exist constants $c\,(>0)$ and $z_{\text{th}}\,(>0)$ such that
\begin{align}\label{eqn:appendix:LF:ccdf}
\p\{Z > z\} = \frac{c}{z^\alpha}, \quad \text{ for all } z \ge z_{\text{th}}.
\end{align}
Then, for all $z \ge 2z_{\text{th}}$ we have
\begin{align*}
\frac{c_l}{z^\alpha} \le \p\{Z_1\cos\theta_1 - Z_2\cos\theta_2 > z\} \le \frac{c_u}{z^\alpha},
\end{align*}
where
\begin{align*}
c_l &\deq \frac{c}{2\pi} \int_{0}^{\frac{\pi}{2}} (\cos \vartheta)^\alpha \,\text{d}\vartheta \,(>0), \\
c_u &\deq \frac{2^{1+\alpha}c}{\pi} \int_{0}^{\frac{\pi}{2}} (\cos \vartheta)^\alpha \,\text{d}\vartheta \,(>0).
\end{align*}
\end{lemma}

\noindent\textit{Proof:} First, we will show that the distribution of $Z\cos\theta$ is of the following power-law form:
\begin{align}\label{eqn:appendix:LF:ccdf(1)}
\p\{Z\cos\theta > x\} = \frac{c_1}{x^\alpha}, \quad \text{ for } x \ge z_{\text{th}},
\end{align}
where $c_1 \deq \frac{c}{\pi}\int_{0}^{\frac{\pi}{2}} (\cos \vartheta)^\alpha \, \text{d}\vartheta \,(>0)$. By conditioning on the values of the random variable $\theta\sim\text{Uniform}[0, 2\pi]$, the probability $\p\{Z \cos\theta > x\}$ can be rewritten as
\begin{align}\label{eqn:appendix:LF:int}
\p\{Z\cos\theta > x\}
&= \frac{1}{2\pi} \int_{0}^{2\pi} \p\{Z\cos\vartheta > x\} \, \text{d}\vartheta \nonumber \\
&= \frac{1}{\pi} \int_{0}^{\pi} \p\{Z\cos\vartheta > x\} \, \text{d}\vartheta,
\end{align}
where the second equality comes from the symmetry of the function $\cos\vartheta$ with respect to $\vartheta = \pi$. For $x \ge 0$, the integral in (\ref{eqn:appendix:LF:int}) can be expressed as
\begin{align}\label{eqn:appendix:LF:int_two}
& \int_{0}^{\pi} \p\{Z\cos\vartheta > x\} \, \text{d}\vartheta \nonumber \\
&= \int_{0}^{\frac{\pi}{2}\!-\!\epsilon} \!\! \p\{Z\cos\vartheta > x\} \, \text{d}\vartheta + \int_{\frac{\pi}{2}\!-\!\epsilon}^{\frac{\pi}{2}} \!\! \p\{Z\cos\vartheta > x\} \, \text{d}\vartheta,
\end{align}
where $\epsilon \in (0, \frac{\pi}{2})$. The first integral in (\ref{eqn:appendix:LF:int_two}) becomes
\begin{align}\label{eqn:appendix:LF:int_1st}
\int_{0}^{\frac{\pi}{2}-\epsilon} \p\{Z\cos\vartheta > x\} \, \text{d}\vartheta
&= \int_{0}^{\frac{\pi}{2}-\epsilon} \p\{Z > \frac{x}{\cos\vartheta}\} \, \text{d}\vartheta \nonumber \\
&= \frac{c}{x^\alpha} \int_{0}^{\frac{\pi}{2}-\epsilon} (\cos \vartheta)^\alpha \, \text{d}\vartheta,
\end{align}
where the first equality comes from $\cos\vartheta > 0$ for $\vartheta \in [0, \frac{\pi}{2}-\epsilon]$ and the second equality comes from (\ref{eqn:appendix:LF:ccdf}) since $\frac{x}{\cos\vartheta} \ge z_{\text{th}}$ for $x \ge z_{\text{th}}$. The second integral in (\ref{eqn:appendix:LF:int_two}) is bounded by
\begin{align}\label{eqn:appendix:LF:int_2nd}
0 \le \int_{\frac{\pi}{2}-\epsilon}^{\frac{\pi}{2}} \p\{Z\cos\vartheta > x\} \, \text{d}\vartheta \le \int_{\frac{\pi}{2}-\epsilon}^{\frac{\pi}{2}} 1\, \text{d}\vartheta = \epsilon.
\end{align}
Combining (\ref{eqn:appendix:LF:int}), (\ref{eqn:appendix:LF:int_two}), (\ref{eqn:appendix:LF:int_1st}), and (\ref{eqn:appendix:LF:int_2nd}) gives
\begin{align}\label{eqn:appendix:LF:prob_main}
\frac{c}{\pi x^\alpha} \int_{0}^{\frac{\pi}{2}-\epsilon} \!\!(\cos \vartheta)^\alpha \, \text{d}\vartheta
&\le \p\{Z \cos\theta > x \} \nonumber \\
&\le \frac{c}{\pi x^\alpha} \int_{0}^{\frac{\pi}{2}-\epsilon}\!\! (\cos \vartheta)^\alpha \, \text{d}\vartheta + \frac{\epsilon}{\pi}.
\end{align}
Letting $\epsilon \to 0$ on (\ref{eqn:appendix:LF:prob_main}) yields
\begin{align*}
\p\{Z \cos\theta > x \}
&= \frac{c}{\pi x^\alpha} \int_{0}^{\frac{\pi}{2}} (\cos \vartheta)^\alpha \, \text{d}\vartheta.
\end{align*}
Hence, we have
\begin{align*}
\p\{Z \cos\theta > x \}
&= \frac{c_1}{x^\alpha}, \quad \text{ for } x \ge z_{\text{th}},
\end{align*}
where $c_1 \deq \frac{c}{\pi}\int_{0}^{\frac{\pi}{2}} (\cos \vartheta)^\alpha \, \text{d}\vartheta \,(>0)$. This proves~(\ref{eqn:appendix:LF:ccdf(1)}).

In the following, we derive the distribution of the random variable $Z_1\cos\theta_1 - Z_2\cos\theta_2$ by using (\ref{eqn:appendix:LF:ccdf(1)}). Since the event $\{Z_1\cos\theta_1 \le \frac{z}{2}\} \cap \{Z_2\cos\theta_2 \ge -\frac{z}{2}\}$ implies the event $\{Z_1\cos\theta_1 - Z_2\cos\theta_2 \le z\}$, we have
\begin{align}\label{eqn:appendix:LF:ccdf-1}
&\p\{Z_1\cos\theta_1 - Z_2\cos\theta_2 > z\} \nonumber \\
&\quad \le \p\{Z_1\cos\theta_1 > \frac{z}{2}\text{ or } Z_2\cos\theta_2 < -\frac{z}{2}\} \nonumber \\
&\quad \le 2 \p\{Z\cos\theta > \frac{z}{2}\},
\end{align}
where the last inequality comes from the union bound and the symmetry of $Z\cos\theta$ (i.e., $Z\cos\theta \ed - Z\cos\theta$). Suppose $z \ge 2 z_{\text{th}}$. Then, by applying (\ref{eqn:appendix:LF:ccdf(1)}) to~(\ref{eqn:appendix:LF:ccdf-1}), we further have
\begin{align}\label{eqn:appendix:LF:ccdf-2}
\p\{Z_1\cos\theta_1 - Z_2\cos\theta_2 > z\} \le \frac{2^{\alpha+1}c_1}{z^\alpha} = \frac{c_u}{z^\alpha}.
\end{align}
Similarly, since the event $\{Z_1\cos\theta_1 > z\} \cap \{Z_2\cos\theta_2 < 0\}$ implies the event $\{Z_1\cos\theta_1 - Z_2\cos\theta_2 > z\}$, we have
\begin{align}\label{eqn:appendix:LF:ccdf-3}
&\p\{Z_1\cos\theta_1 - Z_2\cos\theta_2 > z\} \nonumber \\
&\quad \ge \p\{Z_1\cos\theta_1 > z \text{ and } Z_2\cos\theta_2 < 0\} \nonumber \\
&\quad = \p\{Z_1\cos\theta_1 > z\}\p\{Z_2\cos\theta_2 < 0\} \nonumber \\
&\quad = \frac{c_1}{2z^\alpha} = \frac{c_l}{z^\alpha},
\end{align}
where the first equality comes from the independence between $Z_1\cos\theta_1$ and $Z_2\cos\theta_2$, and the second equality comes from (\ref{eqn:appendix:LF:ccdf(1)}) and the symmetry of $Z_2\cos\theta_2$. Combining (\ref{eqn:appendix:LF:ccdf-2}) and (\ref{eqn:appendix:LF:ccdf-3}) gives the lemma. \hfill $\blacksquare$

%------------------------------------------------
%
% Lemma
%
%------------------------------------------------
\smallskip\smallskip
\section*{Proof of Lemma~\ref{lemma:analysis:LF:property H(k,l)}}
\subsection*{A.~Proof of (i)} By choosing $k=1$, $l = l_0$, and $\mathcal{L}(0) = (r, 2\sqrt{n}]$ in Lemma~\ref{lemma:appendix:LF:markov}, we have
\begin{align*}
\p\{I(1) = 0 \,|\, L(0) = l_0, L(0) \in (r, 2\sqrt{n}]\} = \p\{\Delta\bm{V} \in \bm{\mathcal{S}}(l_0)\}.
\end{align*}
Since $\{L(0) = l_0\} \cap \{L(0) \in (r, 2\sqrt{n}]\} = \{L(0) = l_0\}$, we further have
\begin{align}\label{eqn:appendix:LF:cha-(i)}
\p\{I(1) = 0 \,|\, L(0) = l_0\} & = \p\{\Delta\bm{V} \in \bm{\mathcal{S}}(l_0)\}.
\end{align}
By definition, $H(1,l_0) = \p\{I(1) = 0 \,|\, L(0) = l_0\}$. Thus, we have from~(\ref{eqn:appendix:LF:cha-(i)}) that $H(1,l_0) = \p\{\Delta\bm{V} \in \bm{\mathcal{S}}(l_0)\}.$

\subsection*{B.~Proof of (ii)}
Suppose $r < l_0 \le l_1 \le 2\sqrt{n}$. Then, it is obvious from the definition of $\bm{\mathcal{S}}(\cdot)$ in~(\ref{eqn:analysis:LF:definition S(l)}) that $\bm{\mathcal{S}}(l_0) \subseteq \bm{\mathcal{S}}(l_1)$ (See Fig.~\ref{fig:example_S}). Hence, we have
\begin{align*}
\p\{\Delta\bm{V} \in \bm{\mathcal{S}}(l_0)\} \le
\p\{\Delta\bm{V} \in \bm{\mathcal{S}}(l_1)\},
\end{align*}
which is equivalent to $H(1,l_0) \le H(1,l_1)$ by (i) in Lemma~\ref{lemma:analysis:LF:property H(k,l)}.

\subsection*{C.~Proof of (iii)}
By (ii) in Lemma~\ref{lemma:analysis:LF:property H(k,l)}, we have $H(1,l_0) \le H(1, 2\sqrt{n}) \, (= \hat{P})$ for any $l_0\in(r,2\sqrt{n}]$.

\subsection*{D.~Proof of (iv)}
Recall the definition of $H(k,l_0)$ for $k=2,3,\ldots$:
\begin{align*}
H\!(k,l_0\!) \deq \p\{I\!(k\!) = 0 | I\!(k\!-\!1\!) =  \ldots  = I\!(1\!) = 0, L\!(0\!) = l_0\}.
\end{align*}
By conditioning on the values of $L(k-1)$, the probability $H(k,l_0)$ can be rewritten as
\begin{align}\label{eqn:appendix:LF:H(k,l) -1}
&H(k,l_0) \nonumber \\
&\!=\! \int_{r^+}^{2\sqrt{n}} \!\! \p\{I(k)= 0 | L(k\!-\!1) = l, I(k\!-\!1) = \ldots = I(1) = 0, \nonumber \\
&\hspace{13 mm} L(0) = l_0\} \,\text{d} F_{L\!(\!k\!-\!1\!) | I\!(\!k\!-\!1\!) = \ldots =I\!(\!1\!)= 0, L\!(\!0\!) = l_0}(l),
\end{align}
where $F_{L(k-1) | I(k-1) = \ldots =I(1)= 0, L(0) = l_0}(\cdot)$ denotes the CDF of $L(k-1)$ conditioned that $I(k-1) = \ldots =I(1)= 0$ and $L(0) = l_0$. Here, we integrate $l \deq L(k-1)$ over $(r, 2\sqrt{n}]$ due to the condition $I(k-1) = 0$. By using Lemma~\ref{lemma:appendix:LF:markov}, the probability in the integral in~(\ref{eqn:appendix:LF:H(k,l) -1}) is simplified as follows:
\begin{align}\label{eqn:appendix:LF:H(k,l) -2}
& \p\{I(k) \!= \!0 | L(k-1) \!=\! l, I(k-1) \!=\! \ldots \!=\! I(1) \!=\! 0, L(0) \!=\! l_0\} \nonumber \\
& = \p\{I(k) = 0 | L(k\!-\!1) = l, L(t) \in (r,2\sqrt{n}], \nonumber \\
& \hspace{9mm} 0 < t \le k-1, L(0) = l_0\} \nonumber \\
& = \p\{\Delta\bm{V} \in \bm{\mathcal{S}}(l)\}.
\end{align}
By (i) and (iii) in Lemma~\ref{lemma:analysis:LF:property H(k,l)}, the probability $\p\{\Delta\bm{V} \in \bm{\mathcal{S}}(l)\}$ is bounded for all $l\in(r,2\sqrt{n}]$ by
\begin{align}\label{eqn:appendix:LF:H(k,l) -3}
\p\{\Delta\bm{V} \in \bm{\mathcal{S}}(l)\} &= H(1,l) \le \hat{P}.
\end{align}
By substituting the combined result of (\ref{eqn:appendix:LF:H(k,l) -2}) and (\ref{eqn:appendix:LF:H(k,l) -3}) into (\ref{eqn:appendix:LF:H(k,l) -1}), we have for all $k=2,3,\ldots$ and $l_0 \in (r, 2\sqrt{n}]$ the following:
\begin{align*}
H(k,l_0) &\le  \hat{P} \cdot 1 = \hat{P}.
\end{align*}
This proves (iv) in Lemma~\ref{lemma:analysis:LF:property H(k,l)}.

\begin{figure}[t!]
\centering
{\epsfig{figure=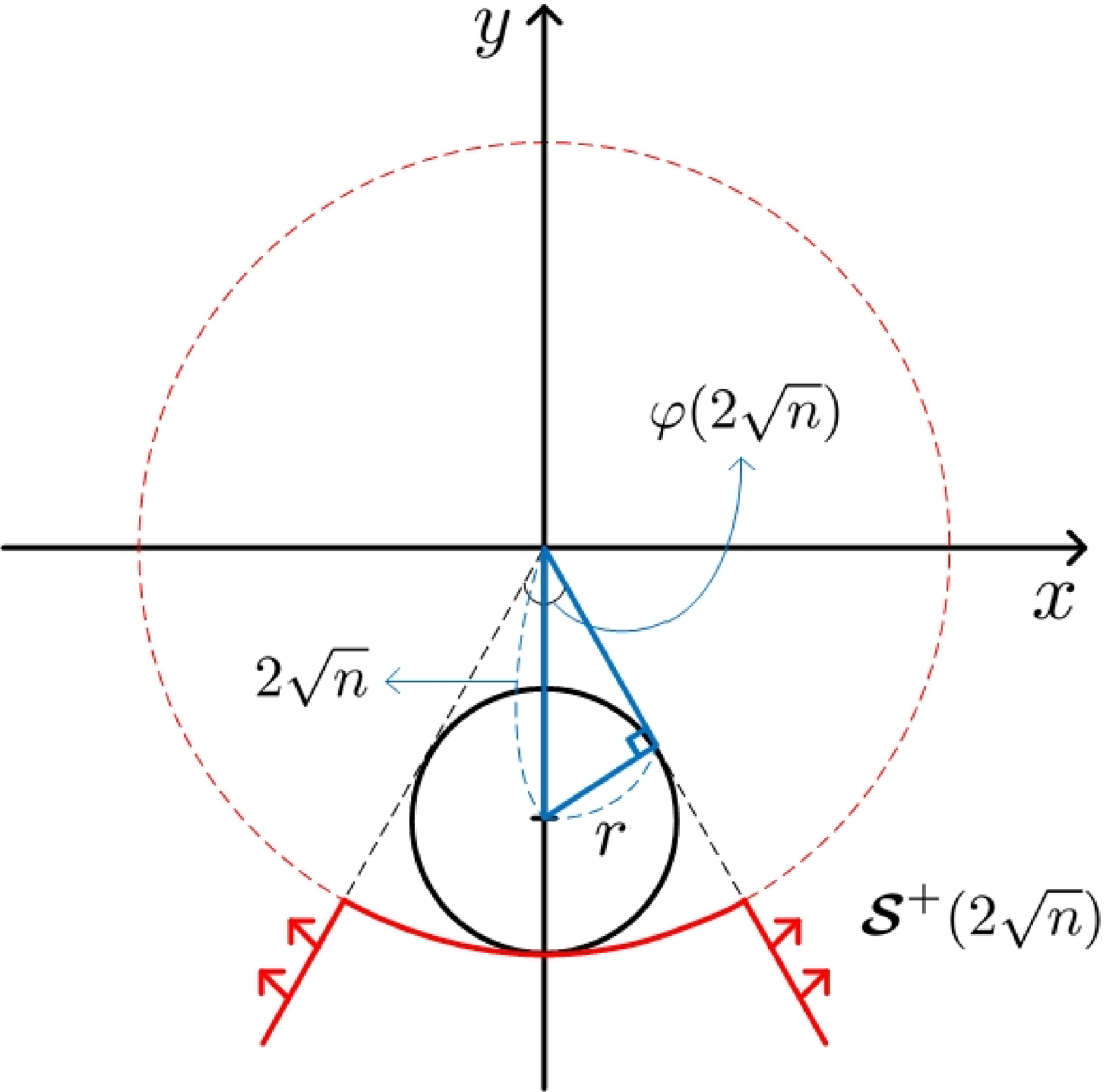,width=0.22\textwidth}}
{\epsfig{figure=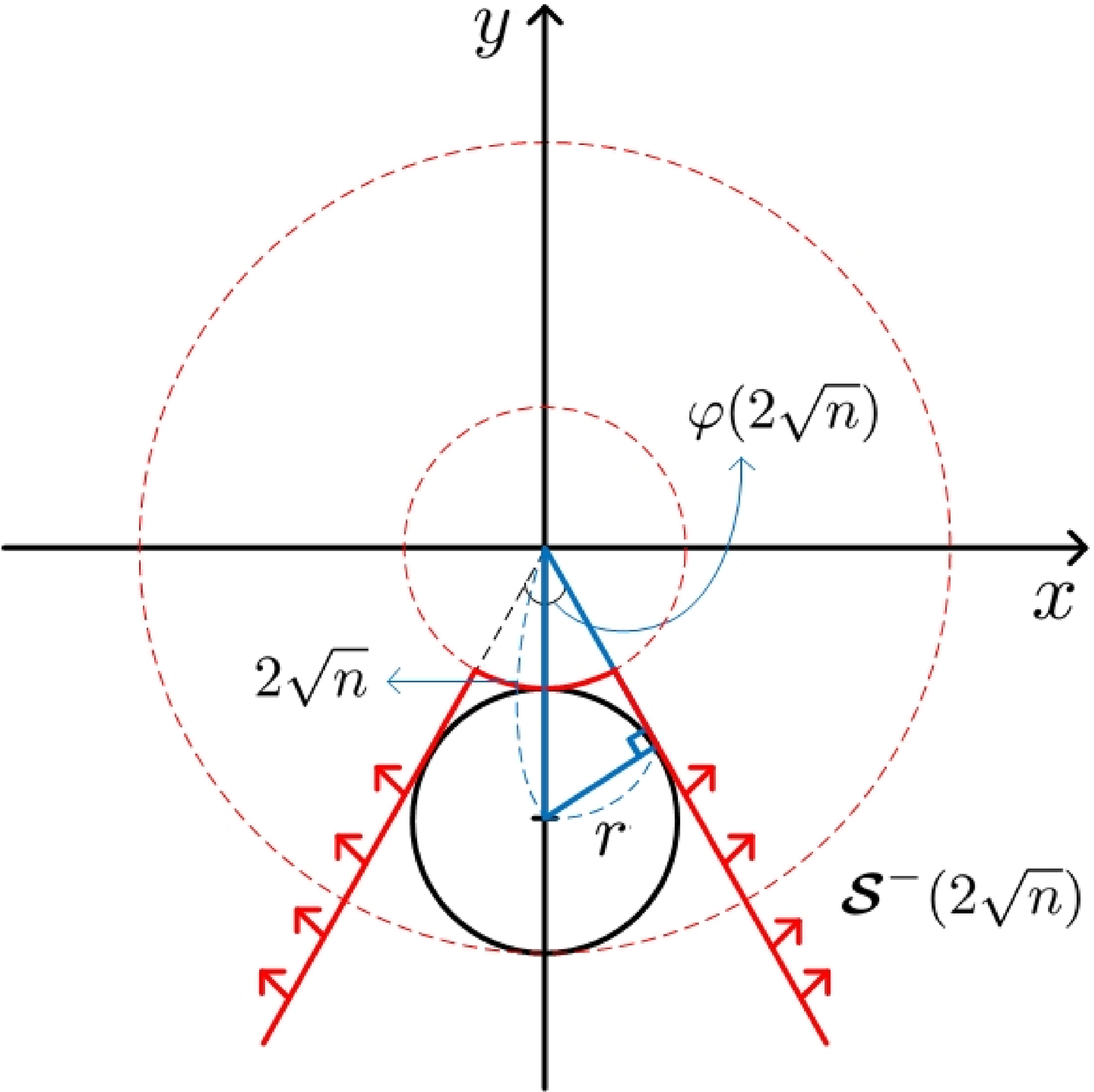,width=0.215\textwidth}}
\caption{The geometric definitions of the superset $\bm{\mathcal{S}}^+(2\sqrt{n})$ (Left) and the subset $\bm{\mathcal{S}}^-(2\sqrt{n})$ (Right) of the set $\bm{\mathcal{S}}(2\sqrt{n})$ used in the proof of (v) in Lemma~\ref{lemma:analysis:LF:property H(k,l)}.}
\label{fig:example_S_sub_sup}
\end{figure}

\subsection*{E.~Proof of (v)}
By (i) in Lemma~\ref{lemma:analysis:LF:property H(k,l)}, $\hat{P} = \p\{\Delta\bm{V} \in \bm{\mathcal{S}}(2\sqrt{n})\}$. To derive a lower and upper bound on $\hat{P}$, we define a subset $\bm{\mathcal{S}}^-(2\sqrt{n})$ and a superset $\bm{\mathcal{S}}^+(2\sqrt{n})$ of the set $\bm{\mathcal{S}}(2\sqrt{n})$ as depicted in Fig.~\ref{fig:example_S_sub_sup}. Then, we have
\begin{align}\label{eqn:appendix:LF:P^o bound}
\p\{\Delta\bm{V} \in \bm{\mathcal{S}}^-(2\sqrt{n}) \} \le \hat{P}  \le \p\{\Delta\bm{V} \in \bm{\mathcal{S}}^+(2\sqrt{n}) \}.
\end{align}
By (iii) in Lemma~\ref{lemma:appendix:LF:property of delta V}, the probabilities $\p\{\Delta\bm{V} \in \bm{\mathcal{S}}^{\pm}(2\sqrt{n}) \}$ are obtained by (double sings in same order)
\begin{align}\label{eqn:appendix:LF:P^o bound -1}
& \p\{\Delta\bm{V} \in \bm{\mathcal{S}}^{\pm}(2\sqrt{n})\} \nonumber \\
&\quad = 1- \p\{\Delta\bm{V} \notin \bm{\mathcal{S}}^{\pm}(2\sqrt{n})\} \nonumber \\
&\quad = 1- \p\{|\Delta\bm{V}| \ge 2\sqrt{n} \pm r\}\cdot\frac{\varphi(2\sqrt{n})}{2\pi},
\end{align}
where $\varphi(2\sqrt{n})$ is the central angle associated with $\bm{\mathcal{S}}^{\pm}(2\sqrt{n})$ (See Fig.~\ref{fig:example_S_sub_sup}). From the geometry in Fig.~\ref{fig:example_S_sub_sup}, the angle $\varphi(2\sqrt{n})$ is given by
\begin{align}\label{eqn:appendix:LF:P^o central angle}
\varphi(2\sqrt{n}) = 2\sin^{-1}\left(\frac{r}{2\sqrt{n}}\right).
\end{align}
We now consider the probabilities $\p\{|\Delta\bm{V}| \ge 2\sqrt{n} \pm r\}$ in (\ref{eqn:appendix:LF:P^o bound -1}). For notational simplicity, we denote $\Delta\bm{V} = (\Delta V_x, \Delta V_y)$. Then, by (ii) in Lemma~\ref{lemma:appendix:LF:property of delta V},
\begin{align*}
\Delta V_x &\ed Z_i(k)\cos\theta_i(k) - Z_j(k)\cos\theta_j(k), \\
\Delta V_y &\ed Z_i(k)\sin\theta_i(k) - Z_j(k)\sin\theta_j(k).
\end{align*}
Note that for any $\bm{v} = (v_x, v_y)\in\mathbb{R}^2$ and $\eta \ge 0$, $|v_x| \ge \eta$ implies $|\bm{v}| \ge \eta$, and $|\bm{v}| \ge \eta$ implies $|v_x| \ge \eta/\sqrt{2}$ or $|v_y| \ge \eta/\sqrt{2}$. Hence, $\p\{|\bm{v}| > \eta\}$ is bounded by
\begin{subequations}
\begin{align}
\p\{|v_x| \ge \eta\} &\le \p\{|\bm{v}| \ge \eta\} \label{eqn:appendix:LF:P^o bound lower}\\
&\le \p\{|v_x| \ge \eta/\sqrt{2} \text{ or } |v_y| \ge \eta/\sqrt{2}\}. \label{eqn:appendix:LF:P^o bound upper}
\end{align}
\end{subequations}
Since $\theta_i(k)$ and $\theta_j(k)$ are independent and uniformly distributed over~$(0,2\pi]$, $\Delta V_x$ is symmetric, i.e., $\Delta V_x \ed - \Delta V_x$. Thus, applying (\ref{eqn:appendix:LF:P^o bound lower}) with $\bm{v} = \Delta\bm{V}$ and $\eta = 2\sqrt{n} + r$ yields
\begin{align*}
\p\{|\Delta\bm{V}| \ge 2\sqrt{n} + r \}
& \ge \p\{|\Delta V_x| \ge 2\sqrt{n} + r\} \\
&= 2 \p\{\Delta V_x \ge 2\sqrt{n} + r\}.
\end{align*}
Since $z_{\text{th}}$ in Lemma~\ref{lemma:appendix:LF:app power law} is a constant independent of $n$, there exists a constant $n_{\text{th},l} \in\mathbb{N}$ such that $2\sqrt{n} + r \ge 2z_{\text{th}}$ for all $n \ge n_{\text{th},l}$. Hence, by Lemma~\ref{lemma:appendix:LF:app power law}, we have for all $n \ge n_{\text{th},l}$
\begin{align}\label{eqn:appendix:LF:P^o bound lower -1}
\p\{|\Delta\bm{V}| \ge 2\sqrt{n} + r \}
&\ge 2 c_l \left(\frac{1}{2\sqrt{n} + r}\right)^\alpha.
\end{align}
Since $\cos\theta \ed \sin\theta$ for $\theta\sim\text{Uniform}[0,2\pi]$, $|\Delta V_x| \ed |\Delta V_y|$. Thus, applying (\ref{eqn:appendix:LF:P^o bound upper}) with $\bm{v} = \Delta\bm{V}$ and $\eta = 2\sqrt{n} - r$ yields
\begin{align*}
&\p\{|\Delta\bm{V}| \ge 2\sqrt{n} - r\} \\
&\le \p\{|\Delta V_x| \ge (2\sqrt{n} - r)/\sqrt{2} \text{ or } |\Delta V_y| \ge (2\sqrt{n} - r)/\sqrt{2}\} \\
&\le 2 \p\{|\Delta V_x| \ge (2\sqrt{n} - r)/\sqrt{2}\} \\
&= 4 \p\{\Delta V_x \ge (2\sqrt{n} - r)/\sqrt{2}\}.
\end{align*}
By the same reason as above, there exists a constant $n_{\text{th},u} \in\mathbb{N}$ such that $(2\sqrt{n} - r)/\sqrt{2} \ge 2z_{\text{th}}$ for all $n \ge n_{\text{th},u}$. Hence, by Lemma~\ref{lemma:appendix:LF:app power law}, we have for all $n \ge n_{\text{th},u}$
\begin{align}\label{eqn:appendix:LF:P^o bound upper -1}
\p\{|\Delta\bm{V}| \ge 2\sqrt{n} - r\}
&\le 2^{\alpha/2+2} c_u \left(\frac{1}{2\sqrt{n} - r}\right)^\alpha.
\end{align}
Combining (\ref{eqn:appendix:LF:P^o bound}), (\ref{eqn:appendix:LF:P^o bound -1}), (\ref{eqn:appendix:LF:P^o central angle}), (\ref{eqn:appendix:LF:P^o bound lower -1}), and (\ref{eqn:appendix:LF:P^o bound upper -1}) yields
\begin{align*}
\hat{P} &\le 1- \frac{2c_l}{\pi}  \left(\frac{1}{2\sqrt{n} + r}\right)^\alpha \sin^{-1}\left(\frac{r}{2\sqrt{n}}\right), \\
\hat{P} &\ge 1- \frac{2^{\alpha/2+2}c_u}{\pi}  \left(\frac{1}{2\sqrt{n} - r}\right)^\alpha \sin^{-1}\left(\frac{r}{2\sqrt{n}}\right),
\end{align*}
for all $n \ge n_{\text{th}} \deq \max(n_{\text{th},l},n_{\text{th},u})$.

%------------------------------------------------
%
% Lemma
%
%------------------------------------------------
\smallskip\smallskip
\section*{Proof of Lemma~\ref{lemma:analysis:LF:delay scaling}}
To complete the proof of Lemma~\ref{lemma:analysis:LF:delay scaling}, it remains to show that (i) $(1-\hat{P})^{-1} = \Theta(n^{(1+\alpha)/2-\beta})$ and (ii) $\bar{D}_{\hat{\pi}} = O(n)$. Without loss of generality, we assume $r = n^\beta$ $(0\le \beta\le1/4)$.

\subsection*{A.~Proof of (i)}
To prove (i), we need the following: for any $x \in [0,1]$,
\begin{align}\label{eqn:appendix:LF:scaling 1-P^o - inv sin}
x \le \sin^{-1}\left(x\right) \le \frac{\pi}{2}x.
\end{align}
The proof of (\ref{eqn:appendix:LF:scaling 1-P^o - inv sin}) is given at the end of this section. From (\ref{eqn:analysis:LF:property P^o-2}) in Lemma~\ref{lemma:analysis:LF:property H(k,l)} with $r= n^\beta$, we have for all $n\ge n_{\text{th}}$ the following:
\begin{align*}
1 - \hat{P} &\ge \frac{2c_l}{\pi} \left(\frac{1}{2\sqrt{n} + n^\beta}\right)^\alpha \sin^{-1}\left(\frac{n^{\beta-1/2}}{2}\right).
\end{align*}
Since $\frac{n^{\beta-1/2}}{2}\in[0,1]$ for any $\beta\in[0,1/4]$ and $n\in\mathbb{N}$, we further have from the lower inequality in (\ref{eqn:appendix:LF:scaling 1-P^o - inv sin}) that
\begin{align*}
1- \hat{P} &\ge \frac{c_l}{\pi} \left(\frac{1}{2\sqrt{n} + n^\beta}\right)^\alpha n^{\beta-1/2}.
\end{align*}
Hence, we have
\begin{align*}
\limsup_{n\to\infty}\frac{(1- \hat{P})^{-1}}{n^{(1+\alpha)/2-\beta}} &\le  \frac{2^\alpha \pi}{c_l} <\infty,
\end{align*}
which gives
\begin{align}\label{eqn:appendix:LF:scaling 1-P^o:lower}
(1- \hat{P})^{-1} = O(n^{(1+\alpha)/2-\beta}).
\end{align}
Using a similar approach as above, from (\ref{eqn:analysis:LF:property P^o-1}) in Lemma~\ref{lemma:analysis:LF:property H(k,l)} and the upper inequality in (\ref{eqn:appendix:LF:scaling 1-P^o - inv sin}), we have for all $n\ge n_{\text{th}}$ the following:
\begin{align*}
1- \hat{P} &\le 2^{\alpha/2}c_u \left(\frac{1}{2\sqrt{n} - n^\beta}\right)^\alpha n^{\beta-1/2}.
\end{align*}
Hence, we have
\begin{align*}
\limsup_{n\to\infty} \frac{n^{(1+\alpha)/2-\beta}}{(1- \hat{P})^{-1}} &\le  2^{-\alpha/2}c_u <\infty,
\end{align*}
which gives
\begin{align}\label{eqn:appendix:LF:scaling 1-P^o:upper}
(1- \hat{P})^{-1} = \Omega(n^{(1+\alpha)/2-\beta}).
\end{align}
Combining (\ref{eqn:appendix:LF:scaling 1-P^o:lower}) and (\ref{eqn:appendix:LF:scaling 1-P^o:upper}) proves (i).

\smallskip
\noindent\textit{\underline{Proof of (\ref{eqn:appendix:LF:scaling 1-P^o - inv sin}):}}
For $|x| \le 1$, the function $\sin^{-1}\left(x\right)$ can be calculated using the following infinite series:
\begin{align*}
\sin^{-1}\left(x\right) = \sum_{l=0}^{\infty} d_l \, x^{2l+1},
\end{align*}
where $d_l \deq \frac{(2l)!}{4^l (l!)^2 (2l+1)}\,(> 0)$.
Hence, for any $x \in [0,1]$, we have a lower bound on $\sin^{-1}\left(x\right)$ as
\begin{align}\label{eqn:appendix:LF:sin(1)}
\sin^{-1}\left(x\right) \ge d_0 \, x =  x.
\end{align}
Since $x^{2l+1} \le x$ for all $l=0,1,\ldots$ and $x\in[0,1]$, we have
\begin{align*}
\sin^{-1}\left(x\right) \le  x \sum_{l=0}^{\infty} d_l.
\end{align*}
Note that $\sum_{l=0}^{\infty}d_l = \sin^{-1}\left(1\right) = \frac{\pi}{2}$. Hence, for any $x \in [0,1]$, we have an upper bound on $\sin^{-1}\left(x\right)$ as
\begin{align}\label{eqn:appendix:LF:sin(2)}
\sin^{-1}\left(x\right) \le \frac{\pi}{2} x.
\end{align}
Combining (\ref{eqn:appendix:LF:sin(1)}) and (\ref{eqn:appendix:LF:sin(2)}) proves (\ref{eqn:appendix:LF:scaling 1-P^o - inv sin}).

\subsection*{B.~Proof of (ii)}
Without loss of generality, we assume $\p\{Z_\alpha > z_{\text{th}}\} = 1$. (In this proof, subscript $\alpha$ is added to all random variables to specify the underlying parameter~$\alpha$ of the L\'{e}vy flight model.) Then, from~(\ref{eqn:model:power law ccdf}), we have $\p\{Z_{\alpha} > z\}  = (\frac{z_{\text{th}}}{z})^{\alpha}$ for all $z \ge z_{\text{th}}$, which gives for any $0 < \alpha_1 \le \alpha_2 \le 2$ and $z \ge z_{\text{th}}$ the following:
\begin{align}\label{eqn:appendix:LF:flight dominance}
\p\{Z_{\alpha_1} > z\}  = \Big(\frac{z_{\text{th}}}{z}\Big)^{\alpha_1} \ge \Big(\frac{z_{\text{th}}}{z}\Big)^{\alpha_2} = \p\{Z_{\alpha_2} > z\}.
\end{align}
The inequality in (\ref{eqn:appendix:LF:flight dominance}) shows that for any $t_2>t_1 \ge 0$ having a sufficiently small difference $\epsilon \deq t_2 - t_1 >0$, we get
\begin{align*}
\p\{L_{\alpha_1}\!(t_2) > r | L_{\alpha_1}\!(t_1) > r\} \le \p\{L_{\alpha_2}\!(t_2) > r | L_{\alpha_2}\!(t_1) > r\},
\end{align*}
which results in
\begin{align}\label{eqn:appendix:LF:trivial scaling(1)}
\p\{T_{\alpha_1} > t_2 \,|\, T_{\alpha_1} > t_1\} \le \p\{T_{\alpha_2} > t_2\,|\, T_{\alpha_2} > t_1\}.
\end{align}
Note that since $\p\{T_{\alpha_1} > t\} = \p\{T_{\alpha_1} > t, T_{\alpha_1} > t-\epsilon\}$ for $t \ge \epsilon$, we can express $\p\{T_{\alpha_1} > t\}$ in a nested form as
\begin{align*}
\p\{T_{\alpha_1} > t\} &= \p\{T_{\alpha_1} > t \,|\, T_{\alpha_1} > t-\epsilon\} \p\{T_{\alpha_1} > t-\epsilon\}.
\end{align*}
Using the nested form continuously, we have
\begin{align}\label{eqn:appendix:LF:trivial scaling(2)}
\p\{T_{\alpha_1} > t\}
&= \p\{T_{\alpha_1} > t \,|\, T_{\alpha_1} > t-\epsilon\} \nonumber \\
& \qquad \times\p\{T_{\alpha_1} > t -\epsilon \,|\, T_{\alpha_1} > t-2\epsilon\} \nonumber \\
& \qquad \times \ldots \nonumber \\
& \qquad \times \p\{T_{\alpha_1} > t- \lfloor t/\epsilon \rfloor \epsilon \,|\, T_{\alpha_1} > 0 \} \nonumber \\
& \qquad \times \p\{T_{\alpha_1} > 0\}.
\end{align}
Hence, by applying (\ref{eqn:appendix:LF:trivial scaling(1)}) to (\ref{eqn:appendix:LF:trivial scaling(2)}), we have \begin{align}\label{eqn:appendix:LF:trivial scaling(3)}
\p\{T_{\alpha_1} > t\}
&\le \p\{T_{\alpha_2} > t \,|\, T_{\alpha_2} > t-\epsilon\} \nonumber \\
& \qquad \times\p\{T_{\alpha_2} > t -\epsilon \,|\, T_{\alpha_2} > t-2\epsilon\} \nonumber \\
& \qquad \times \ldots \nonumber \\
& \qquad \times \p\{T_{\alpha_2} > t-\lfloor t/\epsilon \rfloor \epsilon \,|\, T_{\alpha_2} > 0 \} \nonumber \\
& \qquad \times \p\{T_{\alpha_1} > 0\}.
\end{align}
Note that $\p\{T_{\alpha_1} > 0\} = \p\{L_{\alpha_1}(0) > r\}$. In addition, since $\bm{X}_i(0)\sim\text{Uniform}(\bm{\mathcal{D}})$ for all $i=1,\ldots,n$ regardless of $\alpha$, we have $\p\{T_{\alpha_1} > 0\} = \p\{T_{\alpha_2} > 0\}$. Thus, the right-hand side of (\ref{eqn:appendix:LF:trivial scaling(3)}) boils down to $\p\{T_{\alpha_2} > t\}$, and consequently
\begin{align}\label{eqn:appendix:LF:ccdf dominance}
\p\{T_{\alpha_1} > t\}
&\le \p\{T_{\alpha_2} > t\} \quad \text{for all } t \ge 0.
\end{align}
Due to the property in (\ref{eqn:appendix:LF:ccdf dominance}), the average delay under the L\'{e}vy flight model with a parameter~$\alpha \in (0,2)$ is dominated by the one under Brownian motion $(\alpha = 2)$, which is shown to be $O(n)$~\cite{SL21Lin}, i.e.,
\begin{align*}
\bar{D}_{\hat{\pi}} = O(n) \quad \text{for all } \alpha\in(0,2].
\end{align*}

\section{Proofs of Lemmas for the \emph{i.i.d.} Mobility Model}
Here, we give detailed proofs of Lemmas~\ref{lemma:analysis:iid:property H(k,l)} and~\ref{lemma:analysis:iid:delay scaling}, which are used for analyzing the optimal delay under the \emph{i.i.d.} mobility model. To prove Lemma~\ref{lemma:analysis:iid:property H(k,l)}, we need the following Lemmas~\ref{lemma:appendix:iid:property of delta X},~\ref{lemma:appendix:iid:markov}, and~\ref{lemma:appendix:iid:pdf}.

%------------------------------------------------
%
% Lemma
%
%------------------------------------------------
\begin{lemma}\label{lemma:appendix:iid:property of delta X}
For $i \neq j$ and $t \ge 0$, let
\begin{align*}
\Delta\bm{X}_{(i,j)}(t) \deq \bm{X}_i(t) - \bm{X}_j(t),
\end{align*}
where $\bm{X}_{\cdot}(t)$ denotes the location of a node $\cdot$ at time $t$. Then, under the \emph{i.i.d.} mobility model, $\Delta\bm{X}_{(i,j)}(t)$ has the following properties: \\
(i) $\Delta\bm{X}_{(i,j)}(k)\,(k\in\mathbb{N})$ is independent of $\bm{X}_u(t)$ for all $u=1,\ldots,n$ and $t\in[0,k-1]$. \\
(ii) $\Delta\bm{X}_{(i,j)}(t)$ is identically distributed across pair index $(i,j)$ and time~$t\,(\ge 0)$. Hence, we use $\Delta\bm{X}$ to denote a generic random variable for $\Delta\bm{X}_{(i,j)}(t)$. \\
(iii) The angle $\angle\Delta\bm{X}$ is a uniform random variable on the interval $(0,2\pi]$ and is independent of the length $|\Delta\bm{X}|$.
\end{lemma}

\noindent\textit{Proof:} (i) For any $u=1,\ldots,n$, $\bm{X}_u(t)\,(0 \le t \le k-1)$ under the \emph{i.i.d.} mobility model is completely determined by $\bm{G}_u(k-1) \deq \{\bm{X}_u(0),\ldots,\bm{X}_u(k-1)\}$ (by the relation~(\ref{eqn:model:LF:location any time})). Since $\bm{X}_i(k)$ is independent of $\bm{G}_u(k-1)$, it is independent of $\bm{X}_u(t)$. By the same reason, $\bm{X}_j(k)$ is independent of $\bm{X}_u(t)$. Therefore, the difference $\bm{X}_i(k) - \bm{X}_j(k)$ is independent of $\bm{X}_u(t)$. \\
(ii) For any $i \neq j$ and $t\ge 0$, $\bm{X}_i(t)$ and $\bm{X}_j(t)$ are independent and identically distributed. Therefore, the difference $\bm{X}_i(t)-\bm{X}_j(t)$ is identically distributed across pair index $(i,j)$ and time~$t$. However, it is not necessarily independent neither across~$(i,j)$ nor across $t$. \\
(iii) To prove (iii), it suffices to show that for any $x\ge 0$,
\begin{align}\label{eqn:appendix:iid:circular sym}
\p\big\{\angle\Delta\bm{X} \le \theta \,\big|\, |\Delta\bm{X}| = x\big\} = \frac{\theta}{2\pi},
\end{align}
where $0 < \theta \le 2\pi$. By noting that $\angle\bm{X}_i(t)\sim\text{Uniform}[0,2\pi]$ for any $i=1,\ldots,n$ and $t\ge 0$ and using a similar approach as in the proof of (iii) in Lemma~\ref{lemma:appendix:LF:property of delta V}, we can prove (iii) in Lemma~\ref{lemma:appendix:iid:property of delta X}. Due to similarities, we omit the details. \hfill $\blacksquare$

%------------------------------------------------
%
% Lemma
%
%------------------------------------------------
\begin{lemma}\label{lemma:appendix:iid:markov} Suppose $k\in\mathbb{N}$ and $l\in(r,2\sqrt{n}]$. Then, for any sets $\mathcal{L}(\cdot) \subset [0, 2\sqrt{n}]$ satisfying
\begin{align*}
\p\{L(k-1) = l, L(t) \in \mathcal{L}(t), 0 \le t \le k-1\} > 0,
\end{align*}
we have under the \emph{i.i.d.} mobility model the following:
\begin{align}\label{eqn:appendix:iid:markov}
& \p\{I(k) = 0 \,|\, L(k-1) = l, L(t) \in \mathcal{L}(t), 0 \le t \le k-1\} \nonumber \\
& \quad = \p\{\Delta\bm{X} \in \bm{\mathcal{S}}^\star(l)\}.
\end{align}
The definitions of $\Delta\bm{X}$ and $\bm{\mathcal{S}}^\star(l)$ can be found in Lemma~\ref{lemma:analysis:iid:property H(k,l)}.
\end{lemma}

\begin{remark}
Before proving the lemma, we give a remark. As Lemma~\ref{lemma:appendix:LF:markov} for the L\'{e}vy flight model, Lemma~\ref{lemma:appendix:iid:markov} implies that the future states of a meeting process under the \emph{i.i.d.} mobility model depend only on the state at the beginning of the current slot, not on the sequence of events that preceded it. In addition, the conditional probability distribution of the future state described above is time homogeneous (i.e., the probability in~(\ref{eqn:appendix:iid:markov}) does not depend on the slot index $k$). This restricted time homogeneous memoryless property enables us to derive a bound on the first meeting time distribution as a \emph{geometric form} (See Lemma~\ref{lemma:analysis:iid:ccdf}).
\end{remark}

\noindent\textit{Proof:} Using a similar approach as in the proof of Lemma~\ref{lemma:appendix:LF:markov}, we can prove Lemma~\ref{lemma:appendix:iid:markov}. The difference is that the key idea of this proof is to use the following equality: \emph{for any} $k\in\mathbb{N}$, $l\in(r,2\sqrt{n}]$, $\theta\in(0,2\pi]$, and $\mathcal{F}(k-1)$, we have
\begin{align}\label{eqn:appendix:iid:markov key}
&\p\{I(k) = 0 \,|\, \Delta\bm{X}(k-1) = l\bm{e}^{j\theta}, \mathcal{F}(k-1)\} \nonumber \\
&\quad = \p\{\Delta\bm{X}\in\bm{\mathcal{S}}^\star(l)\},
\end{align}
where the definition of $\mathcal{F}(k-1)$ can be found in (\ref{eqn:appendix:LF:def_history}). Then, similarly to the proof of Lemma~\ref{lemma:appendix:LF:markov}, using the key equality in (\ref{eqn:appendix:iid:markov key}) we can prove Lemma~\ref{lemma:appendix:iid:markov}. Due to similarities, we omit the details.

\begin{figure}[t!]
\centering
{\epsfig{figure=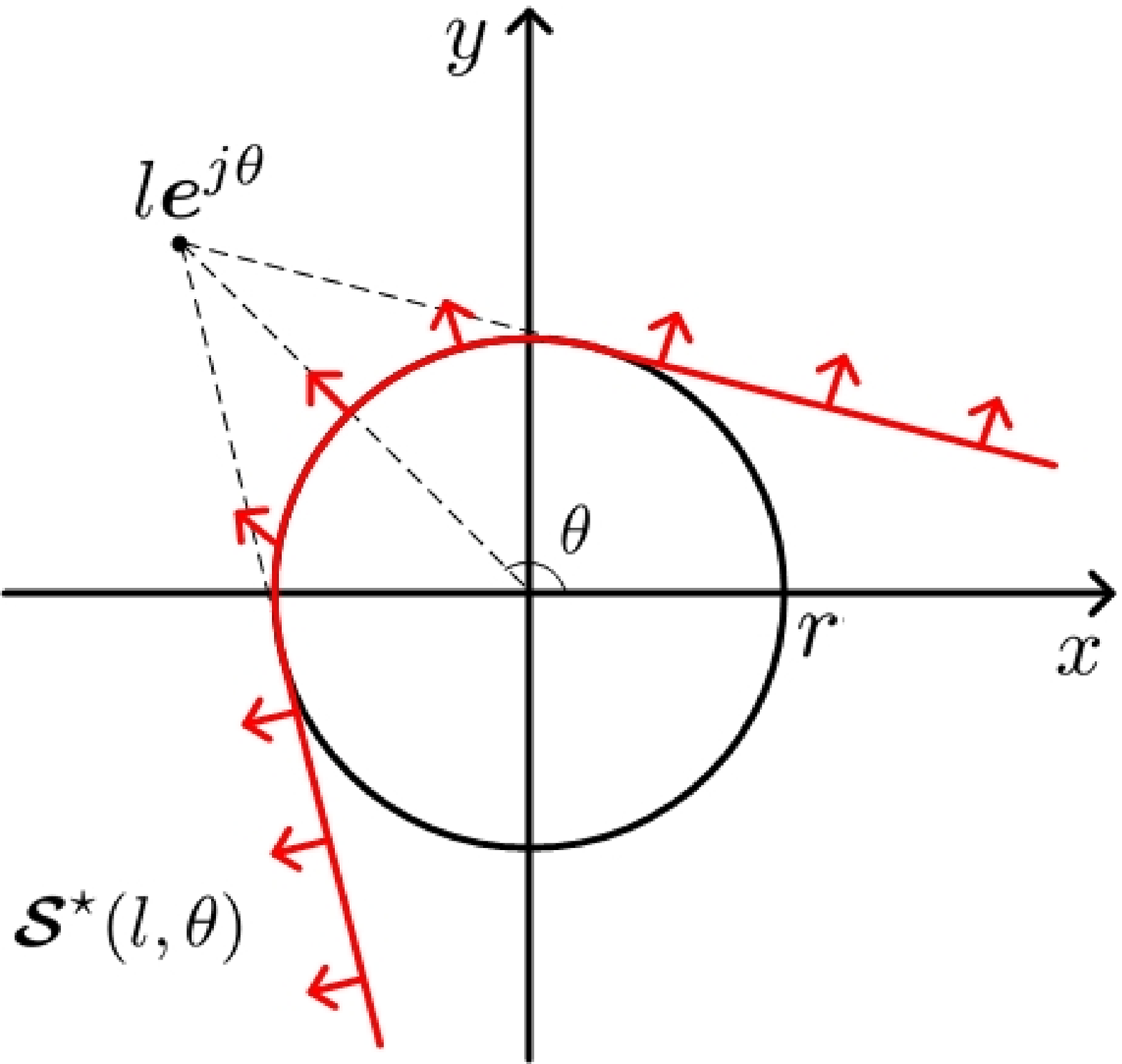,width=0.25\textwidth}}
\caption{An example of $\bm{\mathcal{S}}^\star(l,\theta)$: when $\theta = \frac{\pi}{2}$, $\bm{\mathcal{S}}^\star(l,\theta)$ is identical to $\bm{\mathcal{S}}^\star(l)$.}
\label{fig:example_S_star_theta}
\end{figure}

In the following, we show (\ref{eqn:appendix:iid:markov key}). We first consider the event $\{I(k) = 0\}$. Since (\ref{eqn:appendix:LF:diff location}) also holds for the \emph{i.i.d.} mobility model, by the same reason in the proof of Lemma~\ref{lemma:appendix:LF:markov}, the event $\{I(k) = 0\}$ occurs if and only if the following event occurs:
\begin{align}\label{eqn:appendix:iid:markov event}
\Big\{\overline{\big(\Delta\bm{X}(k-1), \Delta\bm{X}(k)\big)} \cap \bm{\mathcal{D}}_{r}(\bm{0}) = \varnothing\Big\}.
\end{align}
We next consider the event $\{I(k) = 0\}$ conditioned by $\Delta\bm{X}(k-1) = l \bm{e}^{j\theta}$ and $\mathcal{F}(k-1)$. Under these conditions, (\ref{eqn:appendix:iid:markov event}) is reduced to the following:
\begin{align*}
&\Big\{\overline{\big(\Delta\bm{X}(k-1), \Delta\bm{X}(k)\big)} \cap \bm{\mathcal{D}}_{r}(\bm{0})= \varnothing\Big\} \\
&\quad = \Big\{\overline{\big(l \bm{e}^{j\theta}, \Delta\bm{X}(k)\big)} \cap \bm{\mathcal{D}}_{r}(\bm{0}) = \varnothing\Big\} \\
&\quad = \Big\{\Delta\bm{X}(k)\in\bm{\mathcal{S}}^\star(l,\theta)\Big\},
\end{align*}
where
\begin{align*}
\bm{\mathcal{S}}^\star(l,\theta)
&\deq \big\{\bm{x}\in\mathbb{R}^2 \,\big|\, \overline{\big(l \bm{e}^{j\theta}, \bm{x}\big)} \cap \bm{\mathcal{D}}_{r}(\bm{0}) = \varnothing\big\}.
\end{align*}
An example of $\bm{\mathcal{S}}^\star(l,\theta)$ is shown in Fig.~\ref{fig:example_S_star_theta}. Hence, the probability on the left-hand side of (\ref{eqn:appendix:iid:markov key}) becomes
\begin{align}\label{eqn:appendix:iid:markov key -1}
&\p\{I(k) = 0 \,|\, \Delta\bm{X}(k\!-\!1) = l \bm{e}^{j\theta}, \mathcal{F}(k\!-\!1) \} \nonumber \\
&= \p\{\Delta\bm{X}(k) \!\in\!  \bm{\mathcal{S}}^\star(l,\theta)| \Delta\bm{X}(k\!-\!1) = l \bm{e}^{j\theta}, \mathcal{F}(k\!-\!1)\}.
\end{align}
By (i) in Lemma~\ref{lemma:appendix:iid:property of delta X}, $\Delta\bm{X}(k)$ is independent of $\Delta\bm{X}(k-1)$ and $\mathcal{F}(k-1)$, and thus we have
\begin{align}\label{eqn:appendix:iid:markov (i)}
&\p\{\Delta\bm{X}(k)\in\bm{\mathcal{S}}^\star(l,\theta) \,|\, \Delta\bm{X}(k-1) = l \bm{e}^{j\theta}, \mathcal{F}(k-1)\} \nonumber \\
&\quad = \p\{\Delta\bm{X}(k)\in\bm{\mathcal{S}}^\star(l,\theta)\}.
\end{align}
In addition, by (ii) in Lemma~\ref{lemma:appendix:iid:property of delta X},
\begin{align}\label{eqn:appendix:iid:markov (ii)}
\p\{\Delta\bm{X}(k)\in\bm{\mathcal{S}}^\star(l,\theta)\}
&= \p\{\Delta\bm{X}\in\bm{\mathcal{S}}^\star(l,\theta)\}.
\end{align}
Finally, by (iii) in Lemma~\ref{lemma:appendix:iid:property of delta X}, the probability in (\ref{eqn:appendix:iid:markov (ii)}) is invariant for any $\theta\in(0,2\pi]$. When $\theta = \frac{\pi}{2}$, we have $\bm{\mathcal{S}}^\star(l,\frac{\pi}{2}) = \bm{\mathcal{S}}^\star(l)$. Hence, the following holds for any $\theta\in(0,2\pi]$:
\begin{align}\label{eqn:appendix:iid:markov (iii)}
\p\{\Delta\bm{X}\in\bm{\mathcal{S}}^\star(l,\theta)\}
&= \p\{\Delta\bm{X}\in\bm{\mathcal{S}}^\star(l)\}.
\end{align}
Combining (\ref{eqn:appendix:iid:markov key -1}), (\ref{eqn:appendix:iid:markov (i)}), (\ref{eqn:appendix:iid:markov (ii)}), and (\ref{eqn:appendix:iid:markov (iii)}) gives (\ref{eqn:appendix:iid:markov key}). This completes the proof. \hfill $\blacksquare$

%------------------------------------------------
%
% Lemma
%
%------------------------------------------------
\begin{lemma}\label{lemma:appendix:iid:pdf}
Let $f_{|\Delta\bm{X}|}(\cdot)$ denote the probability density function of the random variable $|\Delta\bm{X}|$ under the \emph{i.i.d.} mobility model. Then, it is bounded by
\begin{align*}
f_{|\Delta\bm{X}|}(x) \le \frac{2x}{n}, \quad \text{ for all } x\in[0, 2\sqrt{n}].
\end{align*}
\end{lemma}

\noindent\textit{Proof:} We will prove this lemma by showing the following:
\begin{align}\label{eqn:appendix:iid:pdf_main}
\limsup_{\epsilon\downarrow 0}\frac{\p\{x-\frac{\epsilon}{2} \le |\Delta\bm{X}| \le x +\frac{\epsilon}{2}\}}{\epsilon} \le \frac{2x}{n}.
\end{align}
From (ii) in Lemma~\ref{lemma:appendix:iid:property of delta X}, we have $|\Delta\bm{X}| \ed |\Delta\bm{X}_{(i,j)}(0)|$. Hence, by conditioning on the values of $\bm{X}_i(0)$, the probability in (\ref{eqn:appendix:iid:pdf_main}) can be rewritten as
\begin{align*}
&\p\Big\{x-\frac{\epsilon}{2} \le |\Delta\bm{X}| \le x +\frac{\epsilon}{2}\Big\} \\
&= \int_{\bm{\mathcal{D}}} \p\Big\{x\!-\!\frac{\epsilon}{2} \!\le\! |\Delta\bm{X}_{(i,j)}(0)| \!\le\! x \!+\!\frac{\epsilon}{2}\Big| \bm{X}_i(0) \!=\! \bm{u}\Big\}\text{d}F_{\!\bm{X}_i(0)}(\bm{u}) \\
&= \int_{\bm{\mathcal{D}}} \p\{\bm{X}_j(0)\in\bm{\mathcal{R}}_{(x,\epsilon)}(\bm{u})\,|\, \bm{X}_i(0) = \bm{u}\}\,\text{d}F_{\bm{X}_i(0)}(\bm{u}),
\end{align*}
where $\bm{\mathcal{R}}_{(x,\epsilon)}(\bm{u}) \deq \{\bm{v}\in\mathbb{R}^2 \,|\,x-\frac{\epsilon}{2} \le |\bm{v}-\bm{u}|\le x+\frac{\epsilon}{2}\}$. By independence between $\bm{X}_i(0)$ and $\bm{X}_j(0)$, we further have
\begin{align}\label{eqn:appendix:iid:pdf_inside}
\begin{split}
&\p\Big\{x-\frac{\epsilon}{2} \le |\Delta\bm{X}| \le x +\frac{\epsilon}{2}\Big\} \\
&\quad = \int_{\bm{\mathcal{D}}} \p\{\bm{X}_j(0)\in\bm{\mathcal{R}}_{(x,\epsilon)}(\bm{u})\} \,\text{d}F_{\bm{X}_i(0)}(\bm{u}).
\end{split}
\end{align}
Note that, since $\bm{X}_j(0)\in\bm{\mathcal{D}}$ with probability 1 and $\bm{X}_j(0)\sim\text{Uniform}(\bm{\mathcal{D}})$, the probability $\p\{\bm{X}_j(0)\in\bm{\mathcal{R}}_{(x,\epsilon)}(\bm{u})\}$ in the integral in (\ref{eqn:appendix:iid:pdf_inside}) is given by
\begin{align}\label{eqn:appendix:iid:area(1)}
\p\{\bm{X}_j(0)\in\bm{\mathcal{R}}_{(x,\epsilon)}(\bm{u})\}
&=\frac{\text{Area}(\bm{\mathcal{D}}\cap\bm{\mathcal{R}}_{(x,\epsilon)}(\bm{u}))} {\text{Area}(\bm{\mathcal{D}})} \nonumber \\
&\le \frac{\text{Area}(\bm{\mathcal{R}}_{(x,\epsilon)}(\bm{u}))} {\pi n}.
\end{align}
In addition, for any $\bm{u}\in\bm{\mathcal{D}}$ and sufficiently small $\epsilon\,(>0)$, the area $\text{Area}(\bm{\mathcal{R}}_{(x,\epsilon)}(\bm{u}))$ is calculated as
\begin{align}\label{eqn:appendix:iid:area(2)}
\text{Area}(\bm{\mathcal{R}}_{(x,\epsilon)}(\bm{u}))
&= \begin{cases}
\pi(x+\frac{\epsilon}{2})^2 - \pi(x-\frac{\epsilon}{2})^2, & \text{ if } x >0, \\
\pi(x+\frac{\epsilon}{2})^2, & \text{ if } x = 0,
\end{cases} \nonumber \\
&= \begin{cases}
2\pi x \epsilon, & \text{ if } x >0, \\
\frac{\pi \epsilon^2}{4}, & \text{ if } x = 0.
\end{cases}
\end{align}
By applying the combined result of (\ref{eqn:appendix:iid:area(1)}) and (\ref{eqn:appendix:iid:area(2)}) to (\ref{eqn:appendix:iid:pdf_inside}), we have
\begin{align*}
&\p\Big\{x-\frac{\epsilon}{2} \le |\Delta\bm{X}| \le x +\frac{\epsilon}{2}\Big\}
\le \begin{cases}
\frac{2 x \epsilon}{n}, & \text{ if } x >0, \\
\frac{\epsilon^2}{4n}, & \text{ if } x = 0,
\end{cases}
\end{align*}
which gives
\begin{align*}
\limsup_{\epsilon\downarrow 0}\frac{\p\{x-\frac{\epsilon}{2} \le |\Delta\bm{X}| \le x +\frac{\epsilon}{2}\}}{\epsilon} &
\le \begin{cases}
\frac{2 x }{n} , & \text{ if } x >0, \\
0, & \text{ if } x = 0,
\end{cases} \\
&=\frac{2x}{n}.
\end{align*}
This proves the lemma. \hfill $\blacksquare$

%------------------------------------------------
%
% Lemma
%
%------------------------------------------------
\section*{Proof of Lemma~\ref{lemma:analysis:iid:property H(k,l)}}
\subsection*{A.~Proof of (i)}
Similarly to the proof of (i) in Lemma~\ref{lemma:analysis:LF:property H(k,l)}, we can prove (i) in Lemma~\ref{lemma:analysis:iid:property H(k,l)} by applying Lemma~\ref{lemma:appendix:iid:markov} with $k=1$, $l = l_0$, and $\mathcal{L}(0) = (r, 2\sqrt{n}]$. Due to similarities, we omit the details.

\subsection*{B.~Proof of (ii)}
Suppose $r < l_0 \le l_1 \le 2\sqrt{n}$. Then, it is obvious from the definition of $\bm{\mathcal{S}}^\star(\cdot)$ in (\ref{eqn:analysis:iid:definition S(l)}) that $\bm{\mathcal{S}}^\star(l_0) \subseteq \bm{\mathcal{S}}^\star(l_1)$ (See Fig.~\ref{fig:example_S_star}). Hence, we have
\begin{align*}
\p\{\Delta\bm{X}\in\bm{\mathcal{S}}^\star(l_0)\}
&\le \p\big\{\Delta\bm{X}\in\bm{\mathcal{S}}^\star(l_1)\big\},
\end{align*}
which is equivalent to $H(1,l_0) \le H(1,l_1)$ by (i) in Lemma~\ref{lemma:analysis:iid:property H(k,l)}.

\subsection*{C.~Proof of (iii)}
By (ii) in Lemma~\ref{lemma:analysis:iid:property H(k,l)}, we have $H(1,l_0) \le H(1, 2\sqrt{n}) \, (= \hat{P})$ for any $l_0\in(r,2\sqrt{n}]$.

\subsection*{D.~Proof of (iv)}
By following the approach in the proof of (iv) in Lemma~\ref{lemma:analysis:LF:property H(k,l)}, we can prove (iv) in Lemma~\ref{lemma:analysis:iid:property H(k,l)} based on Lemma~\ref{lemma:appendix:iid:markov} and (i) and (iii) in Lemma~\ref{lemma:analysis:iid:property H(k,l)}. Due to similarities, we omit the details.

\begin{figure}[t!]
\centering
{\epsfig{figure=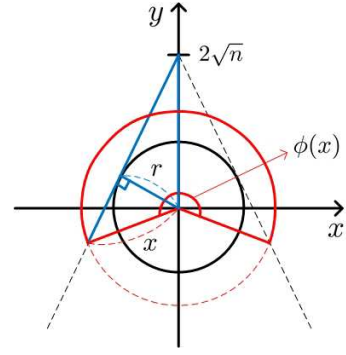,width=0.245\textwidth}}
\caption{The geometric definition of $\phi(x)$: it is the central angle of the arc $\{\bm{x} \in \bm{\mathcal{S}}^\star(2\sqrt{n}) | |\bm{x}|  = x\}$ depicted in red.}
\label{fig:definition_of_phi}
\end{figure}

\subsection*{E.~Proof of (v)}
By (i) in Lemma~\ref{lemma:analysis:iid:property H(k,l)} and (iii) in Lemma~\ref{lemma:appendix:iid:property of delta X}, we have
\begin{align}\label{eqn:appendix:iid:P^o:key}
\hat{P} &= \p\{\Delta\bm{X}\in\bm{\mathcal{S}}^\star(2\sqrt{n})\} \nonumber \\
&=\int_{r^+}^{2\sqrt{n}} \frac{\phi(x)}{2\pi}f_{|\Delta\bm{X}|}(x) \, \text{d}x,
\end{align}
where $\phi(x)$ is the central angle of the arc $\{\bm{x} \!\in\! \bm{\mathcal{S}}^\star(2\sqrt{n}) | |\bm{x}|  = x\}$ (See Fig.~\ref{fig:definition_of_phi}), and $f_{|\Delta\bm{X}|}(\cdot)$ is defined in Lemma~\ref{lemma:appendix:iid:pdf}. From the geometry in Fig.~\ref{fig:definition_of_phi}, the angle $\phi(x)$ is given by
\begin{align}\label{eqn:appendix:iid:P star bound:1:angle}
\phi(x) &= 2\cos^{-1}\Big(\frac{r}{2\sqrt{n}}\Big) + 2\cos^{-1}\Big(\frac{r}{x}\Big) \nonumber \\
&= 2\pi - 2\sin^{-1}\Big(\frac{r}{2\sqrt{n}}\Big) - 2\sin^{-1}\Big(\frac{r}{x}\Big),
\end{align}
where the second equality comes from the identity $\cos^{-1}(\theta) = \frac{\pi}{2} - \sin^{-1}(\theta)$ $(-\frac{\pi}{2} \le \theta \le \frac{\pi}{2})$. By substituting (\ref{eqn:appendix:iid:P star bound:1:angle}) into (\ref{eqn:appendix:iid:P^o:key}), we have
\begin{align}\label{eqn:appendix:iid:p^o:(1)}
\begin{split}
\hat{P} &= \Big(1-\frac{1}{\pi}\sin^{-1}\Big(\frac{r}{2\sqrt{n}}\Big)\Big)\cdot \p\{|\Delta\bm{X}| > r\} \\
&\qquad - \frac{1}{\pi}\int_{r^+}^{2\sqrt{n}} \sin^{-1}\Big(\frac{r}{x}\Big) f_{|\Delta\bm{X}|}(x) \, \text{d}x.
\end{split}
\end{align}
Based on (\ref{eqn:appendix:iid:p^o:(1)}), we derive an upper bound on $\hat{P}$ as follows:
\begin{align*}
\hat{P} &\le \Big(1-\frac{1}{\pi}\sin^{-1}\Big(\frac{r}{2\sqrt{n}}\Big)\Big)\cdot \p\{|\Delta\bm{X}| > r\} \\
&\le 1-\frac{1}{\pi}\sin^{-1}\Big(\frac{r}{2\sqrt{n}}\Big).
\end{align*}
This proves the upper bound in (\ref{eqn:analysis:iid:property P^o-1}).

Using (\ref{eqn:appendix:iid:p^o:(1)}) again, we derive a lower bound on $\hat{P}$ as follows: since $|\Delta\bm{X}| \ed |\Delta\bm{X}(0)|$ by (ii) in Lemma~\ref{lemma:appendix:iid:property of delta X} and $|\Delta\bm{X}(0)| = L(0)$ by definition, we have $\p\{|\Delta\bm{X}| > r\} =\p\{L(0) > r\}$. Hence, by Lemma~\ref{lemma:analysis:CCDF of T(0)}, the probability $\p\{|\Delta\bm{X}| > r\}$ in (\ref{eqn:appendix:iid:p^o:(1)}) is bounded by
\begin{align}\label{eqn:appendix:iid:p^o:(4)}
\p\{|\Delta\bm{X}| > r\} = P_{o} \ge 1 - \frac{r}{n^2}.
\end{align}
By Lemma~\ref{lemma:appendix:iid:pdf}, the integral in (\ref{eqn:appendix:iid:p^o:(1)}) is bounded by
\begin{align}\label{eqn:appendix:iid:p^o:(2)}
\int_{r^+}^{2\sqrt{n}} \sin^{-1}\Big(\frac{r}{x}\Big) f_{|\Delta\bm{X}|}(x) \, \text{d}x
&\le \frac{2}{n}\int_{r^+}^{2\sqrt{n}} \sin^{-1}\Big(\frac{r}{x}\Big) x\, \text{d}x.
\end{align}
Let $y \deq r/x$. By the change of variables, the integral on the right-hand side of (\ref{eqn:appendix:iid:p^o:(2)}) is solved as
\begin{align}\label{eqn:appendix:iid:p^o:(3)}
&\frac{2}{n}\int_{r^+}^{2\sqrt{n}} \sin^{-1}\Big(\frac{r}{x}\Big) x\, \text{d}x \nonumber \\
&\quad = -\frac{\pi r^2}{2n} + \frac{2r}{\sqrt{n}} \sqrt{1-\frac{r^2}{4n}} + 4\sin^{-1}\Big(\frac{r}{2\sqrt{n}}\Big).
\end{align}
By applying (\ref{eqn:appendix:iid:p^o:(4)}), (\ref{eqn:appendix:iid:p^o:(2)}), and (\ref{eqn:appendix:iid:p^o:(3)}) to (\ref{eqn:appendix:iid:p^o:(1)}), we have
\begin{align*}
\hat{P} &\ge \Big(1-\frac{1}{\pi}\sin^{-1}\Big(\frac{r}{2\sqrt{n}}\Big)\Big)\cdot \Big(1-\frac{r^2}{n}\Big) \\
& \qquad + \frac{r^2}{2n} -\frac{2r}{\pi\sqrt{n}}\sqrt{1-\frac{r^2}{4n}} - \frac{4}{\pi}\sin^{-1}\Big(\frac{r}{2\sqrt{n}}\Big) \\
& \ge 1 - \frac{r^2}{2n}- \frac{2r}{\pi\sqrt{n}} - \frac{5}{\pi}\sin^{-1}\Big(\frac{r}{2\sqrt{n}}\Big).
\end{align*}
This proves the lower bound in (\ref{eqn:analysis:iid:property P^o-2}).

%------------------------------------------------
%
% Lemma
%
%------------------------------------------------
\smallskip\smallskip
\section*{Proof of Lemma~\ref{lemma:analysis:iid:delay scaling}}
\noindent\emph{\underline{Order of $\bar{U}(1)$:}} To complete the proof of (\ref{eqn:analysis:iid:order2}), it remains to show $(1-\hat{P})^{-1} = \Theta(n^{1/2-\beta})$. For this, we will show the followings:
\begin{align*}
& \text{(i)}~(1-\hat{P})^{-1} = O(n^{1/2-\beta}), \\
& \text{(ii)}~(1-\hat{P})^{-1} = \Omega(n^{1/2-\beta}).
\end{align*}
Without loss of generality, we assume $r = n^\beta\,(0\le \beta \le 1/4)$ in the rest of this appendix. From (\ref{eqn:analysis:iid:property P^o-1}) in Lemma~\ref{lemma:analysis:iid:property H(k,l)} with $r = n^\beta$, we have for all $n\in\mathbb{N}$ the following:
\begin{align*}
1-\hat{P} \ge \frac{1}{\pi} \sin^{-1}\Big(\frac{n^{\beta-1/2}}{2}\Big).
\end{align*}
Since $\frac{n^{\beta-1/2}}{2} \in [0,1]$ for any $\beta\in[0,1/4]$ and $n\in\mathbb{N}$, we further have from the lower inequality in (\ref{eqn:appendix:LF:scaling 1-P^o - inv sin}) (i.e., $x \le \sin^{-1}\left(x\right)$ for any $x\in[0,1]$) that
\begin{align}\label{eqn:analysis:iid:scaling U(1):P 0}
1-\hat{P} \ge \frac{n^{\beta-1/2}}{2\pi}.
\end{align}
Hence, we have
\begin{align*}
\limsup_{n\to\infty} \frac{(1-\hat{P})^{-1}}{n^{1/2-\beta}}
&\le 2\pi < \infty,
\end{align*}
which proves (i) $(1-\hat{P})^{-1} = O(n^{1/2-\beta})$.

Using a similar approach as above, from (\ref{eqn:analysis:iid:property P^o-2}) in Lemma~\ref{lemma:analysis:iid:property H(k,l)} and the upper inequality in (\ref{eqn:appendix:LF:scaling 1-P^o - inv sin}) (i.e., $\sin^{-1}\left(x\right)\le \frac{\pi}{2}x$ for any $x\in[0,1]$), we have for all $n\in\mathbb{N}$ the following:
\begin{align}\label{eqn:analysis:iid:scaling U(1):P star}
1-\hat{P}
&\le \frac{n^{2\beta-1}}{2} + \frac{2 n^{\beta-1/2}}{\pi} + \frac{5}{\pi}\sin^{-1}\Big(\frac{n^{\beta-1/2}}{2}\Big) \nonumber \\
&\le \frac{n^{\beta-1/2}}{2} + \frac{2 n^{\beta-1/2}}{\pi} + \frac{5n^{\beta-1/2}}{4} \nonumber \\
&= \Big(\frac{2}{\pi}+\frac{7}{4}\Big) n^{\beta-1/2}.
\end{align}
Hence, we have
\begin{align*}
\limsup_{n\to\infty} \frac{n^{1/2-\beta}}{(1-\hat{P})^{-1}}
&\le \frac{2}{\pi}+\frac{7}{4} < \infty,
\end{align*}
which proves (ii) $(1-\hat{P})^{-1} = \Omega(n^{1/2-\beta})$.

\smallskip\smallskip
\noindent\emph{\underline{Order of $\bar{U}(\lceil\gamma r^2\rceil)$:}} To complete the proof of (\ref{eqn:analysis:iid:order3}), it remains to show $(1-(\hat{P})^{\lceil\gamma r^2\rceil})^{-1} = \Theta(n^{\max(0,1/2-3\beta)})$. For this, we will show the followings:
\begin{align*}
& \text{(iii)}~(1-(\hat{P})^{\lceil\gamma r^2\rceil})^{-1} =
O(n^{\max(0,1/2-3\beta)}), \\
& \text{(iv)}~(1-(\hat{P})^{\lceil\gamma r^2\rceil})^{-1} =
\Omega(n^{\max(0,1/2-3\beta)}).
\end{align*}
From (\ref{eqn:analysis:iid:scaling U(1):P 0}), we have
\begin{align}\label{eqn:analysis:iid:scaling U(2):P 0}
1-(\hat{P})^{\lceil\gamma r^2\rceil}
&\ge 1-\Big(1-\frac{n^{\beta-1/2}}{2\pi}\Big)^{\lceil\gamma r^2\rceil}.
\end{align}
To simplify (\ref{eqn:analysis:iid:scaling U(2):P 0}), we will use the following bound: for any $x\in[0,1]$ and $y > 0$,
\begin{align}\label{eqn:analysis:iid:scaling U(2):trick1}
1-x^{\lceil y \rceil}
&= (1-x)(1+x+\ldots + x^{\lceil y \rceil-1}) \nonumber \\
&\ge (1-x) \lceil y \rceil x^{\lceil y \rceil-1} \nonumber \\
&\ge (1-x) y x^{y}.
\end{align}
By applying (\ref{eqn:analysis:iid:scaling U(2):trick1}) with $x = 1-\frac{n^{\beta-1/2}}{2\pi}\,(\in[0,1])$ and $y = \gamma r^2 = \gamma n^{2\beta}\,(>0)$ to the right-hand side of (\ref{eqn:analysis:iid:scaling U(2):P 0}), we have
\begin{align*}
1-(\hat{P})^{\lceil\gamma r^2\rceil}
&\ge \frac{\gamma n^{3\beta-1/2}}{2\pi} \Big(1-\frac{n^{\beta-1/2}}{2\pi}\Big)^{\gamma n^{2\beta}}.
\end{align*}
Hence, we have
\begin{align}\label{eqn:analysis:iid:scaling U(2):P 0:limsup}
\limsup_{n\to\infty} \frac{\big(\!1-(\hat{P})^{\lceil\gamma r^2\rceil}\!\big)^{-1}}{n^{1/2-3\beta}}
&\le \frac{2\pi}{\gamma} \limsup_{n\to\infty} \Big(\!1\!-\!\frac{n^{\beta-1/2}}{2\pi}\!\Big)^{-\gamma n^{2\beta}}.
\end{align}
To obtain the order of $(1-\frac{n^{\beta-1/2}}{2\pi})^{-\gamma n^{2\beta}}$, we take a logarithm function on it and then analyze the limiting behavior:
\begin{align}\label{eqn:analysis:iid:scaling U(2):log}
&\lim_{n\to\infty} \log \Big(1\!-\!\frac{n^{\beta-1/2}}{2\pi}\Big)^{-\gamma n^{2\beta}} \nonumber \\
&\quad = \lim_{n\to\infty} \log \Big(1-\frac{n^{\beta-1/2}}{2\pi}\Big)^{n^{1/2-\beta} (-\gamma)n^{3\beta-1/2}} \nonumber \\
&\quad = - \gamma \lim_{n\to\infty} \bigg(n^{3\beta-1/2} \log \Big(1-\frac{n^{\beta-1/2}}{2\pi}\Big)^{n^{1/2-\beta}} \bigg) \nonumber \\
&\quad = - \gamma \lim_{n\to\infty} n^{3\beta-1/2} \cdot \lim_{n\to\infty} \log \Big(1-\frac{n^{\beta-1/2}}{2\pi} \Big)^{n^{1/2-\beta}} \nonumber \\
&\quad = \frac{\gamma}{2\pi}\lim_{n\to\infty} n^{3\beta-1/2}.
\end{align}
Hence, we have $\lim_{n\to\infty} \log (1-\frac{n^{\beta-1/2}}{2\pi})^{-\gamma n^{2\beta}} = 0$ for $\beta \in [0,1/6)$. That is,
\begin{align}\label{eqn:analysis:iid:scaling U(2):P 0:limsup:1}
& \lim_{n\to\infty} \Big(1-\frac{n^{\beta-1/2}}{2\pi}\Big)^{-\gamma n^{2\beta}} = 1.
\end{align}
By combining (\ref{eqn:analysis:iid:scaling U(2):P 0:limsup}) and (\ref{eqn:analysis:iid:scaling U(2):P 0:limsup:1}), for $\beta \in [0,1/6)$ we obtain
\begin{align*}
\limsup_{n\to\infty} \frac{\big(1-(\hat{P})^{\lceil\gamma r^2\rceil}\big)^{-1}}{n^{1/2-3\beta}}
&\le \frac{2\pi}{\gamma} < \infty,
\end{align*}
which results in
\begin{align}\label{eqn:analysis:iid:scaling U(2):P 0:1:final}
\big(1-(\hat{P})^{\lceil\gamma r^2\rceil}\big)^{-1} \!=\! O(n^{1/2-3\beta}), \quad \text{for } \beta \in [0,1/6).
\end{align}
From (\ref{eqn:analysis:iid:scaling U(2):P 0}), we have
$1-(\hat{P})^{\lceil\gamma r^2\rceil} \ge 1-(1-\frac{n^{\beta-1/2}}{2\pi})^{\gamma n^{2\beta}}.$
Hence, we have
\begin{align}\label{eqn:analysis:iid:scaling U(2):P 0:2:limsup}
\begin{split}
&\limsup_{n\to\infty} \frac{\big(1-(\hat{P})^{\lceil\gamma r^2\rceil}\big)^{-1}}{n^{0}} \\
&\quad \le \frac{1}{\lim_{n\to\infty} 1-\Big(1-\frac{n^{\beta-1/2}}{2\pi}\Big)^{\gamma n^{2\beta}}}.
\end{split}
\end{align}
The limit $\lim_{n\to\infty}(1-\frac{n^{\beta-1/2}}{2\pi})^{\gamma n^{2\beta}}$ is obtained from (\ref{eqn:analysis:iid:scaling U(2):log}) as follows:
\begin{align*}
\lim_{n\to\infty} \log \Big(1\!-\!\frac{n^{\beta\!-\!1/2}}{2\pi}\Big)^{\gamma n^{2\beta}}
&= -\lim_{n\to\infty} \log \Big(1\!-\!\frac{n^{\beta\!-\!1/2}}{2\pi}\Big)^{-\gamma n^{2\beta}} \\
&= \begin{cases}
-\frac{\gamma}{2\pi}, & \text{if } \beta = 1/6,\\
-\infty, & \text{if } \beta\in(1/6, 1/4].
\end{cases}
\end{align*}
That is,
\begin{align}\label{eqn:analysis:iid:scaling U(2):P 0:2:limsup:1}
\lim_{n\to\infty} \Big(1-\frac{n^{\beta-1/2}}{2\pi}\Big)^{\gamma n^{2\beta}}
&= \begin{cases}
\exp(-\frac{\gamma}{2\pi}), & \text{if } \beta = 1/6, \\
0, & \text{if } \beta\in(1/6, 1/4].
\end{cases}
\end{align}
By substituting (\ref{eqn:analysis:iid:scaling U(2):P 0:2:limsup:1}) into (\ref{eqn:analysis:iid:scaling U(2):P 0:2:limsup}), for $\beta \in [1/6,1/4]$ we obtain
\begin{align*}
\limsup_{n\to\infty} \frac{\big(\!1 \!-\! (\hat{P})^{\lceil\gamma r^2\rceil}\!\big)^{-1}}{n^{0}}
&\le \begin{cases}
\frac{1}{1-\exp(\!-\!\frac{\gamma}{2\pi})}, & \text{if } \beta = 1/6,\\
1, & \text{if } \beta\in(1/6, 1/4],
\end{cases} \\
& < \infty,
\end{align*}
which results in
\begin{align}\label{eqn:analysis:iid:scaling U(2):P 0:2:final}
\big(1-(\hat{P})^{\lceil\gamma r^2\rceil}\big)^{-1} = O(n^{0}), \quad \text{ for } \beta\in [1/6,1/4].
\end{align}
Combining (\ref{eqn:analysis:iid:scaling U(2):P 0:1:final}) and (\ref{eqn:analysis:iid:scaling U(2):P 0:2:final}) proves (iii).

From (\ref{eqn:analysis:iid:scaling U(1):P star}), we have
\begin{align}\label{eqn:analysis:iid:scaling U(2):P star}
1-(\hat{P})^{\lceil\gamma r^2\rceil}
&\le 1-\Big(1-\Big(\frac{2}{\pi}+\frac{7}{4}\Big)n^{\beta-1/2}\Big)^{\lceil\gamma r^2\rceil}.
\end{align}
To simplify (\ref{eqn:analysis:iid:scaling U(2):P star}), we will use the following bound: for any $x\in[0,1]$ and $y > 0$,
\begin{align}\label{eqn:analysis:iid:scaling U(2):trick2}
1-x^{\lceil y \rceil}
&= (1-x)(1+x+\ldots + x^{\lceil y \rceil-1}) \nonumber \\
&\le (1-x) \lceil y \rceil.
\end{align}
By applying (\ref{eqn:analysis:iid:scaling U(2):trick2}) with $x = 1-(\frac{2}{\pi}+\frac{7}{4})n^{\beta-1/2}\,(\in[0,1])$ and $y = \gamma r^2 = \gamma n^{2\beta}\,(>0)$ to the right-hand side of (\ref{eqn:analysis:iid:scaling U(2):P star}), we have
\begin{align*}
1-(\hat{P})^{\lceil \gamma r^2 \rceil}
&\le \Big(\frac{2}{\pi}+\frac{7}{4}\Big) n^{\beta-1/2} \lceil \gamma n^{2\beta} \rceil.
\end{align*}
Hence, we have
\begin{align*}
\limsup_{n\to\infty} \frac{n^{1/2-3\beta}} {\big(1-(\hat{P})^{\lceil \gamma r^2 \rceil}\big)^{-1}}
&\le \Big(\frac{2}{\pi}+\frac{7}{4}\Big) \lceil \gamma \rceil < \infty,
\end{align*}
which results in
\begin{align}\label{eqn:analysis:iid:scaling U(2):P star:1:final}
\big(1-(\hat{P})^{\lceil\gamma r^2\rceil}\big)^{-1} = \Omega(n^{1/2-3\beta}), \quad \text{for } \beta\in[0,1/4].
\end{align}
In addition, since $1-(\hat{P})^{\lceil \gamma r^2 \rceil} \le 1$, we have
\begin{align*}
\limsup_{n\to\infty} \frac{n^{0}} {\big(1-(\hat{P})^{\lceil \gamma r^2 \rceil}\big)^{-1}}
&\le 1 < \infty,
\end{align*}
which results in
\begin{align}\label{eqn:analysis:iid:scaling U(2):P star:2:final}
\big(1-(\hat{P})^{\lceil \gamma r^2 \rceil}\big)^{-1} = \Omega(n^{0}), \quad \text{ for } \beta\in[0,1/4].
\end{align}
Combining (\ref{eqn:analysis:iid:scaling U(2):P star:1:final}) and (\ref{eqn:analysis:iid:scaling U(2):P star:2:final}) proves (iv).

\smallskip\smallskip
\noindent\emph{\underline{Order of $\p\{B_{(n-2,P_{o}^c)} \le \lceil\gamma r^2\rceil-2\}$:}} By Chernoff's inequality, the lower tail of the distribution function of the binomial random variable $B_{(n-2,P_{o}^c)}$ for $x \le (n-2)P_{o}^c$ is bounded by
\begin{align}\label{eqn:appendix:iid:chernoff}
\p\{B_{(n-2,P_{o}^c)} \le x\}
&\le \exp\Big(-\frac{((n-2)P_{o}^c-x)^2}{2 (n-2)P_{o}^c}\Big).
\end{align}
From Lemma~\ref{lemma:analysis:CCDF of T(0)}, we have $P_{o}^c=1-P_{o}\ge \frac{r^2}{3n}$. Suppose $0 < \gamma < \frac{1}{3}$ and $n \ge \frac{2}{1-3\gamma}$ (or, equivalently, $n-2 \ge 3\gamma n$). Then, we have $(n-2)P_{o}^c \ge (n-2)\frac{r^2}{3n} \ge 3\gamma n\frac{r^2}{3n} = \gamma r^2$. Hence, we can apply $x = \gamma r^2$ to (\ref{eqn:appendix:iid:chernoff}) under the conditions $0 < \gamma < \frac{1}{3}$ and $n \ge \frac{2}{1-3\gamma}$, and we obtain
\begin{align}\label{eqn:appendix:iid:chernoff:1}
\p\{B_{(n-2,P_{o}^c)} \!\le\! \gamma r^2\}
&\le \exp\Big(\!-\!\frac{((n-2)P_{o}^c\!-\!\gamma r^2)^2}{2(n-2)P_{o}^c}\Big).
\end{align}
Since $P_{o}^c \ge \frac{r^2}{3n}$ and $n-2 \ge 3\gamma n$, the term $(n-2)P_{o}^c-\gamma r^2$ in (\ref{eqn:appendix:iid:chernoff:1}) is bounded below by
\begin{align*}
(n\!-\!2)P_{o}^c\!-\!\gamma r^2
&\ge (n\!-\!2)\frac{r^2}{3n}\!-\!\gamma r^2 \!=\! \frac{n\!-\!2\!-\!3\gamma n}{3n}r^2 \,(\ge 0),
\end{align*}
from which we have
\begin{align}\label{eqn:appendix:iid:chernoff:arg1}
\big((n-2)P_{o}^c-\gamma r^2\big)^2 \ge \Big(\frac{n-2-3\gamma n}{3n}\Big)^2r^4.
\end{align}
From Lemma~\ref{lemma:analysis:CCDF of T(0)}, we also have $P_{o}^c=1-P_{o} \le \frac{r^2}{n}$. Hence, the term $2(n-2)P_{o}^c$ in (\ref{eqn:appendix:iid:chernoff:1}) is bounded above by $2(n-2)P_{o}^c \le 2(n-2)\frac{r^2}{n} \le 2r^2$,
from which we have
\begin{align}\label{eqn:appendix:iid:chernoff:arg2}
\frac{1}{2(n-2)P_{o}^c} \ge \frac{1}{2 r^2}.
\end{align}
Thus, by (\ref{eqn:appendix:iid:chernoff:arg1}) and (\ref{eqn:appendix:iid:chernoff:arg2}), the argument of the exponential function in (\ref{eqn:appendix:iid:chernoff:1}) is bounded below by
\begin{align*}
\frac{\left((n-2)P_{o}^c-\gamma r^2\right)^2}{2(n-2)P_{o}^c} \ge \frac{1}{2}\Big(\frac{n-2-3\gamma n}{3n}\Big)^2 r^2,
\end{align*}
which gives an upper bound on (\ref{eqn:appendix:iid:chernoff:1}) as
\begin{align*}
\p\{B_{(n-2,P_{o}^c)} \le \gamma r^2\}
&\le \exp\left(-\frac{1}{2}\Big(\frac{n-2-3\gamma n}{3n}\Big)^2 r^2\right).
\end{align*}
Since $\exp(-x) \le \frac{1}{x}$ for all $x> 0$, we further have
\begin{align*}
\p\{B_{(n-2,P_{o}^c)} \le \gamma r^2\}
&\le 2\Big(\frac{3n}{n-2-3\gamma n}\Big)^2 r^{-2}.
\end{align*}
Therefore, for $r = n^\beta$ we have
\begin{align*}
\limsup_{n\to\infty} \frac{\p\{B_{(n-2,P_{o}^c)} \le \gamma r^2\}}{n^{-2\beta}} \le \frac{18}{(1-3\gamma)^2} < \infty,
\end{align*}
which results in $\p\{B_{(n-2,P_{o}^c)} \le \gamma r^2\} = O(n^{-2\beta})$.
Since $\p\{B_{(n-2,P_{o}^c)} \le \lceil\gamma r^2\rceil-2\}
\le \p\{B_{(n-2,P_{o}^c)} \le \gamma r^2\}$, we have
\begin{align*}
\p\{B_{(n-2,P_{o}^c)} \le \lceil\gamma r^2\rceil-2\} &= O(n^{-2\beta}).
\end{align*}
This completes the proof.

\end{document}